\documentclass{article}
\usepackage[a4paper,top=4cm,bottom=4cm,left=2.5cm,right=2.5cm,bindingoffset=5mm]{geometry}
\usepackage[dvipsnames]{xcolor}

\usepackage[hyphens]{url}
\usepackage[breaklinks=true,colorlinks,linkcolor={blue},citecolor={red},urlcolor={purple},]{hyperref}

\usepackage[textwidth=2cm]{todonotes}

\usepackage[font=small,format=default,indention=2em,labelsep = period]{caption} 
\usepackage[subrefformat=parens,labelformat=parens]{subfig}
\usepackage{enumitem}
\usepackage{multirow}
\usepackage{pict2e,overpic}
\usepackage{booktabs}
\usepackage{tikz}
\tikzset{>=latex}
\usetikzlibrary{shapes.geometric, arrows}

\makeatletter
\def\namedlabel#1#2{\begingroup
    #2%
    \def\@currentlabel{#2}%
    \phantomsection\label{#1}\endgroup
}
\makeatother

\newlist{mylist}{enumerate*}{1}
\setlist[mylist]{label=(\alph*)}

\usepackage{amssymb,bm,amsmath,amsthm,accents}
\usepackage{graphicx,mathtools}	
\usetikzlibrary{calc}

\newcommand{\disp}{\displaystyle}     
\newcommand{\R}{\mathbb R}
\newcommand{\dt}{\Delta t}
\newcommand{\de}{\partial}        
\def\rhoc{\rho_{c}}
\def\rhof{\rho_{f}}
\def\rhom{\rho^{\mathrm{max}}}
\def\vmax{V^{\mathrm{max}}}
\def\qmax{Q^{\mathrm{max}}}
\def\dx{\Delta x}
\def\nx{N_{x}}
\def\nt{N_{t}}
\def\nr{N_{r}}
\def\dt{\Delta t}
\def\wl{w_{L}}
\def\wr{w_{R}}

\def\vr{V_\rho}

\def\umeno{U_{-}}
\def\upiu{U_{+}}
\def\luno{\lambda_{1}}
\def\ldue{\lambda_{2}}

\def\duno{d_{1}}
\def\ddue{d_{2}}
\def\stre{s_{3}}
\def\rettar{\mathrm{r}}
\def\rettas{\mathrm{s}}
\newcommand{\dem}[2]{d(#1,#2)}
\newcommand{\supp}[2]{s(#1,#2)}
\def\pos{\mathcal{P}}
\def\neg{\mathcal{N}}
\def\dex{\partial_{x}}
\def\det{\partial_{t}}
\def\betad{\beta_{d}}
\def\betaS{\hat\beta}
\def\bb{\bar\beta}

\def\edge{\mathcal{I}}
\def\vert{\mathcal{J}}

\newcommand{\rhor}[2]{\rho^{#1}_{#2}}
\newcommand{\wroad}[2]{w^{#1}_{#2}}

\def\zuno{z_{1}}
\def\zdue{z_{2}}
\def\quno{q_{1}}
\def\qdue{q_{2}}

\def\m{\frac{\vmax}{\rhom}}

\def\umeno{U^{-}}
\def\wumeno{\widetilde U^{-}}
\def\rhomeno{\rho^{-}}
\def\wrhomeno{\widetilde\rho^{-}}

\def\wmeno{w^{-}}

\def\upiu{U^{+}}

\def\rhopiu{\rho^{+}}

\def\wpiu{w^{+}}
\def\vpiu{v^{+}}
\def\umorto{U^{\dagger}}
\def\wumorto{\widetilde U^{\dagger}}
\def\rhomorto{\rho^{\dagger}}
\def\wrhomorto{\widetilde\rho^{\dagger}}

\def\wmorto{w^{\dagger}}
\def\umenoS{\hat U}
\def\rhomenoS{\hat\rho}
\def\qmenoS{\hat q}
\def\wmenoS{\hat w}

\def\upiuS{\hat U}
\def\rhopiuS{\hat\rho}
\def\qpiuS{\hat q}
\def\wpiuS{\hat w}

\def\funE{\mathcal{F}_{E}}
\def\funT{\mathcal{F}_{T}}
\def\funET{\mathcal{F}}

\def\km{\mathrm{km}}
\def\kmh{\mathrm{km/h}}
\def\vehkm{\mathrm{veh/km}}
\def\minute{\mathrm{min}}
\def\myhour{\mathrm{h}}

\def\mysecond{\mathrm{s}}

\def\mygram{\mathrm{g}}
\def\mymeter{\mathrm{m}}
\def\nox{\mathrm{NO_{x}}}

\let\oldparagraph=\paragraph
\renewcommand\paragraph[1]{\oldparagraph{#1.}}

\numberwithin{equation}{section}

\newtheorem{remark}{Remark}

\newtheorem{prop}{Proposition}
\newtheorem{defn}{Definition}

\newtheorem{lemma}{Lemma}

\numberwithin{remark}{section}
\numberwithin{theorem}{section}
\numberwithin{prop}{section}
\numberwithin{defn}{section}

\title{\Large\textbf{Emissions minimization on road networks\\
via Generic Second Order Models}}

\author{\normalsize{Caterina Balzotti}\thanks{Istituto per le Applicazioni del Calcolo ``M.\ Picone'', Consiglio Nazionale delle Ricerche, Rome, Italy (\href{mailto:c.balzotti@iac.cnr.it}{c.balzotti@iac.cnr.it}, \href{mailto:m.briani@iac.cnr.it}{m.briani@iac.cnr.it}).}
\and {\normalsize{Maya Briani}\footnotemark[1]}
\and {\setcounter{footnote}{3}\normalsize{Benedetto Piccoli}\thanks{Department of Mathematical Sciences, Rutgers University, Camden, USA (\href{mailto:piccoli@camden.rutgers.edu}{piccoli@camden.rutgers.edu}).}}
}
\date{\vspace{-2em}}

\begin{document}

\maketitle

\begin{abstract}
In this paper we consider the problem of estimating emissions due to vehicular traffic on complex networks, and minimizing their effect by regulating traffic at junctions. 
For the traffic evolution, we consider a Generic Second Order Model, 
which encompasses the majority of two-equations (i.e.\ second-order) models available in the literature, 
and extend it to road networks with merge and diverge junctions. 
The dynamics on the whole network is determined by selecting a solution 
to the Riemann Problems at junctions, i.e.\ the Cauchy problems with constant initial data on each incident road. 
The latter are solved assuming the maximization of the flow and 
assigning a traffic distribution coefficient for outgoing roads of diverges,
and a priority rule for incoming roads of merges.
A general emission model is considered and its parameters are tuned
to the $\nox$ emission rate. 
The minimization of emissions is then formulated in terms
of the traffic distribution and priority parameters,
taking into account travel times.
A comparison is provided between roundabouts with optimized parameters and traffic lights, which correspond to time-varying
traffic priorities.
Our approach can be adapted to manage traffic 
in complex networks in order to reduce emissions while
keeping travel time at acceptable levels.
\end{abstract}

\begin{description}
\item[\textbf{Keywords.}] Second order traffic models; road networks; Riemann problem; emissions.
\item[\textbf{Mathematics Subject Classification.}]  35L65, 90B20, 62P12.
\end{description}

\section{Introduction}\label{sec:intro}
The aim of this paper is to build a model to estimate and minimize traffic emissions  by regulating traffic dynamics. Such regulation corresponds to the choice of suitable model parameters,
which in turn represent traffic signals and traffic light timing. 
Specifically, we extend the Generic Second Order Model (briefly GSOM),
introduced in \cite{TGF07,LebacqueMammarHajSalem2007}, 
to road networks,
pair it to an emission model and then minimize a functional
comprising $\nox$ emissions and travel time.

Estimating traffic emissions is an important and challenging problem. 
First, most emission models are based on the knowledge of
vehicle speed and acceleration.
Thus, at macroscopic level, a first-order system based only
on conservation of cars, such as the Lighthill-Whitham-Richards (briefly LWR) model \cite{LighthillWhitham1955,Richards1956}, 
is not sufficient to feed an emission model.
It is necessary to consider a so-called second-order model, i.e.\ a model with two equations:
a first equation for the conservation of mass and a second for the 
conservation or balance of a modified momentum, which may model drivers' property.
The first second-order model goes back to Payne and Whitham \cite{Payne1971,Whitham1974}. After criticisms to the model,
see \cite{Daganzo1995},
a new line of research originated starting with the Aw-Rascle-Zhang (briefly ARZ) model \cite{AwRascle2000,Zhang2002}, which successfully addressed criticisms to the Payne-Whitham approach.
More recently, various second-order models were proposed
ranging from generalizations of the ARZ, such as in \cite{FanHertySeibold2013,GaravelloHanPiccoli2016}, 
to phase transition models as in \cite{BlandinWorkGoatinPiccoliBayen2011,Colombo2003}
%In particular, the Collapsed Generalized Aw-Rascle-Zhang model 
%\cite{FanYenPiccoliSeiboldWork2017}
%fits in the framework of Generic Second Order Models  \cite{LebacqueMammarHajSalem2007,TGF07}. Such models are characterized by a family of fundamental diagrams (density-flow graphs)
and GSOM in \cite{TGF07,LebacqueMammarHajSalem2007}.
Such models are characterized by a family of fundamental diagrams (density-flow functions) and, due to their multi-faceted nature,
are particularly appropriate to fit real traffic data. We refer to \cite{fan2015TRR,PiccoliKeFrieszYao2012} for more details on data-fitted second order models.

Traffic models on networks have been widely studied in last two decades and authors have considered many different traffic scenarios proposing a rich amount of alternative models at junctions.
The LWR model has been extended to road networks in several papers, see for example \cite{dellemonache2018CMS,GaravelloPiccoli2006,garavello2006AIHP,HoldenRisebro1995}. The ARZ model on networks was considered in 
\cite{GaravelloPiccoli2006AwRascle,herty2006NHM,herty2006SIAM} 
and phase-transition models in \cite{ColomboGoatinPiccoli2010,GaravelloPiccoli2013}.
In this paper we consider a road network with merge (two incoming and
one outgoing roads) and diverge (one incoming and two outgoing roads) junctions.
On each road, we assume that the traffic flow evolution is described by the GSOM%, introduced in \cite{LebacqueMammarHajSalem2007}:
\begin{equation}\label{eq:GSOM}
	\begin{cases}
		\det\rho+\dex(\rho v) = 0\\
		\det w+v\dex w = 0,\\
	\end{cases}
\end{equation}
where $\rho$ is the density of vehicles, $v = V(\rho,w)$ is the velocity function,  and $w$ is a property of drivers. Notice that the first equation in \eqref{eq:GSOM} models the conservation of cars, while the second
is the passive advection of the variable $w$, which gives rise to different fundamental diagrams.
To define the solution on the whole network we follow the approach proposed in \cite{GaravelloPiccoli2006} based on the concept of Riemann Problem at a junction,
which is a Cauchy problem with constant initial data on each road.
Solutions to Riemann Problems are required to maximize the flux while conserving the density $\rho$ 
and total property $y=\rho w$ through the junction.
To determine a unique solution to Riemann Problems, we need
to introduce additional criteria,
which depend on the type of junction.
For diverge junctions, 
a traffic distribution parameter is assigned to outgoing roads
as done in \cite{herty2006SIAM} for the ARZ model.
For merge junctions, a priority rule between incoming roads
is considered, as it was done for the LWR model in
\cite{CocliteGaravelloPiccoli2005}. 
More precisely, for a fixed priority parameter $\beta\in[0,1]$,
given the two incoming fluxes 
$\qpiuS_{1}$, $\qpiuS_{2}$, we require:
\begin{equation}\label{eq:beta-prop}
	(1-\beta)\ \qpiuS_{2} = \beta\ \qpiuS_{1}.
\end{equation}
Equation \eqref{eq:beta-prop}
establishes a proportional relationship between the two incoming fluxes.
For instance, if $\beta=0$ only traffic from the first road
is allowed and vice versa for $\beta=1$. Therefore, traffic lights
can be easily represented by time-varying priority parameters.
This rule, together with the maximization of flux and conservation of $\rho$ and $y$, determines unique values of the variable $w$ on each road. 
In fact, the value $\wmenoS_3$ on the outgoing road  is given by a convex combination of the values $\wpiuS_1$ and $\wpiuS_2$ of the two incoming roads, i.e. 
\begin{equation}\label{eq:w3-intro}
	\wmenoS_{3} = (1-\beta)\wpiuS_{1}+\beta \wpiuS_{2}.
\end{equation}
As a result, the maximal flux that can be received by the outgoing road,
i.e.\ the supply, depends on the priority rule. 
The final solution is determined by maximizing the flow
through the junction respecting the priority rule,
but relaxing the latter in case the supply exceeds the demand
from the road with higher priority.
In rough words, the supply is given to incoming roads according
to the priority rule and redistributed in case of surplus.
%This gives rise to new solutions which can describe different traffic scenarios. 
%Not always the priority and flow maximization can be fully satisfied together and the procedure favors one or the other with respect the traffic condition. 
%Therefore, \cate{for traffic lights} the procedure strictly respects the priority, while this is not the case for roundabouts.   
The complete procedure to build the solution for a merge junction is explained in details in Definition \ref{prop:sol}.

The solution on networks to GSOM is then used
to feed an emission model, focusing on the emission of nitrogen oxides ($\nox$). 
Several studies deal with estimating emissions from dynamic traffic models, see for instance \cite{alvarez2017JCAM, alvarez2018MCRF, balzotti2021DCDSB, garciachan2017ECMI, huang2021EP, bayen2014, vardoulakis2003AE} and references therein. In particular, in \cite{alvarez2018MCRF} the authors deal with minimizing emissions by acting on the parameters of the model, while in \cite{garciachan2017ECMI} the authors analyze the possible benefits on emissions deriving from the limitation of traffic. The interest on $\nox$ gases in our work is due to their negative effects on health \cite{Zhang2013} and to their connection with ozone \cite{atkinson1984CR}.
Minimizing only emissions would result in
extreme solutions blocking traffic, thus we consider a cost function including
a term measuring travel times. Therefore, we express
the cost of emissions and travel time over the whole network as: 
\begin{align*}
	\funET(\gamma) &= \sum_{r}\left(c_{1}\int\int E^{\gamma}_{r}(x,t)dxdt+c_{2}\int\int\frac{1}{\mathcal{V}^{\gamma}_{r}(x,t)}dxdt\right),
\end{align*}
where $E^\gamma_r$, respectively $\mathcal{V}^{\gamma}_{r}$, is the emission rate, respectively velocity, along the road $r$, while
$c_1$ and $c_2$ are weights. The functional $\funET$ depends on the parameter vector $\gamma$ governing the traffic dynamic, which is comprised of the traffic distribution and priority parameters.
Our interest is in minimizing $\funET(\gamma)$ and compare different type of intersections, such as traffic
lights and roundabouts.
%Specifically,$\funET(\gamma)$ changes with respect to the priority rule and to the presence of traffic lights on the network.  
%We then set up a minimization problem aimed at finding the optimal $\gamma$, i.e. priority rule or traffic light timing, to manage traffic. 
Due the the high nonlinearity of $\funET(\gamma)$, 
explicit analytical solutions can not be found in general. Therefore,
we resort to numerical optimization to compute the optimal vectors 
$\gamma$. First, we focus on a merge junction and compare
a priority-based junction with one regulated by a traffic light.
The latter corresponds to alternating the values $\beta=0$ and $\beta=1$ for the green and red phases. These cycles
are parameterized by  the green-phase duration $t_g$ 
and the red-phase duration $t_r$.
The numerical results show that it is possible to find an optimal $\beta$ and an optimal couple $(t_g, t_r)$, and that the two types of junctions perform similarly when minimizing 
emissions and travel time.\\
Next, we analyze how the solution to the minimization problem 
depends on the initial traffic state $(\rho,w)$. 
Here we interpret $w$ as drivers'  preferred speed: low values of $w$ correspond to slow drivers, and high values of $w$ to fast drivers. 
For the priority-ruled junction, the minimum of the functional 
is achieved by giving high priority to the incoming road with higher density and fast drivers. 
Similarly, for the traffic light, the road with higher density 
must have a longer green-phase, except for high congestion
when the opposite happens.
In the latter situation, the sensitivity with respect to $w$ is greater.

We then focus on a more complex situation of a roundabout
with two incoming and two outgoing roads. The roundabout
has four additional stretch of roads to connect incoming
to outgoing roads and form a circle.
As before we compare priority-based junctions with traffic lights,
by choosing optimally the priorities and the traffic light timing.
%The numerical tests show that traffic lights may result in a 70\% increase in emissions compared to the priority-based case. Then time-varying controls are considered: priorities can be modified every 20 minutes as well as the traffic light timing. The total amount of $\nox$ is reduced by 16\% w.r.t. the fixed controls for traffic lights, but increases of 25\% for priority-based case.
The numerical tests show that, when few vehicles enter the network, traffic lights produce lower emissions and travel times compared to the priority-based case. In congested situations, instead, the use of priorities produces higher levels of emissions but with shorter travel times w.r.t. traffic lights dynamics. 
It is worth to notice that traffic light timing can be easily adjusted
in time, while changing priority-based rule would be more challenging.
Overall, traffic lights outperform traffic signals in terms of emissions for roundabouts and perform better also taking into account travel times for low densities. Moreover, the optimal traffic light timing are more robust
for variation of the functional weights.
Interestingly, there is an increasing diffusion of
roundabouts in Europe and US given the expected better performance in terms of output.
This study shows that traffic signals should be added to roundabouts if one aims also at lowering emissions.
This is a first example of how the model can be used
to support decision makers for sustainable traffic management. 

\smallskip

The paper is organized as follows. In Section \ref{sec:cgarz} we define the GSOM and the Riemann problem at junctions. In Section \ref{sec:cgarzNetwork} we describe the solution to the Riemann problem for diverge and merge junctions. In Section \ref{sec:ottimizzazione} a functional is formulated to estimate emission rate and travel time,
while in Section \ref{sec:num_generale} we provide details for the numerical approach.  Sections \ref{sec:numerica} and \ref{sec:rotatoria} are devoted to the numerical tests for optimal controls and estimation of $\nox$ emissions. In Section \ref{sec:conclusioni} we draw our conclusions. Finally, in Appendix \ref{appendice} we report some additional numerical tests for the roundabout.

\section{The Riemann Problem for GSOM at a junction} \label{sec:cgarz}
In order to extend the GSOM model  to networks, one has to 
analyze the Riemann problem at a junction, i.e.\ the Cauchy problem
with constant initial data on each road incident to the junction.

Recall the GSOM model equations \eqref{eq:GSOM}. The variable $w$ parametrizes a family of fundamental diagrams $Q(\rho,w)=\rho V(\rho,w)$. The usual assumptions on $Q$ and $V$ are:
\begin{enumerate}[label=(H\arabic*),ref=\textup{(H\arabic*)}]
	\item\label{q1} $Q(0,w) = 0$ and $Q(\rhom(w),w) = 0$ for each $w$, where $\rhom(w)$ is the maximum density of vehicles for $Q(\cdot,w)$.
	\item\label{q2} $Q(\rho,w)$ is strictly concave with respect to $\rho$, i.e.\ $\frac{\de^{2}Q}{\de\rho^{2}}<0$.
	\item\label{q3} $Q(\rho,w)$ is non-decreasing with respect to $w$, i.e.\ $Q_{w}\geq0$.
	\item\label{v1} $V(\rho,w)\geq0$ for each $\rho$ and $w$.
	\item\label{v2} $V(\rho,w)$ is strictly decreasing with respect to $\rho$, i.e. $V_{\rho}<0$ for each $w$.
	\item\label{v3} $V(\rho,w)$ is non-decreasing with respect to $w$, i.e.\ $V_{w}\geq0$.
\end{enumerate}
From \ref{q2} and \ref{q3}, for every $w$ the curve $\rho\to Q(\cdot,w)$ has a unique point of maximum, denoted by $\sigma(w)$, and we set
$\qmax(w)=Q(\sigma(w),w)$. Moreover, when $\rho=0$ there is not a unique maximum velocity. For every $w$ we set $\vmax(w)=V(0,w)$.

The eigenvalues of \eqref{eq:GSOM} are
\begin{align}
	\lambda_1(\rho,w)&=V(\rho,w)+\rho V_\rho(\rho,w) \label{eq:l1}\\
	\lambda_2(\rho,w)&=V(\rho,w) \label{eq:l2}.
\end{align}
The concavity of the flux implies $\lambda_1\leq\lambda_2$ and $\lambda_1=\lambda_2$ if and only if $\rho=0$, thus for $\rho\ne 0$ the system is strictly hyperbolic. 
The eigenvectors associated to the eigenvalues are
\begin{align*}
	\gamma_1(\rho,w)=(\rho,\rho w) \quad\mbox{ and }\quad
	\gamma_2(\rho,w)=\disp\left(-\frac{1}{\rho}V_{w}(\rho,w),V_{\rho}(\rho,w)-\frac{1}{\rho^{2}}V_{w}(\rho,w)\right). \label{eq:g2}
\end{align*}
The first eigenvalue is genuinely nonlinear, i.e.\ $\nabla\lambda_1\cdot\gamma_1 \neq0$,
while the second one is linearly degenerate, i.e.\ $\nabla\lambda_2\cdot\gamma_2 =0$. Hence, the curves of the first family are 1-shocks or 1-rarefaction waves, while the curves of the second family are 2-contact discontinuities. 
Finally the Riemann invariants are
\begin{equation*}
	\begin{split}
	z_{1}(\rho,w)&=w\\
	z_{2}(\rho,w)&=V(\rho,w).
	\end{split}
\end{equation*}
The first Riemann invariant $\zuno$ is constant along 1-shock and 1-rarefaction waves, while the second Riemann invariant $\zdue$ is constant along the 2-contact discontinuities.

By defining the \emph{total property} $y = \rho w$, system \eqref{eq:GSOM} can be rewritten in conservative form as
$$\label{eq:GSOM2}
\begin{cases}
			\det\rho+\dex(\rho v) = 0\\
			\det y+\dex(y v) = 0\\
		\end{cases}
$$
where $v = V\Big(\rho,\frac{y}{\rho}\Big)$.

\medskip
We recall now the main definitions concerning traffic models on road networks and we refer to \cite{dellemonache2018CMS,GaravelloPiccoli2006,garavello2006AIHP,HoldenRisebro1995} for further details. 
A road is modeled by an interval $I=(a,b)\subset\R$, with possibly
$a=-\infty$ or $b=+\infty$. A junction $J$ is a collection of roads
$((I_1,\ldots, I_n),(I_{n+1},\ldots,I_{n+m}))$ where $I_1,\ldots,I_n$
are the incoming roads and $I_{n+1},\ldots,I_{n+m}$ are the outgoing ones.
We define a network as a couple $(\edge,\vert)$ where $\edge$ is a finite collection of roads $I_{r}$, and $\vert$ is a finite collection of junctions $J$.\\
On each road $I_{r}$, the traffic dynamic is described by a GSOM as
\begin{align}\label{eq:GSOMrete}
\begin{cases}
			\det\rho_{r}+\dex(\rho_{r} v_{r}) = 0\\
			\det y_{r}+\dex(y_{r} v_{r}) = 0\\
\end{cases}
\end{align}
with $v_{r} = V\left(\rho_{r},\frac{y_{r}}{\rho_{r}}\right)$,
for $x\in I_{r}$ and $t\geq0$. 
%Following \cite{herty2006SIAM,HoldenRisebro1995}, we consider a set of smooth test functions $\phi_{r}:I_{r}\times [0,+\infty)\to\R^{2}$ with compact support in $I_{r}=[a_{r},b_{r}]$ which are smooth also across each junction $J$, i.e.
%\begin{equation}\label{eq:phi}
%	\phi_{i}(b_{i})=\phi_{j}(a_{j}) \qquad\text{and}\qquad \dex\phi_{i}(b_{i})=\dex\phi_{j}(a_{j})
%\end{equation}
%for $i=1,\dots,n$ and $j=n+1,\dots,m$. We define a \emph{weak solution} of \eqref{eq:GSOMrete} as a couple of functions $(\rho_{r}(x,t),y_{r}(x,t))$ which satisfy
%\begin{equation}\label{eq:weak}
%\begin{split}
%	&\sum_{r=1}^{n+m}\left(\int_{0}^{\infty}\int_{a_{r}}^{b_{r}}(\rho_{r}(x,t)\det\phi_{r}(x,t)+(\rho_{r}(x,t)v_{r}(x,t))\dex\phi_{r}(x,t))dxdt+\int_{a_{r}}^{b_{r}}\rho_{r}(x,0)\phi_{r}(x,0)dx\right)=0\\
%	&\sum_{r=1}^{n+m}\left(\int_{0}^{\infty}\int_{a_{r}}^{b_{r}}(y_{r}(x,t)\det\phi_{r}(x,t)+(y_{r}(x,t)v_{r}(x,t))\dex y_{r}(x,t))dxdt+\int_{a_{r}}^{b_{r}}y_{r}(x,0)\phi_{r}(x,0)dx\right)=0
%\end{split}
%\end{equation}
%for all the test functions $\phi_{r}$ satisfying \eqref{eq:phi}, where $(\rho_{r}(x,0),y_{r}(x,0))$ is the initial data.
The construction of a solution on the whole network is obtained via wave-front tracking starting from solutions to Riemann problems to \eqref{eq:GSOMrete} 
at each junction. 
%$r$ with the following initial data
%\begin{equation*}
%	(\rho_{r}(x,0),y_{r}(x,0)) = \begin{cases}
%		(\rho^{-},y^{-}) &\quad\text{for $x<x_{0}$}\\
%		(\rho^{+},y^{+}) &\quad\text{for $x>x_{0}$,}
%	\end{cases}
%\end{equation*}
%where only one between the left $((\rho^{-},y^{-}))$ and right state is known. %Depending on $I_{r}$ if it is an incoming or an outgoing road, we 
%We have the following possibilities: if $I_{r}$ is an incoming road then $x_{0}=b_{r}$ and only the left state $(\rho^{-},y^{-})$ is known; in this case we look for solutions of \eqref{eq:GSOMrete} such that the waves have non-positive speed. If $I_{r}$ is an outgoing road then $x_{0}=a_{r}$ and only the right state $(\rho^{+},y^{+})$ is known; in this case we look for solutions of \eqref{eq:GSOMrete} such that the waves have non-negative speed.
More precisely, given constant initial data on each road, we look
for possible waves with negative speed for incoming roads and positive ones 
on outgoing roads. This is necessary to have conservation of mass
through the junction, see \cite{GaravelloPiccoli2006}.
%\note{see \cite{GaravelloPiccoli2016}Quale tra GaravelloPiccoli2006 e GaravelloHanPiccoli2016??}.
%In order to construct waves with non-negative or non-positive speed at the junction, 
To isolate the admissible waves, 
we study the sign of the eigenvalues \eqref{eq:l1} and \eqref{eq:l2}. By the concavity of the flux function, 
the first eigenvalue $\luno(\rho,w)=\rho+\rho\vr(\rho,w)=Q_{\rho}(\rho,w)$ 
satisfies  $\luno\geq0$ for $\rho\leq\sigma(w)$ and $\luno<0$ for $\rho>\sigma(w)$. 
%For each $w\in[\wl,\wr]$, for suitable $\wl$ and $\wr$, 
The second eigenvalue is given by $\ldue(\rho,w)=V(\rho,w)$, thus 
by \ref{v1} the speed of the 2-contact discontinuity is always non-negative. 

In order to describe the flux maximization, let us consider the \emph{supply} and \emph{demand} functions, see \cite{GaravelloHanPiccoli2016} for details and discussion. %For each \note{$w\in[\wl,\wr]$???}, 
The supply function $\supp{\rho}{w}$ is defined as
\begin{equation}\label{eq:supply}
\supp{\rho}{w} = \begin{cases}
\qmax(w) &\quad\text{if $\rho\leq\sigma(w)$}\\		
Q(\rho,w) &\quad\text{if $\rho>\sigma(w)$}
\end{cases} ,
\end{equation}
and the demand function $\dem{\rho}{w}$ as
\begin{equation}\label{eq:demand}
\dem{\rho}{w} = \begin{cases}
Q(\rho,w) &\quad\text{if $\rho\leq\sigma(w)$}\\
\qmax(w) &\quad\text{if $\rho>\sigma(w)$}	
\end{cases}.
\end{equation}

%%%----------------------------------------------------------------------%%%
\subsection{Incoming roads}
Let us consider an incoming road at a junction. 
Only waves with negative speed are admissible. Since $\ldue\geq0$, we can have only 1-shock or 1-rarefaction waves. 

We fix a left state $\umeno=(\rhomeno,\wmeno)$ and look for the set of all admissible right states $\upiuS=(\rhopiuS,\wpiuS)$ that can be connected to $\umeno$ with waves with negative speed.
Along the 1-waves the variable $w$ is conserved, therefore only the density $\rho$ changes. This case is analogous to the definition of admissible solutions on incoming roads for first order traffic models, see for instance \cite{GaravelloPiccoli2006}.

%Let us fix a left state $\umeno=(\rhomeno,\wmeno)$: our aim is to define the set of all possible right states $\upiuS=(\rhopiuS,\wpiuS)$ that can be connected to $\umeno$ with a wave with non-positive speed. We denote by $\neg(\umeno)$ the set of admissible densities. 
%
\begin{prop}\label{prop:incoming}
Let $V$ be a velocity function that verifies properties  \ref{v1}-\ref{v3} and 
let $\umeno=(\rhomeno,\wmeno)$ be a left state on an incoming road.\\
If $\rhomeno=0$, then the only admissible right state is $\upiuS=\umeno$.\\ 
If $\rhomeno\neq0$, then the set of possible right states $\upiuS=(\rhopiuS,\wpiuS)$ 
%that can be connected to the left state lies on the level curve $\{\zuno=\wmeno\}$, i.e. 
verifies $\wpiuS=\wmeno$ and:
\begin{enumerate}
\item If $\rhomeno\leq\sigma(\wmeno)$, then $\rhopiuS\in\neg(\umeno)=\{\rhomeno\}\cup(\wrhomeno(\wmeno),\rhom(\wmeno)]$, where $\wrhomeno(\wmeno)$ is the density such that $Q(\wrhomeno(\wmeno),\wmeno)=Q(\rhomeno,\wmeno)$.
\item If $\rhomeno>\sigma(\wmeno)$, then $\rhopiuS\in\neg(\umeno)=[\sigma(\wmeno),\rhom(\wmeno)]$.
\end{enumerate}
Moreover, denoting by $d$ the demand function defined in \eqref{eq:demand},  it holds
\begin{equation}\label{eq:dem}
Q(\rhopiuS,\wpiuS)\leq \dem{\rhomeno}{\wmeno}.
\end{equation}

\end{prop}

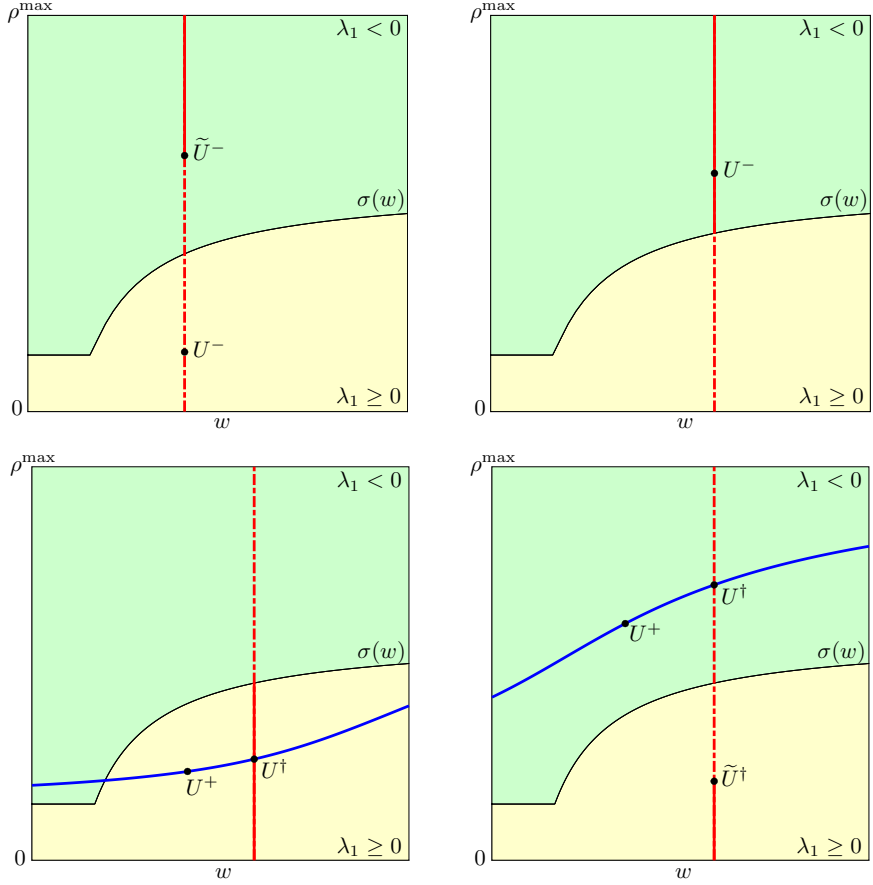
\begin{figure}[h!]
\centering
\normalsize
\scalebox{.85}{\begin{overpic}[width=0.4\columnwidth,trim=-1cm -1cm 0 0, clip]{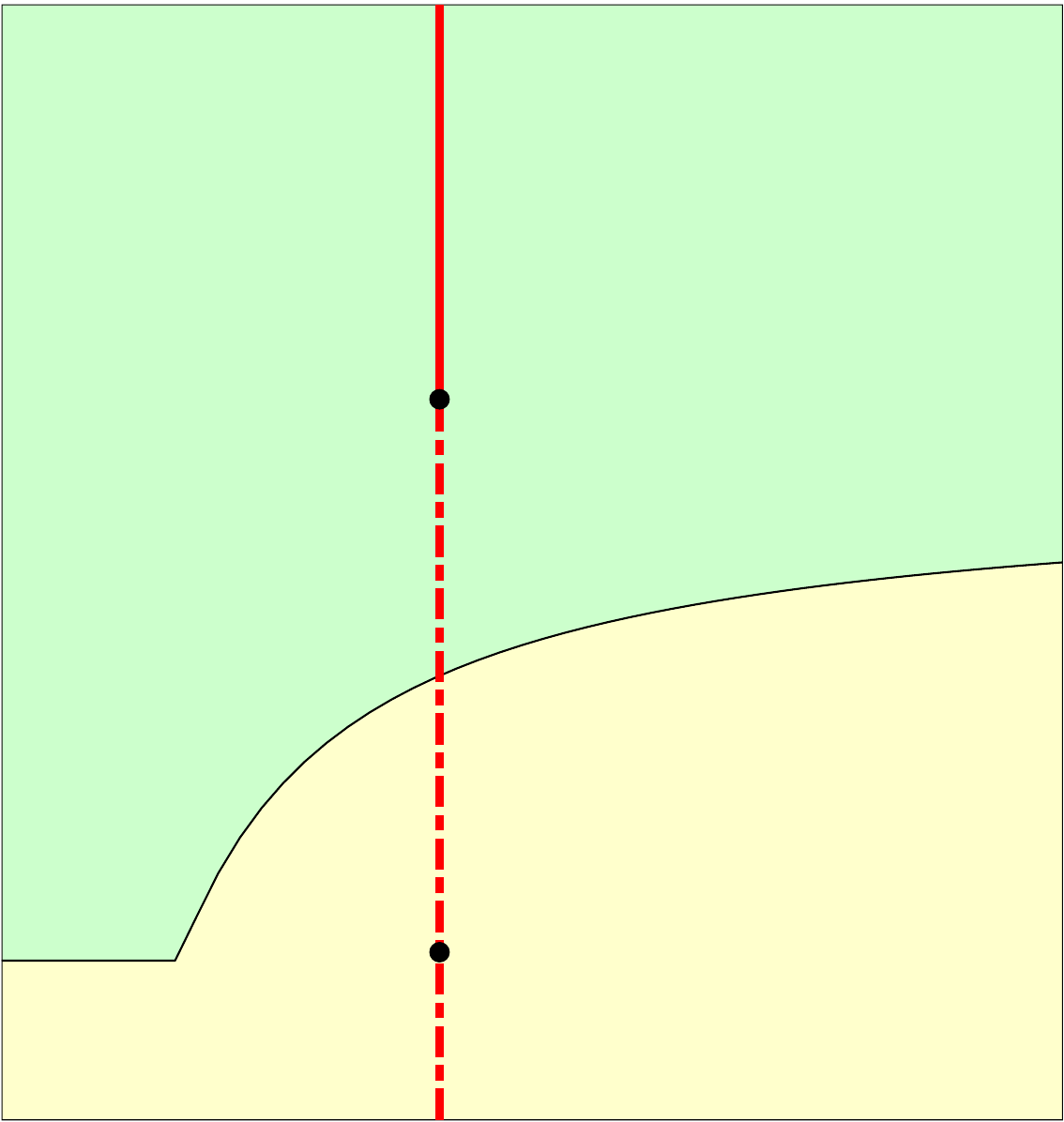}
\put(1,4){$0$} \put(0,99){$\rhom$} %\put(-1,18){$\rhof$}
%\put(4,1){$\wl$} \put(92,1){$\wr$}
\put(49,1){$w$}
\put(78,95){$\luno<0$} \put(78,7){$\luno\geq0$}
\put(44,18){$\umeno$} \put(44,65){$\wumeno$}
\put(83,54){$\sigma(w)$}
\end{overpic}}
\qquad
\scalebox{.85}{\begin{overpic}[width=0.4\columnwidth,trim=-1cm -1cm 0 0, clip]{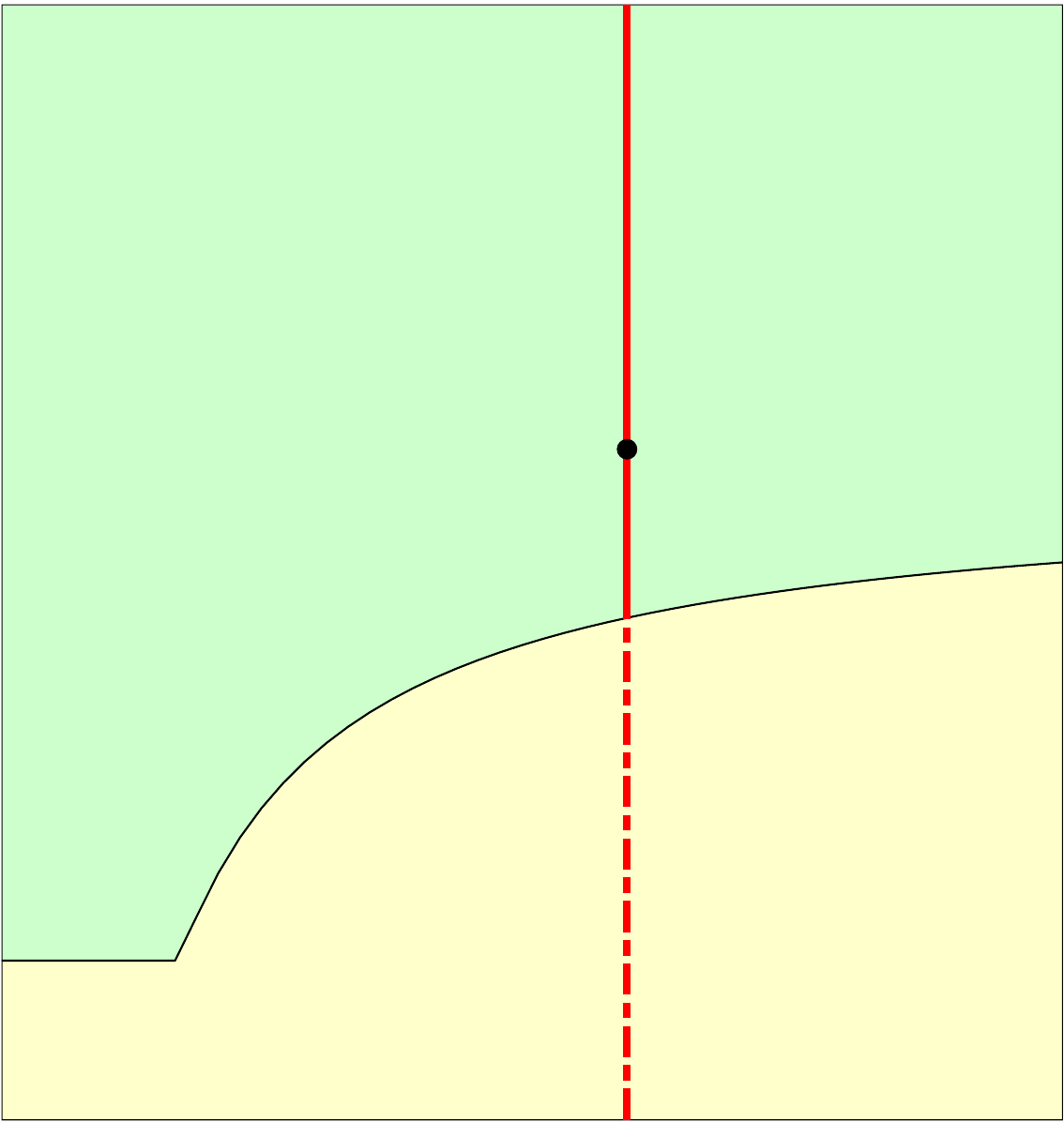}
\put(1,4){$0$} \put(0,99){$\rhom$} %\put(-1,18){$\rhof$}
%\put(4,1){$\wl$} \put(92,1){$\wr$}
\put(49,1){$w$}
\put(78,95){$\luno<0$} \put(78,7){$\luno\geq0$}
\put(60,61){$\umeno$} 
\put(83,54){$\sigma(w)$}
\end{overpic}}\bigskip\\
\scalebox{.85}{\begin{overpic}[width=0.4\columnwidth,trim=-1cm -1cm 0 0, clip]{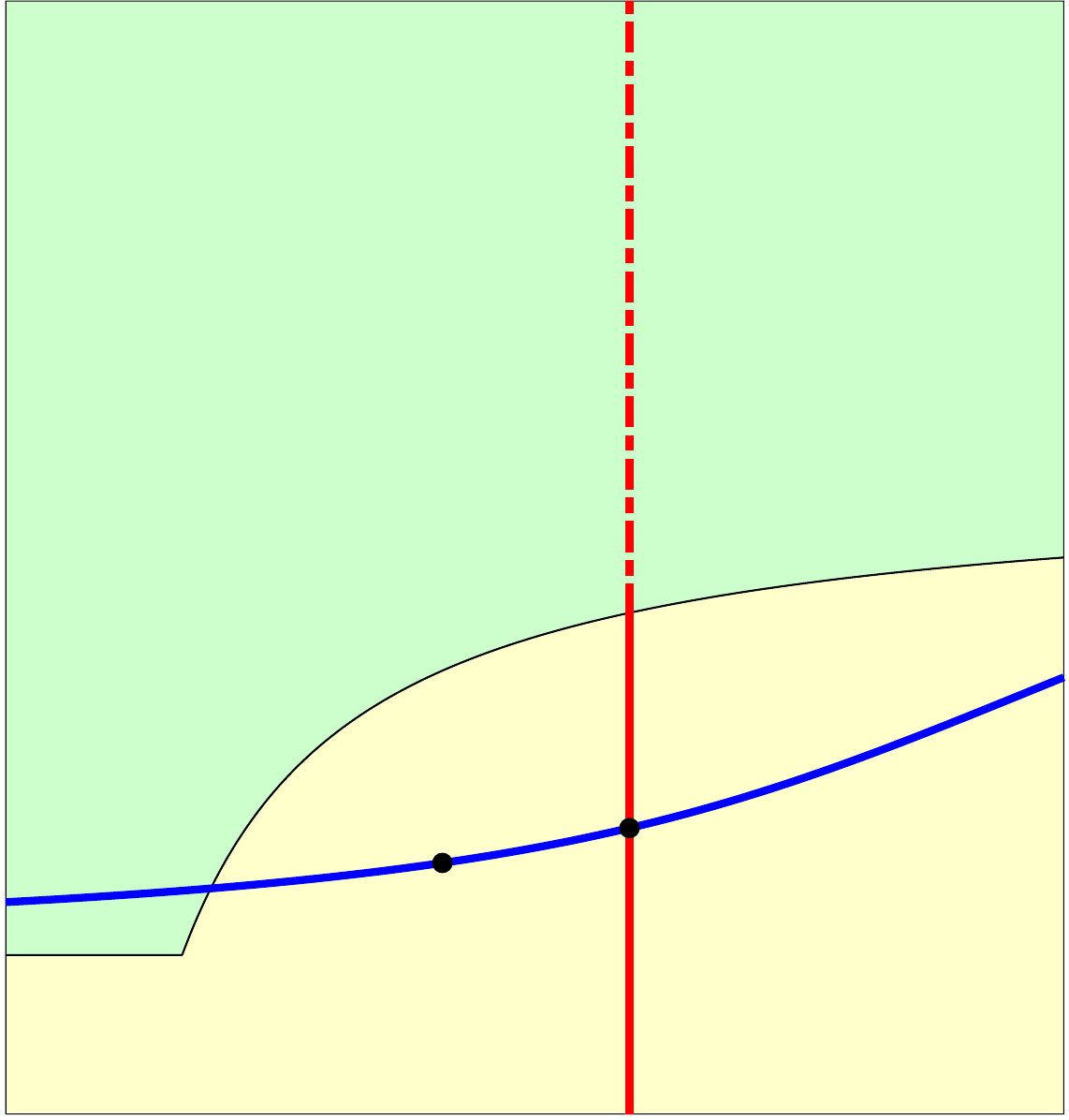}
\put(1,4){$0$} \put(0,99){$\rhom$} %\put(-1,18){$\rhof$}
%\put(4,1){$\wl$} \put(92,1){$\wr$}
\put(49,1){$w$}
\put(78,95){$\luno<0$} \put(78,7){$\luno\geq0$}
\put(42,21){$\upiu$} \put(60,25){$\umorto$}
\put(83,54){$\sigma(w)$}
\end{overpic}}
\qquad
\scalebox{.85}{\begin{overpic}[width=0.4\columnwidth,trim=-1cm -1cm 0 0, clip]{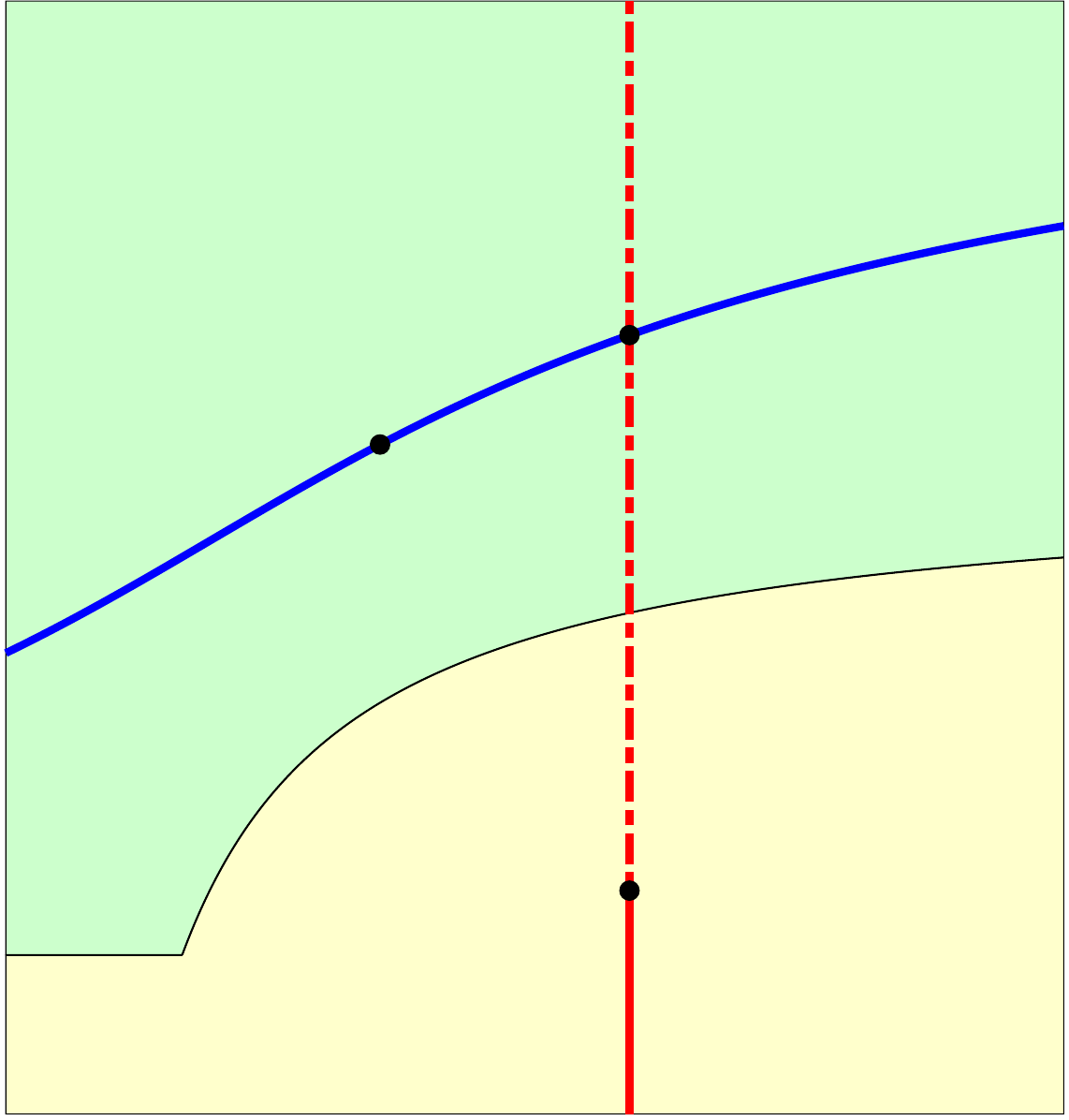}
\put(1,4){$0$} \put(0,99){$\rhom$} %\put(-1,18){$\rhof$}
%\put(4,1){$\wl$} \put(92,1){$\wr$}
\put(49,1){$w$}
\put(78,95){$\luno<0$} \put(78,7){$\luno\geq0$}
\put(38,58){$\upiu$} \put(60,23){$\wumorto$} \put(60,67){$\umorto$}
\put(83,54){$\sigma(w)$}
\end{overpic}}
\caption{The function graphs refer to the CGARZ model \cite{FanSunPiccoliSeiboldWork2017}
with the family of flux functions defined in \eqref{eq:Qcgarz}-\eqref{eq:Qc}. 
Top: two possible configurations of incoming road states. The red solid line identifies the set of all possible right states $\upiuS$ reachable from the left state $\umeno$.
Bottom: two possible configurations of the state on an outgoing road. The red solid line identifies the set of possible left states $\umenoS$ reachable from the right state $\upiu$.%In all plots, the black line represents the function $\sigma(w)$.
}
\label{fig:inEout}
\end{figure}

\begin{proof}
First assume 
%$\umeno=(\rhomeno,\wmeno)$ with 
$\rhomeno\neq0$.
If $\rhomeno\leq\sigma(\wmeno)$ (Figure \ref{fig:inEout} top-left) 
to have $\lambda_1\leq 0$ there are two possibilities: either $\upiuS=\umeno$, or moving above the density value $\wrhomeno(\wmeno)>\sigma(\wmeno)$ by a jump with zero speed. Indeed, since $Q(\wrhomeno(\wmeno),\wmeno)=Q(\rhomeno,\wmeno)$, the Rankine-Hugoniot condition $s(\wumeno-\umeno) = Q(\wrhomeno(\wmeno),\wmeno)-Q(\rhomeno,\wmeno)$ implies that the speed of the discontinuity $s$ is zero. In this case we can move with a 1-shock with negative speed towards any right state $\upiuS$ with $\wpiuS=\wmeno$ and $\wrhomeno(\wmeno)<\rhopiuS\leq\rhom(\wmeno)$. If $\rhomeno=0$ then $\wrhomeno(\wmeno)=\rhom(\wmeno)$, therefore  the solution is $\upiuS=\umeno$.

If $\rhomeno>\sigma(\wmeno)$, every state $\upiuS$ with $\wpiuS=\wmeno$ and $\rhopiuS\in[\sigma(\wmeno),\rhom(\wmeno)]$ is connected to $\umeno$ with waves with negative speed (Figure \ref{fig:inEout} top-right). In particular, we have a 1-rarefaction wave if  $\rhopiuS\leq\rhomeno$ and a 1-shock if $\rhopiuS>\rhomeno$.

\end{proof}

\subsection{Outgoing roads}
Let us consider an outgoing road at a junction. We are interested in the waves with positive speed, thus we can have a 1-shock or 1-rarefaction wave and a 2-contact discontinuity.

We fix a right state $\upiu=(\rhopiu,\wpiu)$ and look for the set of all admissible left states $\umenoS=(\rhomenoS,\wmenoS)$ that can be connected to $\upiu$ with waves with positive speed.
We emphasize that along the 1-waves the $w$ is conserved and only the density $\rho$ changes. We therefore assume that the value $\wmeno$ is given and depends on the states on the incoming roads (see Section \ref{sec:cgarzNetwork}). 
On the other hand, along the 2-wave the velocity $V(\rho,w)$ is conserved. Then, the definition of the admissible states $\umenoS$ depends on the existence of an intermediate point 
$\umorto=(\rhomorto,\wmorto)$ such that $\wmorto=\wmeno$ and $V(\rhomorto,\wmorto)=V(\rhopiu,\wpiu)$.

\begin{prop}\label{prop:Umorto}
Let $V$ be a velocity function that verifies properties  \ref{v1}-\ref{v3}. For a given value $\wmeno$ and a given right state $\upiu=(\rhopiu,\wpiu)$ with associated velocity $\vpiu=V(\rhopiu,\wpiu)$, if $\vpiu\leq\vmax(\wmeno)$ then there exists a unique point $\umorto=(\rhomorto,\wmorto)$ such that $\wmorto=\wmeno$ and $V(\rhomorto,\wmorto)=\vpiu$.
\end{prop}
\begin{proof}
If $\vpiu\leq\vmax(\wmeno)$ then the equation $V(\rho,\wmeno)=\vpiu$ admits a solution. 
By \ref{v2}, $\partial_\rho V<0$ and, by the implicit function theorem,
there exists $\rho(w;\vpiu)$ such that
$V(\rho(w;\vpiu),w)=\vpiu$. Moreover, \ref{v2}-\ref{v3} imply 
$$
\frac{d\rho}{dw}\,(w;\vpiu)=-\partial_w V/\partial_\rho V \geq 0.
$$
We then have $\wmorto=\wmeno$ and $\rhomorto = \rho(\wmeno;\vpiu)$.
\end{proof}

\begin{prop}\label{prop:outgoing}
Let $V$ be a velocity function that verifies properties  \ref{v1}-\ref{v3},
$\upiu=(\rhopiu,\wpiu)$ a right state on an outgoing road, and $\vpiu=V(\rhopiu,\wpiu)$ the associated velocity.
%Let $\{\zuno(U)=\wmeno \}$ be the level curve of the first Riemann invariant given by $w=\wmeno$.
% with $\wmeno$ a given constant value.
A left state $\umenoS=(\rhomenoS,\wmenoS)$, which can be connected to $\upiu$ with positive speed waves,
satisfies $\wmenoS = \wmeno$ and the following.
\begin{itemize}
\item[(i)] If $v^+ \leq \vmax(\wmeno)$, let %there exists the intersection point 
$\umorto=(\rhomorto,\wmorto)$ be the intersection point between 
the level curves $\{\zdue=\vpiu\}$ and $\{\zuno=\wmeno\}$, then $\wmorto=\wmeno$ and
\begin{enumerate}
\item if $\rhomorto\leq\sigma(\wmorto)$, then $\rhomenoS\in\pos(\upiu)=[0,\sigma(\wmorto)]$;
\item if $\rhomorto>\sigma(\wmorto)$, then $\rhomenoS\in\pos(\upiu)=[0,\wrhomorto(\wmorto))\cup\{\rhomorto\}$, where $\wrhomorto(\wmorto)$ is the density such that $Q(\wrhomorto(\wmorto),\wmorto)=Q(\rhomorto,\wmorto)$.
\end{enumerate}
\item[(ii)] If $v^+ > \vmax(\wmeno)$ then $\rhomenoS\in\pos(\upiu)=[0,\sigma(\wmeno)]$.
%$\rhomorto=0$, $\wmorto=\bar w$ and $\rhomenoS\in\pos(\upiu)=[0,\sigma(\wmorto)]$.
\end{itemize}
Moreover, denoting by $s$ the supply function defined in \eqref{eq:supply}, it holds 
\begin{equation}\label{eq:sup}
Q(\rhomenoS,\wmenoS)\leq \supp{\rhomorto}{\wmorto}.
\end{equation}
\end{prop}
\begin{proof}
If $\vpiu\leq\vmax(\wmeno)$, by Proposition \ref{prop:Umorto} there exists a unique point $\umorto$ such that $\wmorto = \wmeno$ and $V(\rho,\wmorto)=\vpiu$. Thus,
if $\rhomorto\leq\sigma(\wmorto)$, then every state $\umenoS$ with $\wmenoS=\wmeno$ and $\rhomenoS\in[0,\sigma(\wmorto)]$ 
can be connected to $\umorto$ by waves with positive speed (Figure \ref{fig:inEout} bottom-left). In particular we have a 1-rarefaction wave if  $\rhomorto\leq\rhomenoS$ and a 1-shock if $\rhomorto>\rhomenoS$. Then, $\umorto$ is connected to $\upiu$ by a 2-contact discontinuity which has positive speed. 

If $\rhomorto>\sigma(\wmorto)$ (Figure \ref{fig:inEout} bottom-right), we have two possibilities: no wave, then $\umenoS=\umorto$, or moving below the density value $\wrhomorto(\wmorto)<\sigma(\wmorto)$ by a jump with positive speed. In this case, a 1-rarefaction connects to an intermediate state $\umenoS$ with $\wmenoS=\wmorto$ and $0\leq\rhomenoS\leq\wrhomorto(\wmorto)$, then a 2-contact discontinuity connects to $\upiu$.

Otherwise, if $\vpiu>\vmax(\wmeno)$ then the equality $V(\rho,\wmeno)=\vpiu$ can not hold. It holds $\rhomorto = 0$ and the admissible left state $\rhomenoS$ has to be in $[0,\sigma(\wmeno)]$.  
\end{proof}

To summarize, we denote 
\begin{equation}\label{eq:rhomorto}
	\rhomorto(\wmeno;\vpiu) = \begin{cases}
		\rho(\wmeno;\vpiu) &\text{if $\vpiu\leq\vmax(\wmeno)$}\\
		0 &\text{if $\vpiu>\vmax(\wmeno)$}
	\end{cases}
\end{equation}
where $\rho(\cdot;\vpiu)$ is the implicit function given by the equation $V(\rho,w)=\vpiu$, which is well defined as stated in Proposition \ref{prop:Umorto}.
\begin{remark}
For numerical purposes, we use the \  Collapsed Generalized Aw-Rascle-Zhang (CGARZ) model, see \cite{FanSunPiccoliSeiboldWork2017} and Section \ref{sec:num_generale}. %$\vmax(w)=\vmax$ does not depend on $w$ 
This model is characterized by a maximum velocity $\vmax$ common to any $w$. Hence, the case of $\vpiu>\vmax(w)$ never holds for the CGARZ model.
\end{remark}

%%%%%%%%%%%%%%%%%%%%%%%%%%%%%%%%%%
\section{The GSOM on networks}\label{sec:cgarzNetwork}
In this section we apply Propositions \ref{prop:incoming} and 
\ref{prop:outgoing} to define the solution to Riemann problems for merge and diverge junctions.
To identify a unique solution we assume the maximization of the flux and the conservation of $\rho$ and $y=\rho w$ across the junction. Moreover, we assume that a distribution parameter on outgoing roads and a priority rule on incoming ones are given.

%%%%%%%%%%%%%%%%%%%%%%%%%%%%%%%%%%%
\subsection{Diverge junction}\label{sec:1in2}
We consider the case of a junction with one incoming and two outgoing roads. 
Given a left state $\umeno_{1}$ for the incoming road and two right states $\upiu_{2}$ and $\upiu_{3}$ for the outgoing roads, our aim is to determine the junction values $\upiuS_{i}=(\hat\rho_i,\hat w_i)$, $i=1,2,3$, giving rise to a boundary-value problem on each road. The solutions to the latter pieced together provide a solution
to the Riemann problem at the junction.\\
First, introduce a traffic distribution parameter $\alpha\in(0,1)$: vehicles are distributed in proportion $\alpha$ and $1-\alpha$ on the roads 2 and 3, respectively. Note that the cases $\alpha=0$ or $\alpha=1$ reduce the problem to a simple 1 to 1 junction, thus in this analysis we exclude the two extreme values.

\noindent
Set $\hat q_i=\hat \rho_i \hat v_i$, $\hat v_i = V(\hat\rho_i,\hat w_i)$, $i=1,2,3$, then the conservation of $\rho$ and $y$ across the junction reads:

\noindent\begin{minipage}{0.5\columnwidth}
\begin{align}
\alpha \hat q_{1}&=\hat q_{2}\label{eq:1in2q2}\\
\alpha \hat q_{1}\wpiuS_{1}&=\hat q_{2}\wmenoS_{2}\label{eq:1in2w2}
\end{align}
\end{minipage}\hfill
\begin{minipage}{0.5\columnwidth}
\begin{align}
(1-\alpha) \hat q_{1}&=\hat q_{3}\label{eq:1in2q3}\\
(1-\alpha) \hat q_{1}\wpiuS_{1}&=\hat q_{3}\wmenoS_{3}.\label{eq:1in2w3}
\end{align}
\end{minipage}

\bigskip
\noindent By Proposition \ref{prop:incoming} we have $\wpiuS_{1}=\wmeno_{1}$, and by
\eqref{eq:1in2q2}-\eqref{eq:1in2w3} we deduce $\wpiuS_{1}=\wmenoS_{2}$ and $\wpiuS_{1}=\wmenoS_{3}$, hence $\wmenoS_{2}=\wmenoS_{3}=\wmeno_{1}$.
Now the states $\upiuS_{i}$ correspond to six unknowns for which we have five equations.
Using the free parameter $q=\hat q_1$ and, by \eqref{eq:dem} and \eqref{eq:sup} we get the constraints
\begin{align}\label{eq:maxConstrDiv}
0&\leq q\leq \dem{\rhomeno_{1}}{\wmeno_{1}} \nonumber \smallskip\\
0&\leq \alpha q\leq \supp{\rhomorto_{2}}{\wmeno_{1}} \smallskip\\
0&\leq (1-\alpha) q\leq \supp{\rhomorto_{3}}{\wmeno_{1}},\nonumber 
\end{align}
where, by Proposition \ref{prop:outgoing},  $\wmorto_{2}=\wmorto_{3}=\wmeno_{1}$ and $\rhomorto_{2}$, $\rhomorto_{3}$ are given by \eqref{eq:rhomorto} with $\wmeno = \wmeno_i$ and $\vpiu = \vpiu_i$, $i=2,3$, respectively. 
To satisfy \eqref{eq:maxConstrDiv} and maximize the outgoing flux, it holds
$$q = \min\{\dem{\rhomeno_{1}}{\wmeno_{1}}, \supp{\rhomorto_{2}}{\wmeno_{1}}/\alpha, \supp{\rhomorto_{3}}{\wmeno_{1}}/(1-\alpha)\}$$ 
and 
$$
\qpiuS_{1}= q, \quad
\qmenoS_{2}=\alpha  q, \quad
\qmenoS_{3}=(1-\alpha) q.
$$
Then, the junction density values are $\rhopiuS_{1}\in\neg(\umeno_{1})$ such that $Q(\rhopiuS_{1},\wmeno_1)=\qpiuS_{1}$ and $\rhomenoS_{j}\in\pos(\upiu_{j})$ such that $Q(\rhomenoS_{j},\wmeno_{j})=\qmenoS_{j}$, $j=2,3$. In \cite{herty2006SIAM,kolb2018SIAM}, the authors obtain the same solution for the ARZ model.

%%%%%%%%%%%%%%%%%%%%%%%%%%%%%%%%%%
\subsection{Merge junction}\label{sec:2in1}
We consider the case of a junction with two incoming and one outgoing roads. 
Given two states $\umeno_{1}$ and $\umeno_2$ for the incoming roads and a state $\upiu_{3}$ for the outgoing road, we look for the junction values $\upiuS_{1}$, $\upiuS_{2}$ and $\umenoS_{3}$.
As done before, we set $\hat q_i=\hat\rho_i \hat v_i$, $i=1,2,3$, and 
we assume that vehicles from roads 1 and 2 enter into the road 3 with the following priority rule
\begin{equation}\label{eq:prec}
	(1-\beta)\qpiuS_{2} = \beta\qpiuS_{1},
\end{equation} 
where $\beta\in[0,1]$. Note that for $\beta=0$ or $\beta=1$, one of the two incoming roads is completely stopped at the junction, and the problem reduces to the 1 to 1 case.\\
The conservation of $\rho$ and $y$ across the junction yields:
\begin{align}
	\qpiuS_{1}+\qpiuS_{2}& = \qmenoS_{3}\label{eq:q2in1}\\
	\qpiuS_{1}\wpiuS_{1}+\qpiuS_{2}\wpiuS_{2}& = \qmenoS_{3}\wmenoS_{3}.\label{eq:w2in1}
\end{align}
By Proposition \ref{prop:incoming}, we have that $\wpiuS_{1}=\wmeno_{1}$ and $\wpiuS_{2}=\wmeno_{2}$. Equation \eqref{eq:q2in1} combined with \eqref{eq:prec} and \eqref{eq:w2in1}, implies
\begin{equation}\label{eq:w3}
	\wmenoS_{3} = (1-\beta)\wmeno_{1}+\beta \wmeno_{2}.
\end{equation}
%
%Hence, it remains to compute the values $\qpiuS_1$, $\qpiuS_2$, $\qmenoS_3$ and $\wmenoS_3$. Relations \eqref{eq:prec}-\eqref{eq:w3} fix some of them and we have to define the remaining free variables.
Hence, $\wpiuS_1$, $\wpiuS_2$ and $\wmenoS_3$ are defined and $\qpiuS_1$, $\qpiuS_2$ and $\qpiuS_3$ have to satisfy equations \eqref{eq:prec} and \eqref{eq:q2in1}. It remains a free parameter and, in order to define a unique solution, we impose the maximization of the flux on the outgoing road. 
By \eqref{eq:dem} and \eqref{eq:sup}, we get the constraints
\begin{align}\label{eq:maxConstrMerge}
0&\leq \hat q_1\leq \dem{\rhomeno_{1}}{\wmeno_{1}} \nonumber \smallskip\\
0&\leq \hat q_2 \leq \dem{\rhomeno_{2}}{\wmeno_{2}} \smallskip\\
0&\leq \hat q_3 \leq \supp{\rhomorto_{3}}{\hat w_3},\nonumber 
\end{align}
where $\rhomorto_{3}$ is given by \eqref{eq:rhomorto} with $\wmeno=\wmenoS_3$ and $\vpiu=V(\rhopiu_3,\wpiu_3)$. From now on, we set
\begin{equation*}
\duno=\dem{\rhomeno_{1}}{\wmeno_{1}} \mbox{ and } \ddue=\dem{\rhomeno_{2}}{\wmeno_{2}}.
\end{equation*}
We assume that both $\duno$ and $\ddue$ are greater than 0. Indeed, the trivial case of $\duno=\ddue=0$ means that no vehicles cross the intersection, and the case of $\duno=0$ or $\ddue=0$ reduces the junction to the 1 to 1 type.
In order maximize the flux on the outgoing road we set
\begin{equation}\label{eq:maxim}
	\qpiuS_3 = \supp{\rhomorto_{3}}{\wmenoS_3}.
\end{equation}
To summarize, the couple $(\hat\quno,\hat\qdue)$ is given by the intersection point $P$ between the following two lines
\begin{align}
	\mathrm{r}: \qdue &= \frac{\beta}{1-\beta}\quno\label{eq:rettar}\\
	\mathrm{s}: \qdue &= \supp{\rhomorto_{3}}{\wmenoS_{3}}-\quno,\label{eq:rettas}
\end{align}
where the first one represents the priority rule \eqref{eq:prec}, while the second one represent the conservation equation \eqref{eq:q2in1} coupled with \eqref{eq:maxim}. In \eqref{eq:rettar}, $\rettar$ coincides with the axis $x=0$ when $\beta=1$. 
Note that, 
since $\rhomorto_{3}=\rhomorto_{3}(\wmenoS_{3};\vpiu_3)$ and $\wmenoS_{3}$ depends on $\beta$, the maximum flux that can be received by the outgoing road is a function of the priority rule, i.e.\ $\supp{\rhomorto_{3}}{\wmenoS_{3}}=\stre(\beta)$.
\\
The intersection point between $\mathrm{r}$ and $\mathrm{s}$ is 
\begin{equation}\label{eq:intP}
	P = ((1-\beta)\stre(\beta),\beta\stre(\beta)).
\end{equation}
If $P\in\Omega=[0,d_{1}]\times[0,d_{2}]$, we can set $\qpiuS_{1}=(1-\beta)\stre(\beta)$ and $\qpiuS_{2}=\beta\stre(\beta)$.
Otherwise, if $P\notin\Omega$, then the point does not satisfy the constraints \eqref{eq:maxConstrMerge}, and we need to \textit{relax} one of our constraints. 
We propose two possible approaches:
\begin{itemize}
	\item[(RP)]  The relation \eqref{eq:prec} is satisfied with $\beta$ fixed a priori, while  the outgoing flow \eqref{eq:maxim} is not maximized. This is the case for instance of a stop sign or a traffic policeman regulating the junction.
	\item[(AP)] The priority parameter $\beta$ is modified, thus allowing to maximize the outgoing flux. This is the case, for instance, of unsupervised junction.
\end{itemize}

\noindent
To detail the procedure to compute the junction densities $\hat \rho_i$, $i=1,2,3$, first recall that $\hat w_1=\hat w_2 = \wmeno_1$ and $\hat w_3=(1-\beta)\wmeno_1+\beta\wmeno_2$ as stated in \eqref{eq:w3}. We introduce the parameter
\begin{equation}\label{eq:bd}
	\betad = \frac{\ddue}{\duno+\ddue}
\end{equation}
which identifies the priority line in \eqref{eq:rettar} that passes through the point $(\duno,\ddue)$. If $P\notin\Omega$, we distinguish two cases:
\begin{itemize}
\item[(i)] $\beta\geq\betad$ then the $y$-coordinate of $P$, $\beta\stre(\beta)$  is greater than the upper bound $\ddue$. Then we fix $\qpiuS_2=\ddue$ and look for an admissible value $\qpiuS_1$;
\item[(ii)] $\beta<\betad$ then the $x$-coordinate $(1-\beta)\stre(\beta)>\duno$. Then we fix $\qpiuS_1=\duno$ and look for an admissible value $\qpiuS_2$.
\end{itemize}

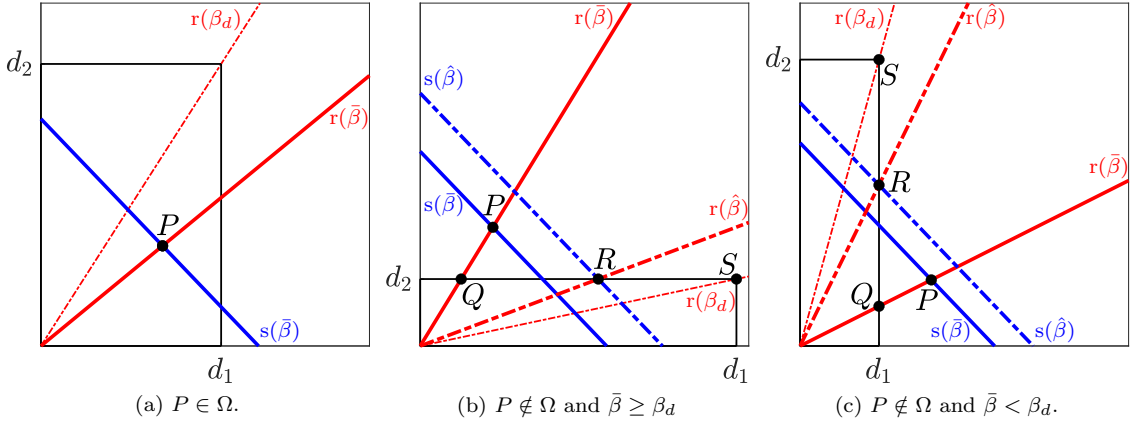
\begin{figure}[h!]
\centering
\normalsize
\subfloat[][$P\in\Omega$.]{\label{P1}
\begin{overpic}[width=0.31\columnwidth]{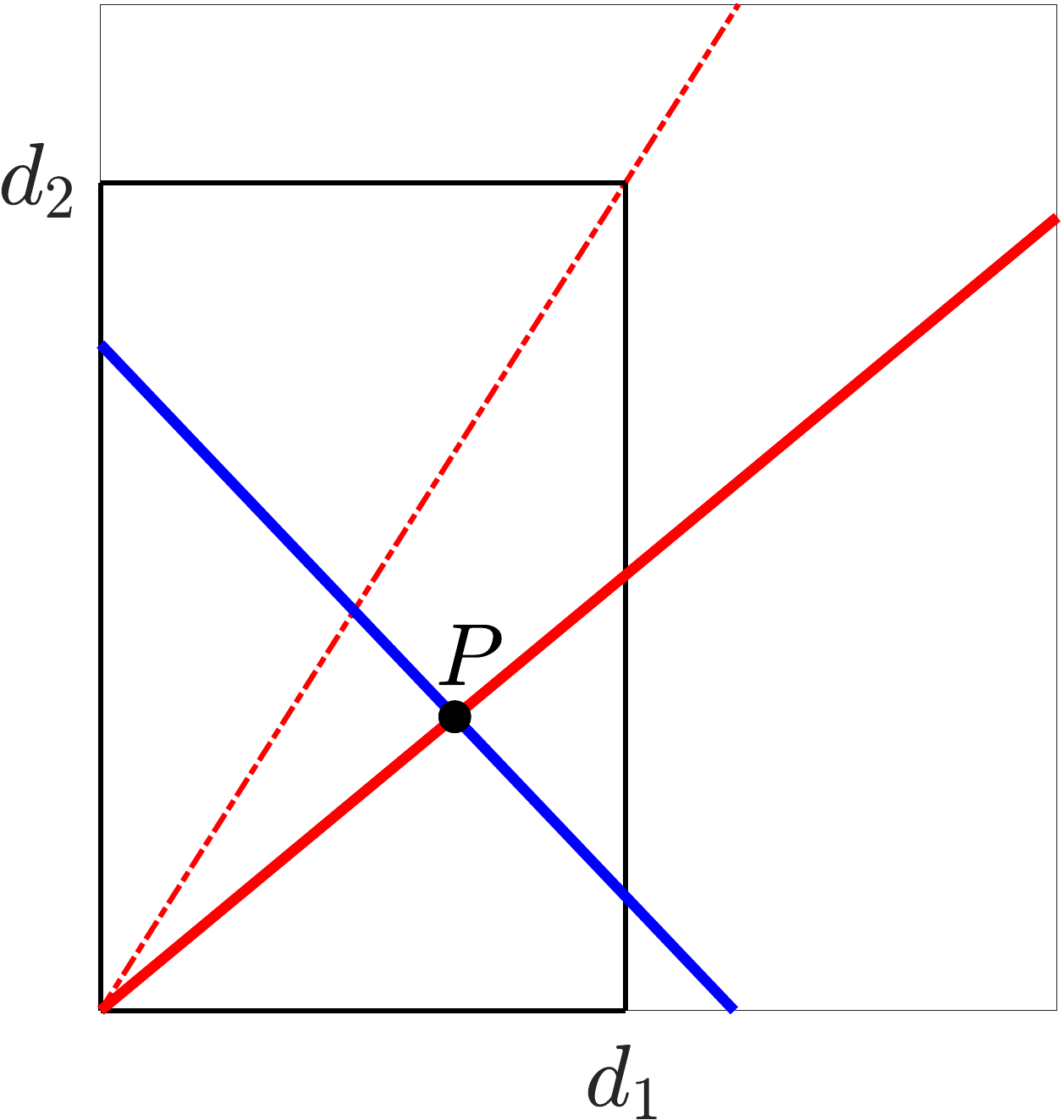}
\put(48,94){\footnotesize\color{red}$\rettar(\betad)$}
\put(84,68){\footnotesize\color{red}$\rettar(\bb)$}
\put(66,12){\footnotesize\color{blue}$\rettas(\bb)$}
\end{overpic}
}
\subfloat[][$P\notin\Omega$ and $\bb\geq\betad$]{\label{P2}
\begin{overpic}[width=0.31\columnwidth]{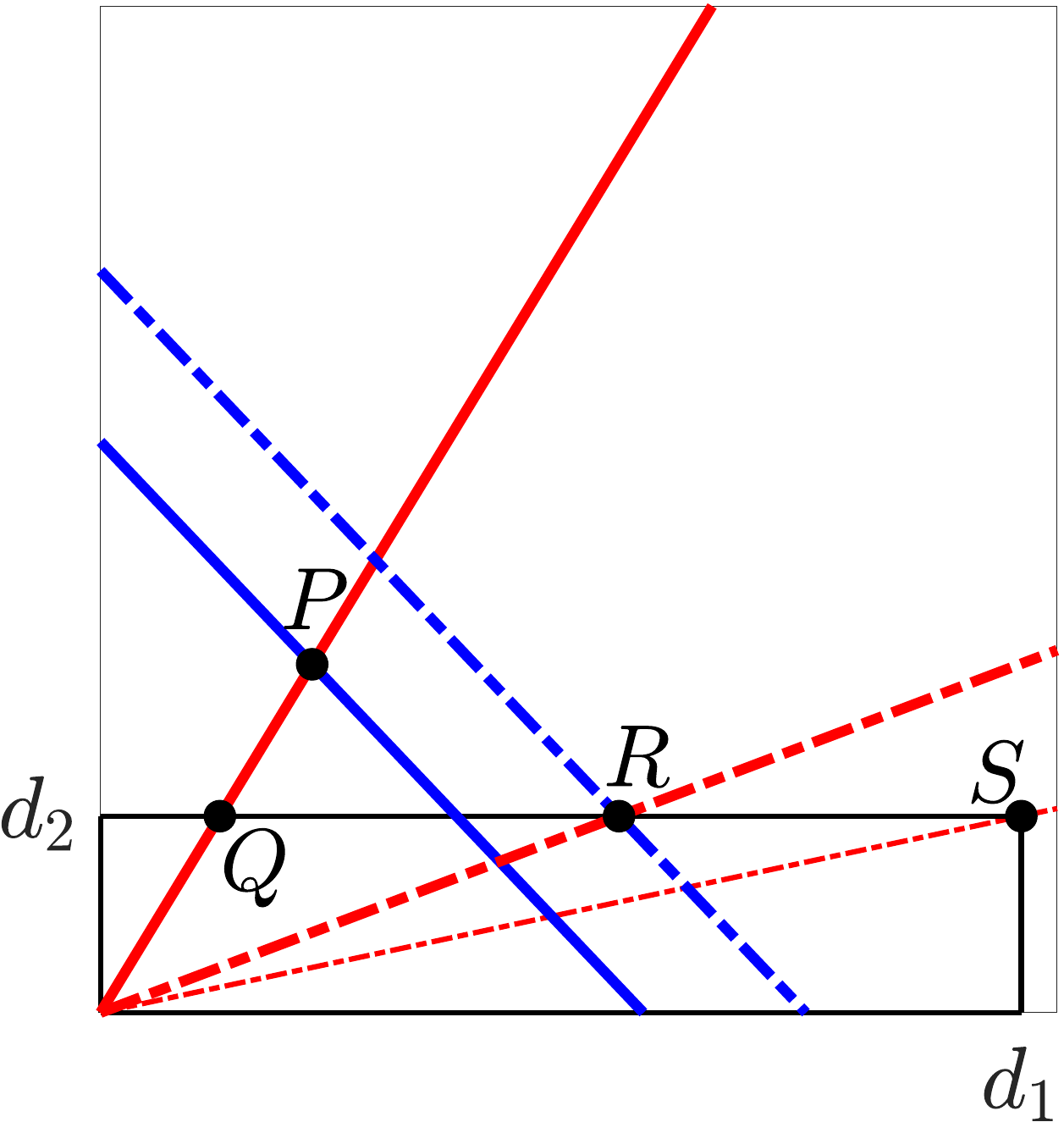}
\put(78,20){\footnotesize\color{red}$\rettar(\betad)$}
\put(84,44){\footnotesize\color{red}$\rettar(\betaS)$}
\put(49,94){\footnotesize\color{red}$\rettar(\bb)$}
\put(10,78){\footnotesize\color{blue}$\rettas(\betaS)$}
\put(10,45){\footnotesize\color{blue}$\rettas(\bb)$}
\end{overpic}
}
\subfloat[][$P\notin\Omega$ and $\bb<\betad$.]{\label{P3}
\begin{overpic}[width=0.31\columnwidth]{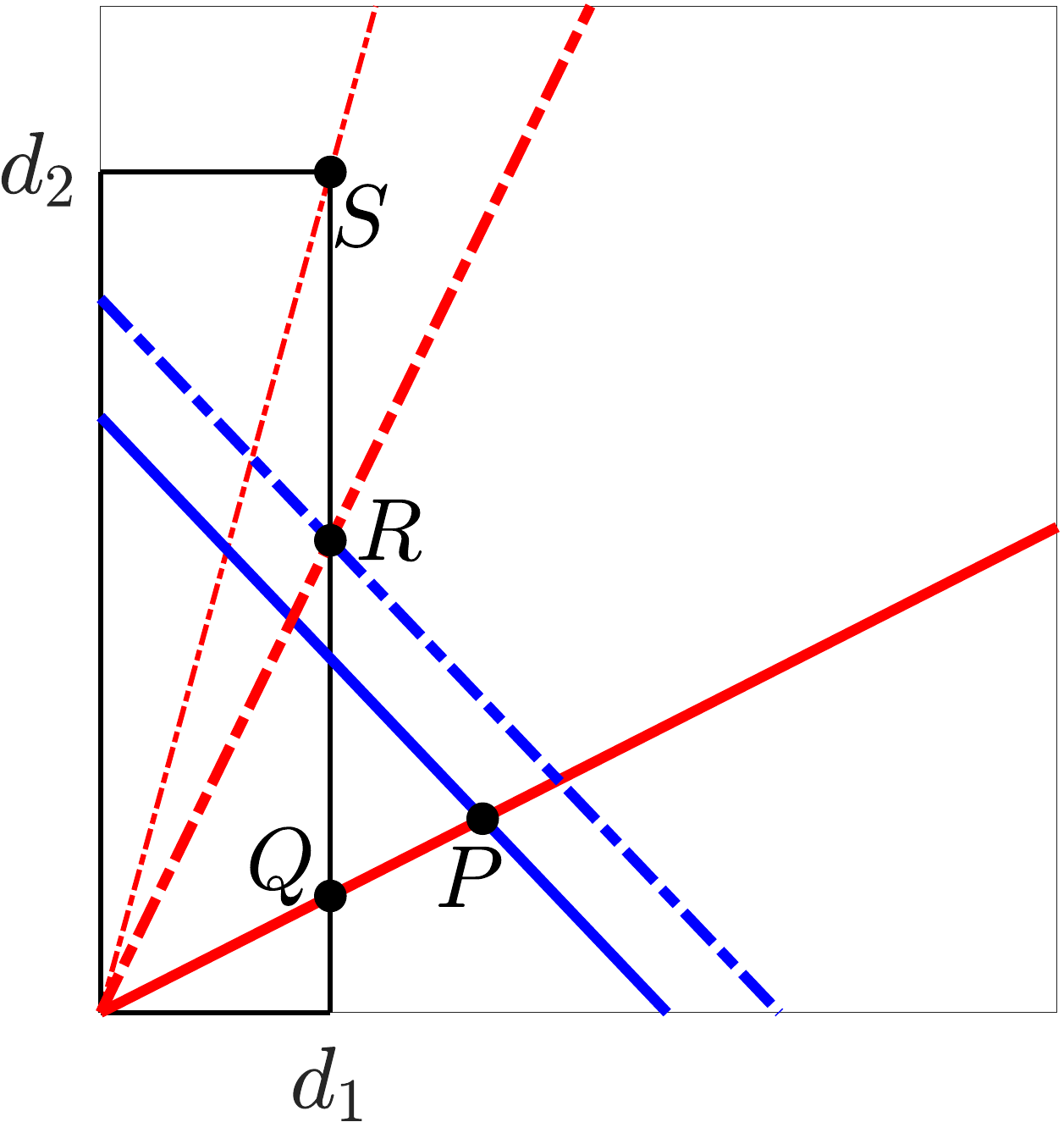}
\put(18,94){\footnotesize\color{red}$\rettar(\betad)$}
\put(52,94){\footnotesize\color{red}$\rettar(\betaS)$}
\put(84,55){\footnotesize\color{red}$\rettar(\bb)$}
\put(70,12){\footnotesize\color{blue}$\rettas(\betaS)$}
\put(43,12){\footnotesize\color{blue}$\rettas(\bb)$}
\end{overpic}
}
\caption{The three possible cases defining the RP and AP algorithms, with $P$ in \eqref{eq:intP}, $\betad$ in \eqref{eq:bd}, $\bb\in[0,1]$ and $\betaS$ in Definition \ref{prop:sol}. The red and blue lines represent $\rettar$ and $\rettas$ defined in \eqref{eq:rettar} and \eqref{eq:rettas}, respectively, for different values of $\beta$. For both algorithms, plot \protect\subref{P1} represents case 1, plot \protect\subref{P2} shows case 2 with $\wmeno_{1}>\wmeno_{2}$ and plot \protect\subref{P3} shows case 3 with $\wmeno_{1}<\wmeno_{2}$.}
\label{fig:rettangolo}
\end{figure}

\noindent
We first describe the RP algorithm. For a given priority parameter $\bb\in[0,1]$, to satisfy the priority rule, the solution must lie on the line $(1-\bb)\qpiuS_2=\bb\qpiuS_1$. For this reason, when $P\notin\Omega$ the couple $(\qpiuS_{1},\qpiuS_{2})$ will be defined by the intersection point between the priority line and the boundary $\de\Omega$, see for instance the point $Q$ in Figures \subref*{P2} and \subref*{P3}.
%Hence the following definition.

\begin{defn}\textup{Algorithm RP.}\label{def:rp} Let $\bb\in[0,1]$ and let $\wmenoS_{3}$ and $P$ be as in \eqref{eq:w3} and \eqref{eq:intP} with $\beta=\bb$, respectively. Assume that \eqref{eq:q2in1} holds. 
Define $(\hat q_1,\hat q_2)\in\Omega$ as follows: 
\begin{enumerate}[label={\arabic*.},ref={\arabic*}]
	\item If $P\in\Omega$, then $\qpiuS_{1}=(1-\bb)\stre(\bb)$ and $\qpiuS_{2}=\bb\stre(\bb)$.
	%%%%%%%%%%
	\item\label{rp2} If $P\notin\Omega$ and $\bb\geq\betad$, then $\qpiuS_{1}=(1-\bb)\ddue/\bb$ and $\qpiuS_{2}=\ddue$.
	%%%%
	\item\label{rp3} If $P\notin\Omega$ and $\bb<\betad$, then $\qpiuS_{1}=\duno$ and $\qpiuS_{2}=\bb\duno/(1-\bb)$.
\end{enumerate}
The density value $\rhopiuS_{i}\in\neg(\umeno_{i})$ is determined by the equality $Q(\rhopiuS_{i},\wpiuS_{i})=\qpiuS_{i}$, $i=1,2$, while  $\rhomenoS_{3}\in\pos(\upiu_{3})$ is determined by $Q(\rhomenoS_{3},\wmenoS_{3})=\qmenoS_{3}$.
\end{defn}

\noindent
To describe the AP algorithm we need some preliminary results.
\begin{lemma}\label{rem:monotonia}
The supply function $s(\rhomorto_3(w;\cdot),w)$ is non-decreasing in $w$. 
\end{lemma}
\begin{proof}
By \eqref{eq:supply} we can have 
$$s(\rhomorto_3(w;\cdot),w)=\qmax(w)\quad\mbox{ or }\quad s(\rhomorto_3(w;\cdot),w)=\rhomorto_3(w;\cdot)V(\rhomorto_3(w;\cdot),w).$$ In the first case, assumption \ref{q3} applies; in the second one, by Proposition \ref{prop:Umorto} we have $V(\rhomorto_3(w;\cdot),w)=\vpiu_3$ and $\rhomorto_3(w;\cdot)$ is non-decreasing in $w$.
\end{proof}
\noindent
To study the function $\stre(\beta)=\supp{\rhomorto_{3}(\wmenoS_{3}(\beta);\cdot)}{\wmenoS_{3}(\beta)}$ with respect to $\beta$, we distinguish two cases:  
\begin{itemize}
\item[(a)] $\wmeno_1 \leq \wmeno_2$ then both $\wmenoS_3$ and $\stre$ are increasing in $\beta$;
\item[(b)] $\wmeno_1 > \wmeno_2$ then both $\wmenoS_3$ and $\stre$ are decreasing in $\beta$.
\end{itemize}
\begin{lemma}\label{lemma}
Let $\bb\in[0,1]$ and $\betad$ given in \eqref{eq:bd}.
\begin{enumerate}[label={\arabic*.},ref={\arabic*}]
	\item\label{lem1} If $\bb\geq\betad$ and $\bb \stre(\bb)>\ddue$, then there exists at least a $\beta\in[0,\bb)$ such that $\beta\stre(\beta)=\ddue$. 
	\item\label{lem2}  If $\bb<\betad$ and $(1-\bb)\stre(\bb)>\duno$, then there exists at least a $\beta\in(\bb,1]$ such that $(1-\beta)\stre(\beta)=\duno$.
\end{enumerate}
\end{lemma}
\begin{proof}
We first prove point $1$. Consider the two cases (a) and (b), i.e. $\wmeno_1<\wmeno_2$ and $\wmeno_1>\wmeno_2$, respectively.\\
If $\wmeno_1<\wmeno_2$ then the function $f(\beta)=\beta\stre(\beta)$ is increasing in $[0,1]$ and such that $f(0)=0$ and $f(\bb)>\ddue$ by hypothesis; therefore, there exists a unique $\beta^*<\bb$ such that $f(\beta^*)=\ddue$.\\
If $\wmeno_1>\wmeno_2$ then $\stre(\beta)$ is decreasing in $\beta$ and the behavior of the function $f(\beta)=\beta\stre(\beta)$ is not known a priori. However the function $f$ is continuous and such that $f(0)=0$ and $f(\bb)>\ddue$ by hypothesis; therefore there exists at least a $\beta<\bb$ such that $f(\beta) =\ddue$.\\
The proof of point $2$ is entirely similar, so we skip the details.
\end{proof}

\noindent
The AP algorithm is described in the following definition. 
As mentioned above, the algorithm adapts the priority parameter to maximize the outgoing flux while keeping the parameter as close as possible to its initial value.

\begin{defn}\textup{Algorithm AP.}\label{prop:sol} Let $\bb\in[0,1]$ and $P$ be as in \eqref{eq:intP} with $\beta=\bb$. Assume that \eqref{eq:q2in1} holds. Define $(\hat q_1,\hat q_2)\in\Omega$ as follows: 
\begin{enumerate}[label={\arabic*.},ref={\arabic*}]
	\item If $P\in\Omega$ then $\qpiuS_{1}=(1-\bb)\stre(\bb)$, $\qpiuS_{2}=\bb\stre(\bb)$ and $\wmenoS_{3}=(1-\betaS)\wmeno_{1}+\betaS\wmeno_{2}$ with $\betaS=\bb$.
	%%%%%%%%%%
	\item\label{pp2} If $P\notin\Omega$ and $\bb\geq\betad$, then for $\beta^{*}=\max\{\beta\in[0,\bb)\,:\,  \beta\stre(\beta)=\ddue\}$, we set $\betaS=\max\{\beta^{*},\betad\}$, $\qpiuS_{1} = \min\{(1-\betaS) \stre(\betaS),\duno\}$, $\qpiuS_{2}=\ddue$ and $\wmenoS_{3}=(1-\betaS)\wmeno_{1}+\betaS\wmeno_{2}$.
	%%%%
	\item\label{pp3}  If $P\notin\Omega$ and $\bb<\betad$, then for $\beta^{*}=\min\{\beta\in(\bb,1] \,:\, (1-\beta)\stre(\beta)=\duno\}$, we set $\betaS=\min\{\beta^{*},\betad\}$, $\qpiuS_{1} = \duno$, $\qpiuS_{2}=\min\{\betaS\stre(\betaS),\ddue\}$ and $\wmenoS_{3}=(1-\betaS)\wmeno_{1}+\betaS\wmeno_{2}$.

\end{enumerate}
The density value $\rhopiuS_{i}\in\neg(\umeno_{i})$ is determined by the equality $Q(\rhopiuS_{i},\wpiuS_{i})=\qpiuS_{i}$, $i=1,2$, while $\rhomenoS_{3}\in\pos(\upiu_{3})$ is determined by $Q(\rhomenoS_{3},\wmenoS_{3})=\qmenoS_{3}$.
\end{defn}
\begin{prop}
The couple $(\qpiuS_{1},\qpiuS_{2})$ in Definition \ref{prop:sol} satisfies the constraints \eqref{eq:maxConstrMerge}.
\end{prop}
\begin{proof}
The given couple $(\qpiuS_1,\qpiuS_2)$ verifies the first two constraints in \eqref{eq:maxConstrMerge} by construction. Therefore, it remains to prove that $\qmenoS_3=\qpiuS_{1}+\qpiuS_{2}\leq\stre(\betaS) = \stre(\rhomorto_{3}(\wmenoS_3;\cdot),\wmenoS_{3})$.

We start from the case \ref{pp2} of Definition \ref{prop:sol}. In light of Lemma \ref{lemma} case \ref{lem1}, the value $\beta^{*}$ is well defined.
Moreover, since the slope of r increases with $\beta$, the point $((1-\beta^*)\stre(\beta^*),\ddue)$ is such that: 
if $\beta^*\geq \betad$ then $(1-\beta^*)\stre(\beta^*)\leq d_1$ and if $\beta^*> \betad$ then $(1-\beta^*)\stre(\beta^*)>d_1$.
Therefore, we focus on these two possibilities:
\begin{itemize}
\item If $\beta^*\geq\betad$ then $\betaS=\beta^*$ and $\qpiuS_1=(1-\betaS)\stre(\betaS)$. Hence, $\qpiuS_1+\qpiuS_2= (1-\betaS)\stre(\betaS) + \ddue = (1-\betaS)\stre(\betaS) + \betaS\stre(\betaS)=\stre(\betaS)$ and the thesis follows. This is the case, for instance, of point $R$ in Figure \subref*{P2}.
\item If $\beta^*<\betad$ then $\betaS = \betad$ and $\qpiuS_1=\duno$. The couple $(\qpiuS_{1},\qpiuS_{2})=(\duno,\ddue)$  is admissible if $\duno+\ddue\leq\stre(\betad)$. Since $\duno=(1-\betad)\ddue/\betad$ we have $\duno+\ddue =\ddue/\betad$. 
From the definition of $\beta^*$, for each $\beta\in(\beta^{*},\bb]$ it holds $\beta\stre(\beta)>\ddue$, and we get the thesis:
\begin{equation*}
	\duno+\ddue=\frac{\ddue}{\betad} < \frac{\betad\stre(\betad)}{\betad}=\stre(\betad).
\end{equation*}
This is the case of point $S$ in Figure \subref*{P2}.
\end{itemize}
The proof of case \ref{pp3} follows similarly, see Figure \subref*{P3} for an example of possible configuration.
\end{proof}

\begin{remark}
Let us consider the particular case of $\wmeno_1=\wmeno_2=\wpiu_3$, i.e.\ the variable $w$ is constant on the roads network. The diverge junction can be treated exactly as the LWR model at junctions, as done in \cite{GaravelloPiccoli2006}. For the merge junction we observe that the assumption of $w$ constant implies that the straight line $\rettas$ defined in \eqref{eq:rettas} coincides for all $\beta$, therefore the solution is limited to the points $P$ or $Q$  in Figure \ref{fig:rettangolo}, excluding the points $R$ and $S$. Thus, we recover again the LWR model on networks, as treated in \cite{GaravelloPiccoli2006}. 
\end{remark}

\section{Minimize emissions and travel time}\label{sec:ottimizzazione}

The emission of pollutants is strictly connected to speed and acceleration of vehicles.
% which depend on traffic dynamics that can be regularized by proper traffic policies.
In this section we set up an optimization problem to minimize the $\nox$ emission rates due to vehicular traffic.
% We link emissions to traffic quantities and the latter to suitable parameters which give different scenarios. 

Consider \eqref{eq:GSOMrete} on a network with roads $I_{r}$, $r=1,\ldots, N_r$, during a time interval $[0,T]$.
Following \cite{balzotti2021DCDSB}, we use the microscopic emission model proposed in \cite{panis2006elsevier} which estimates the emission rate $E_{i}$ of vehicle $i$ at time $t$ using the instantaneous speed $v_i(t)$ and acceleration $a_i(t)$.
In order to work with the macroscopic variables provided by the traffic model, we set the emission rate formula on a portion $\dx_j$ of space at time $t^{n}$ for the road $r$ as
\begin{equation}
E^{n}_{r,j}=\rho^{n}_{r,j}\dx_{j} \max \{ E_0 ,  f_1 + f_2 v^{n}_{r,j} + f_3 (v^{n}_{r,j})^2 + f_4 a^{n}_{r,j} + f_5 (a^{n}_{r,j})^{2} + f_6 v^{n}_{r,j} a^{n}_{r,j}\},
\label{eq:emissioni}
\end{equation}
where $E_0$ is a lower-bound of emission and $f_1$ to $f_6$ are emission constants associated to $\nox$, see \cite[Table 2]{panis2006elsevier}. The vehicles densities $\rho^{n}_{r,j}$ and velocities $v^{n}_{r,j}$ are mean values in $\dx_j$, given by the model \eqref{eq:GSOMrete}, and $a^{n}_{r,j}$ is the acceleration given by computing the total derivative of $V(\rho,w)$, i.e.
$$
a(x,t) = \frac{Dv(x,t)}{Dt} = v_t(x,t) + v(x,t) v_x(x,t),
$$
where
$$
v(x,t)=V(\rho(x,t),w(x,t)), \quad v_t = V_\rho \rho_t + V_w w_t, \quad v_x = V_\rho\rho_x+V_w w_x.
$$
By simple computations, for the GSOM we have
\begin{equation}\label{eq:accAnalitica}
a(x,t) = V_\rho\left(\rho_t + v \rho_x\right) = -V_\rho \rho v_x.
\end{equation}

%By fixing now the set $\Gamma$ of $k$ control parameters $\gamma=(\gamma_1,\ldots,\gamma_k)$ governing the traffic dynamic, we introduce the following operator to estimate the total emission rate on a road network as a function of $\gamma\in\Gamma$,
Let $\Gamma$ be the set of $k$ control parameters $\gamma=(\gamma_1,\ldots,\gamma_k)$ governing the traffic dynamic. These are given by the traffic distribution and priority parameters $\alpha$ and $\beta$
of Section \ref{sec:cgarzNetwork}.
We introduce the following operator to estimate the total emission rate on a road network as a function of $\gamma\in\Gamma$,
\begin{equation}\label{eq:fune}
	\funE(\gamma) = \sum_{r=1}^{\nr}\int_{0}^{T}\int_{0}^{L}E^{\gamma}_{r}(x,t)dxdt,
\end{equation}
where $\nr$ is the number of roads and $E^{\gamma}_{r}(x,t)$ is the emission rate \eqref{eq:emissioni} in $x$ at time $t$ related to $\gamma$ and to road $r$. 
To guarantee acceptable travel times, we include a velocity term thus getting
the objective function  
\begin{align}\label{eq:funet}
	\funET(\gamma) &= \sum_{r=1}^{\nr}\left(c_{1}\int_{0}^{T}\int_{0}^{L}E^{\gamma}_{r}(x,t)dxdt+c_{2}\int_{0}^{T}\int_{0}^{L}\frac{1}{\mathcal{V}^{\gamma}_{r}(x,t)}dxdt\right),
\end{align}
where $c_1$ and $c_2$ are two proper weights %. To remove the singularity in the second integrand we set 
and $\mathcal{V}^{\gamma}_{r}=\max\{V^{\gamma}_{r}(x,t),\varepsilon\}$, $\varepsilon>0$, with $V^\gamma_r$ velocity function of the traffic model, related to control parameter $\gamma$ and to road $r$. The parameter $\varepsilon$ allows to exclude the null speeds in the calculation.
Our goal is to solve the minimization problem
\begin{equation}\label{eq:minProb}
	\min_{\gamma\in\Gamma}\funET(\gamma).
\end{equation}
Due to the complexity and the strictly nonlinear dependence of the functional $\funET$ on the control $\gamma$, we treat the problem numerically using global search.

\section{Numerical setup}\label{sec:num_generale}
We consider the traffic model \eqref{eq:GSOMrete} and we divide each road into $\nx$ cells $[x_{j-1/2},x_{j+1/2})$ of length $\dx$ centered in $x_{j}$, and the time interval into $\nt+1$ steps $t^{n}=n\dt$.

To compute the traffic quantities $\rho^{n}_{r,j}$ and $V^{n}_{r,j}$ in \eqref{eq:emissioni}, we choose the CGARZ model \cite{FanSunPiccoliSeiboldWork2017} among the family of GSOM, see details below. 
The model is then solved numerically with the 2CTM scheme \cite{balzotti2021DCDSB} with suitable boundary conditions at the extremes of the network. We use the theory given in Sections \ref{sec:1in2} and \ref{sec:2in1} to build the numerical solution at junctions.

Once $\rho^{n}_{r,j}$ and $V^{n}_{r,j}$ are known, from \eqref{eq:accAnalitica} we get the discrete acceleration
$$
a^{n}_{r,j} = -V_{\rho}(\rho^{n}_{j},w^{n}_{j})\rho^{n}_{j}\frac{v^{n}_{j+1}-v^{n}_{j-1}}{2\dx}
$$
and we can compute the emission rate with formula \eqref{eq:emissioni} in each cell $x_j$, $j=1,\ldots,\nx$.

The functional $\funET(\gamma)$ in \eqref{eq:funet} is then discretized as
\begin{equation}\label{eq:funET}
    \funET(\gamma) \approx \frac{1}{\nx\nt\nr}\sum_{r=1}^{\nr}\sum_{n=1}^{\nt}\sum_{j=1}^{\nx} \left[ \frac{E^{\gamma}_{r}(x_{j},t^{n})}{E^{\max}} + \frac{\varepsilon}{\mathcal{V}^{\gamma}_{r}(x_{j},t^{n})} \right],
\end{equation}
where $E^{\max}$ is the maximum emission rate, $\varepsilon$ is the  rounded minimum velocity, and,
in order to have comparable quantities for the emission and travel time functional,
the weights $c_{1}$ and $c_{2}$ are given by
\begin{equation}\label{eq:c1c2}
	c_{1} = \frac{1}{E^{\max}\nx\nt\nr} \qquad\text{and}\qquad c_{2} = \frac{\varepsilon}{\nx\nt\nr}.
\end{equation}
From now on we assume $\varepsilon = 1\,\kmh$. 
As shown in Appendix \ref{appendice}, 
this choice of weights does not substantially affect the numerical results described in the following sections. Thus $\funET$ in \eqref{eq:funET}-\eqref{eq:c1c2} is an appropriate  functional to analyze the cost in emission and travel time.

The CGARZ model assumes that there is a unique maximum density $\rhom$ independent of $w$ at which the vehicles stop, i.e.\ $V(\rhom,w)=0$ for all $w$. Furthermore, it assumes given a \emph{free-flow threshold density} $\rho_f$ such that the flux of vehicles is not influenced by $w$ when $\rho\leq\rho_f$ (\emph{free-flow regime}).
Thus, the flux is described
by a single-valued fundamental diagram in free-flow regimes and by a multi-valued function in congestion.
For $\rho\in[0,\rhom]$, we have
\begin{equation}\label{eq:Qcgarz}
	Q(\rho,w) = \begin{cases}
		Q_f(\rho) &\quad\text{if $0\leq\rho\leq\rho_f$}\\
		Q_c(\rho,w) &\quad\text{if $\rho_f<\rho\leq\rhom$}.
	\end{cases}
\end{equation}
Following \cite{balzotti2021DCDSB}, we assume a lower and upper bound for $w$, i.e.\ $0\leq \wl\leq w\leq \wr$, a Greenshields flux function in the free-flow phase, i.e.
\begin{equation}\label{eq:Qf}
	Q_f(\rho) =  \m\rho\left(\rhom-\rho\right),
\end{equation}
and a flux in congested phase given by
\begin{equation}\label{eq:Qc}
	Q_{c}(\rho,w) =\m \left(\rhom-\rho\right)\, \big((1-\theta(w)) \rhof +\theta(w) \rho\big), \quad \theta(w)=\frac{w-\wl}{\wr-\wl},
\end{equation}
where $\wl=Q_{f}(\rho_{f})$, $\wr=Q_{f}(\rhom/2)$ and $\rhom/2$ is the critical density of $Q_{f}(\cdot)$. The velocity function is then given by
\begin{equation*}\label{eq:velocityV}
	V(\rho,w) = \frac{Q(\rho,w)}{\rho}.
\end{equation*} 
With these choices, the property $w$ describes drivers attitude with respect to speed. 
Low values of $w$ describe \textit{slow drivers}, and high values of $w$ \textit{fast drivers}.

\smallskip

\section{Case study of a merge junction}\label{sec:numerica}
\begin{figure}[h!]
\centering
\subfloat[][priority rule]{
\begin{tikzpicture}[scale=0.9]
\draw[thick] (0,1.5) -- node[above = 0.1] {1} (3,0);
\draw[thick,->] (0,1.5) -- (1.5,0.75);
\draw[thick] (0,0) -- node[above=0.1] {2} (3,0);
\draw[thick,->] (0,0) -- (1.5,0);
\draw[thick] (3,0) -- node[above=0.1] {3} (6,0);
\draw[thick,->] (3,0) -- (4.5,0);
\draw[red,fill=red] (3,0) circle (0.1cm) node[below=0.1] {$\funET(\beta)$};
\end{tikzpicture}
}
\quad
\subfloat[][traffic light]{\label{fig:sem}
\begin{tikzpicture}[scale=0.9]
\draw[thick] (0,1.5) -- node[above = 0.1] {1} (3,0);
\draw[thick,->] (0,1.5) -- (1.5,0.75);
\draw[thick] (0,0) -- node[above=0.1] {2} (3,0);
\draw[thick,->] (0,0) -- (1.5,0);
\draw[thick] (3,0) -- node[above=0.1] {3} (6,0);
\draw[thick,->] (3,0) -- (4.5,0);
\draw[red,fill=red] (3,0) circle (0.1cm) node[below=0.1] {$\funET(t_r,t_g)$};
\draw[black, very thick, fill=black] (2.8,0.4) rectangle (3.2,1.2);
\draw[red] (3,1) circle (0.15cm);
\draw[green,fill=green] (3,0.6) circle (0.15cm);
\end{tikzpicture}
}
\caption{Example of merge junction where road 1 joins roads 2 and 3.%Case study of a merge junction: sensitivity from initial data 
}
\label{fig:2in1disegno}
\end{figure}
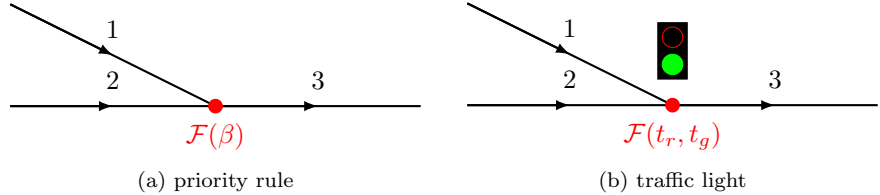

Let us consider the merge junction depicted in Figure \ref{fig:2in1disegno},
where we assume road 1 to be a ramp merging to roads 2 and 3.
We assume the junction to be governed first by a priority rule and then by a traffic light. The latter is modeled by alternating $\beta=0$ and $\beta=1$ in time.

The model parameters in \eqref{eq:Qcgarz} and those for the numerical tests are fixed in Table \ref{tab:parametriTest}.  The initial data is assumed to be constant on all the three roads and is chosen according to Table \ref{tab:merge}.

\begin{table}[h!]
\centering
\small
\begin{tabular}{cccccccc}\toprule
$\rhof$ & $\rhom$ & $\rhoc$ & $\vmax$ & $L$ & $\dx$ & $T$ & $\dt$\\
%$$ & $\vehkm$ & $\vehkm$ & $\kmh$\\
\midrule
$19\,\vehkm$ & $133\,\vehkm$ & $67.5\,\vehkm$ & $70\,\kmh$ & $3\,\km$  & $100\,\mymeter$ & $10\,\min$ & $4\,\mysecond$\\
\bottomrule
\end{tabular}
\caption{Parameters used for the numerical tests.}
\label{tab:parametriTest}
\end{table}

\begin{table}[h!]
\centering
\small
\begin{tabular}{lccc}\toprule
\multicolumn{1}{c}{Road $r$} & 1 & 2 & 3 \\\midrule
$\rhor{0}{r}\,(\vehkm)$ & 12 & 60 & 60 \\
$\wroad{0}{r}$ & $\wr$ & $w_{M}$ & $w_{M}$\\\bottomrule
\end{tabular}
\caption{Initial data for the test on a merge junction, with $w_{M}=(\wr+\wl)/2$. We assume fast drivers coming from road 1 and moderate drivers on road 2 and 3.}
\label{tab:merge}
\end{table}

\begin{figure}[h!]
\centering
\subfloat[][$\funE(\beta)$ with RP algorithm.]{
\includegraphics[width=0.3\linewidth]{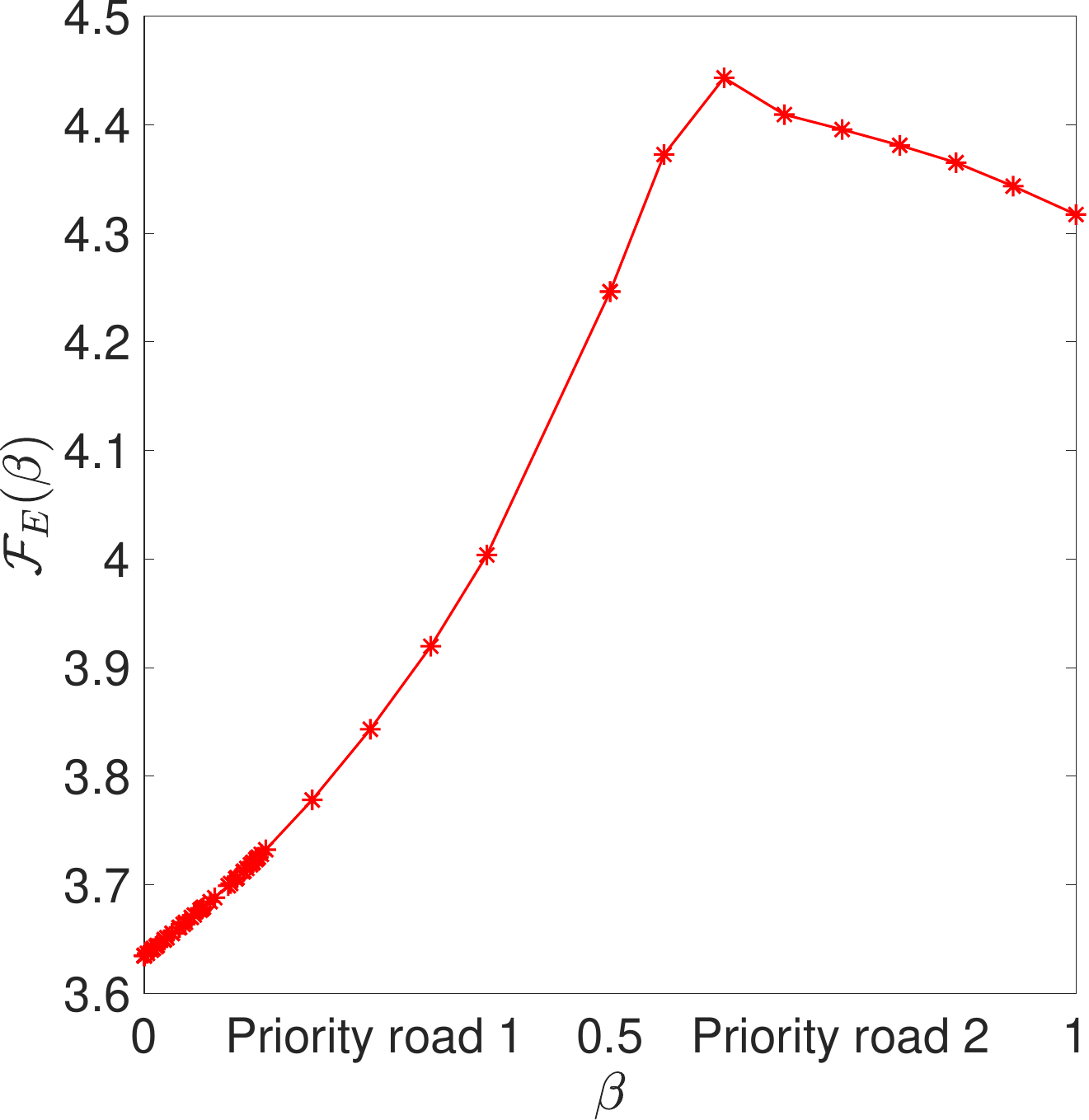}
}\quad
\subfloat[][$\funE(\beta)$ with AP algorithm.]{
\includegraphics[width=0.3\linewidth]{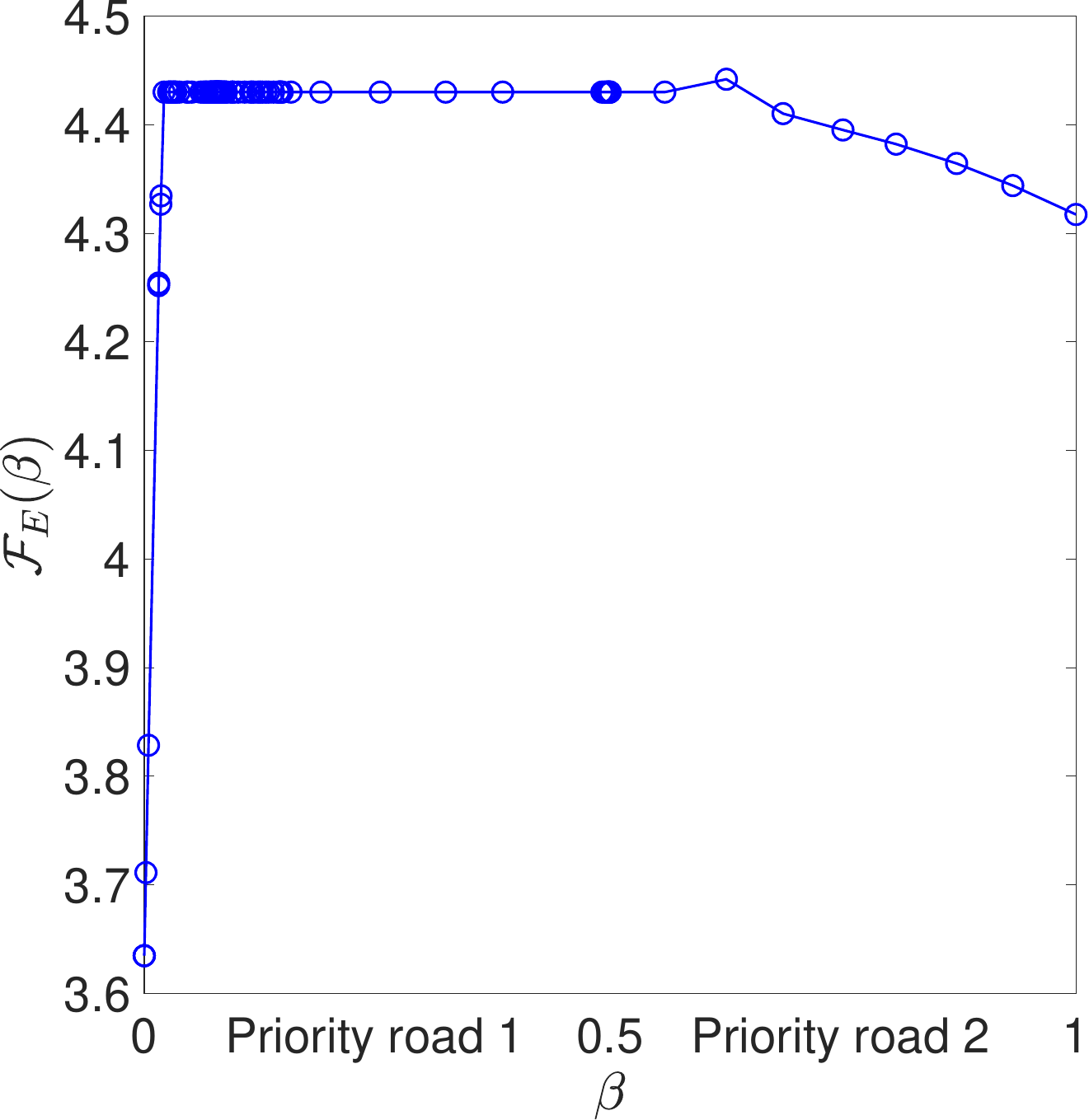}
}\\
\subfloat[][$\funET(\beta)$ with RP algorithm.]{
\includegraphics[width=0.3\linewidth]{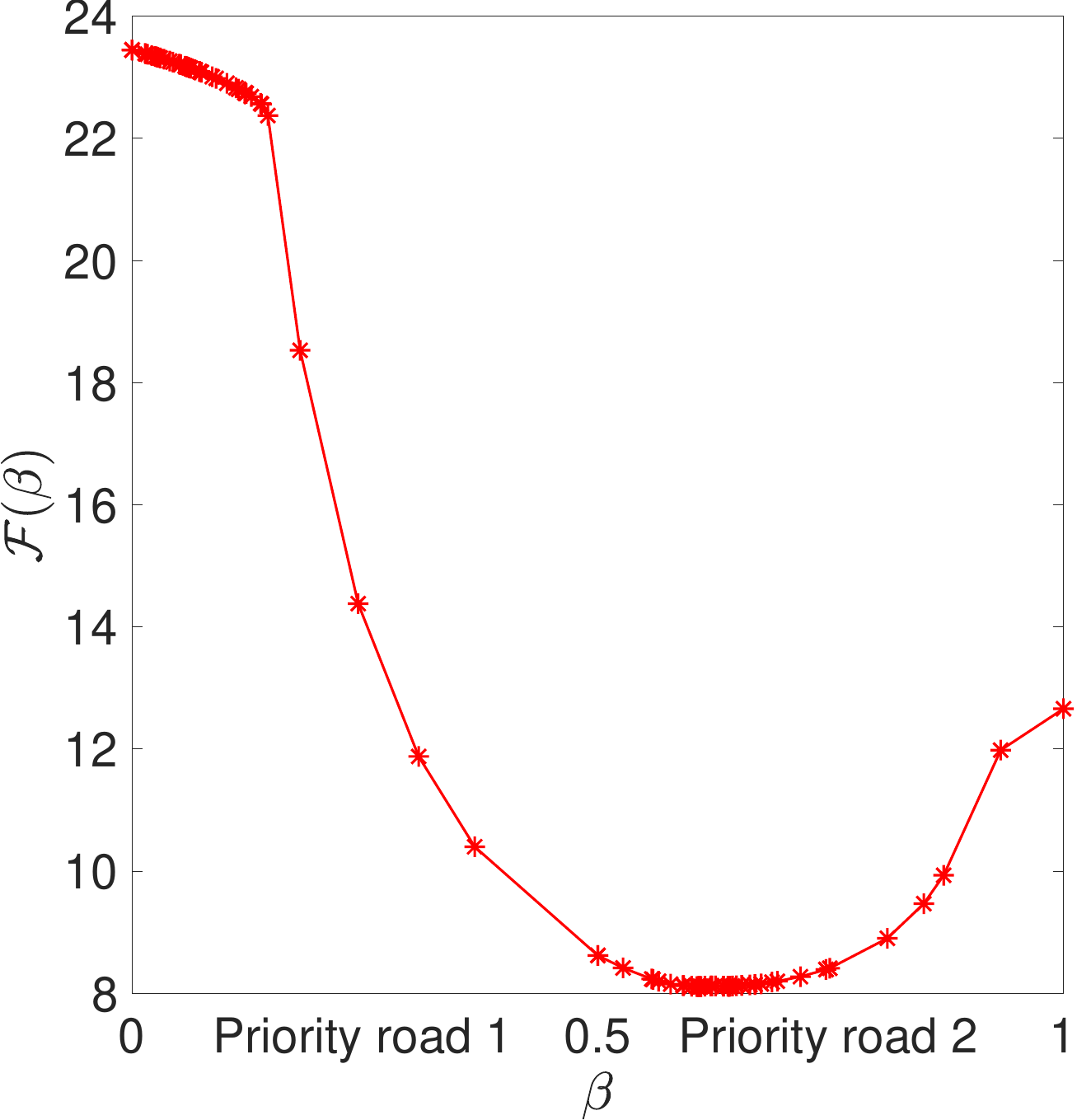}
}\quad
\subfloat[][$\funET(\beta)$ with AP algorithm.]{
\includegraphics[width=0.3\linewidth]{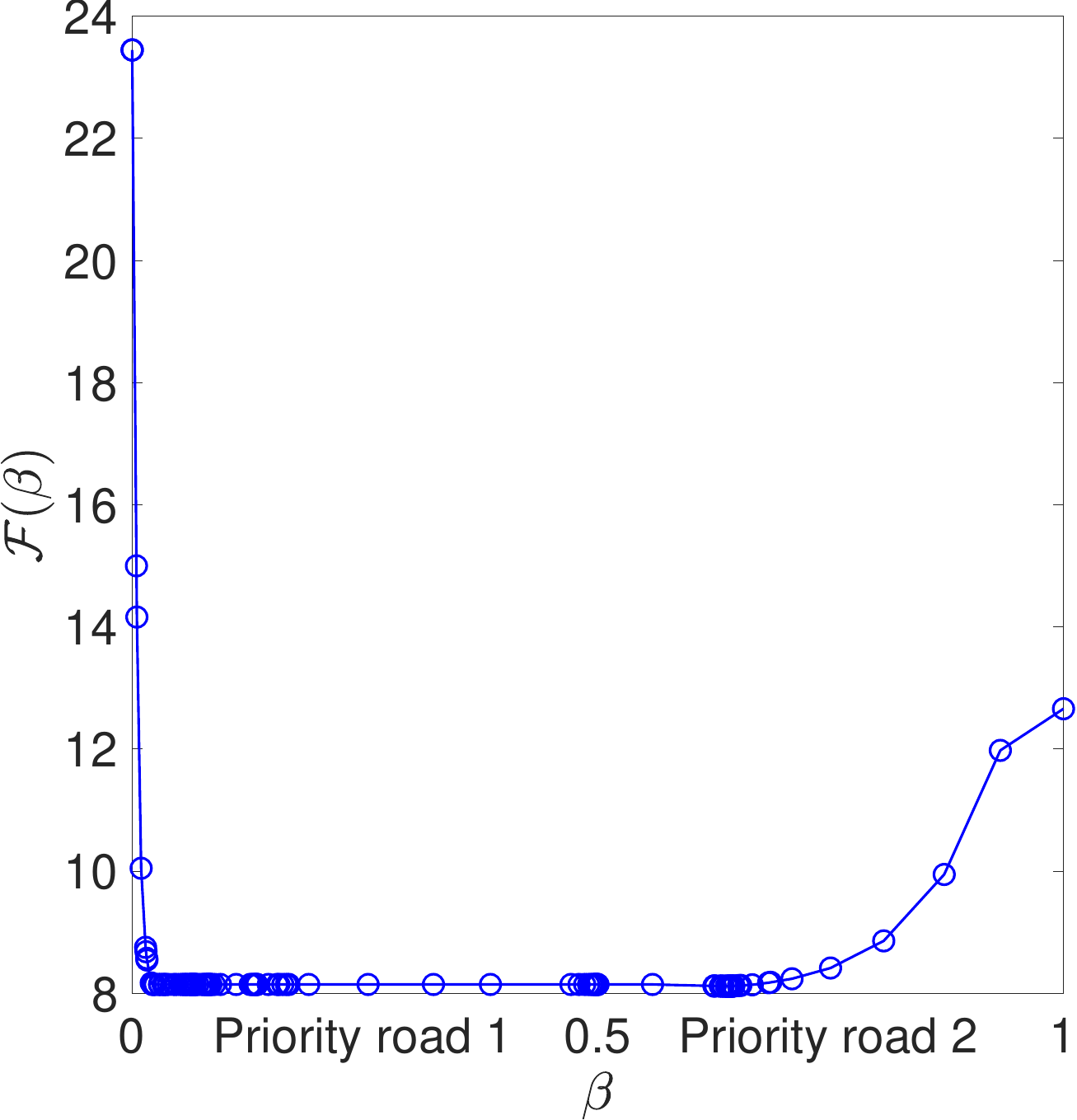}
}
\caption{Case study of a merge junction. $\funE(\beta)$ (top) and $\funET(\beta)$ (bottom) as $\beta$ changes in $[0,1]$ using the  RP and AP algorithm. The initial data is given in Table \ref{tab:merge}.}
\label{fig:test3Fun}
\end{figure}

\paragraph{Optimal priority rule}
We study the optimization problem \eqref{eq:minProb} with 
$\Gamma = [0,1]\subset\R$, where the control $\gamma=\beta$ is the priority parameter defined in \eqref{eq:prec}.
First we focus on the emission functional 
\begin{equation*}
	\funE(\gamma) \approx \frac{1}{\nx\nt\nr E^{\max}}\sum_{r=1}^{\nr}\sum_{n=1}^{\nt}\sum_{j=1}^{\nx}E^{\gamma}_{r}(x_{j},t^{n}).
\end{equation*}
We look for the parameter $\beta\in[0,1]$ which minimizes $\funE(\beta)$, and analyze the two proposed algorithms in Definition \ref{def:rp} and \ref{prop:sol}.
%, hence we simulate the traffic dynamic both respecting (RP algorithm) and adapting (AP algorithm) the priority rule. 
In Figure \ref{fig:test3Fun} top plots, we show $\funE(\beta)$ for $\beta$ varying in $[0,1]$. 
%It is immediately clear that the results are not realistic, since 
In both cases the optimal priority rule is given by $\beta^{opt}=0$, i.e.\ 
no vehicle enters the junction from road 2.
This result motivates the use of the extended functional \eqref{eq:funet} including travel times.
The test results for the functional $\funET$ with $\varepsilon=1\,\kmh$ are shown in
Figure \ref{fig:test3Fun} bottom plots. 
The optimal parameter is $\beta^{opt}=0.64$ and $\funET(\beta^{opt})=8.10$ for both algorithms.
Note that, when we use the AP algorithm, $\funET$ is close to its minimum for a large set of $\beta$ values. %The minimum value of $\funET$ coincides for the two approaches on the priority rule and it is $\funET(\beta^{opt})=8.05$.

\begin{figure}[h!]
\centering
\includegraphics[width=0.4\linewidth]{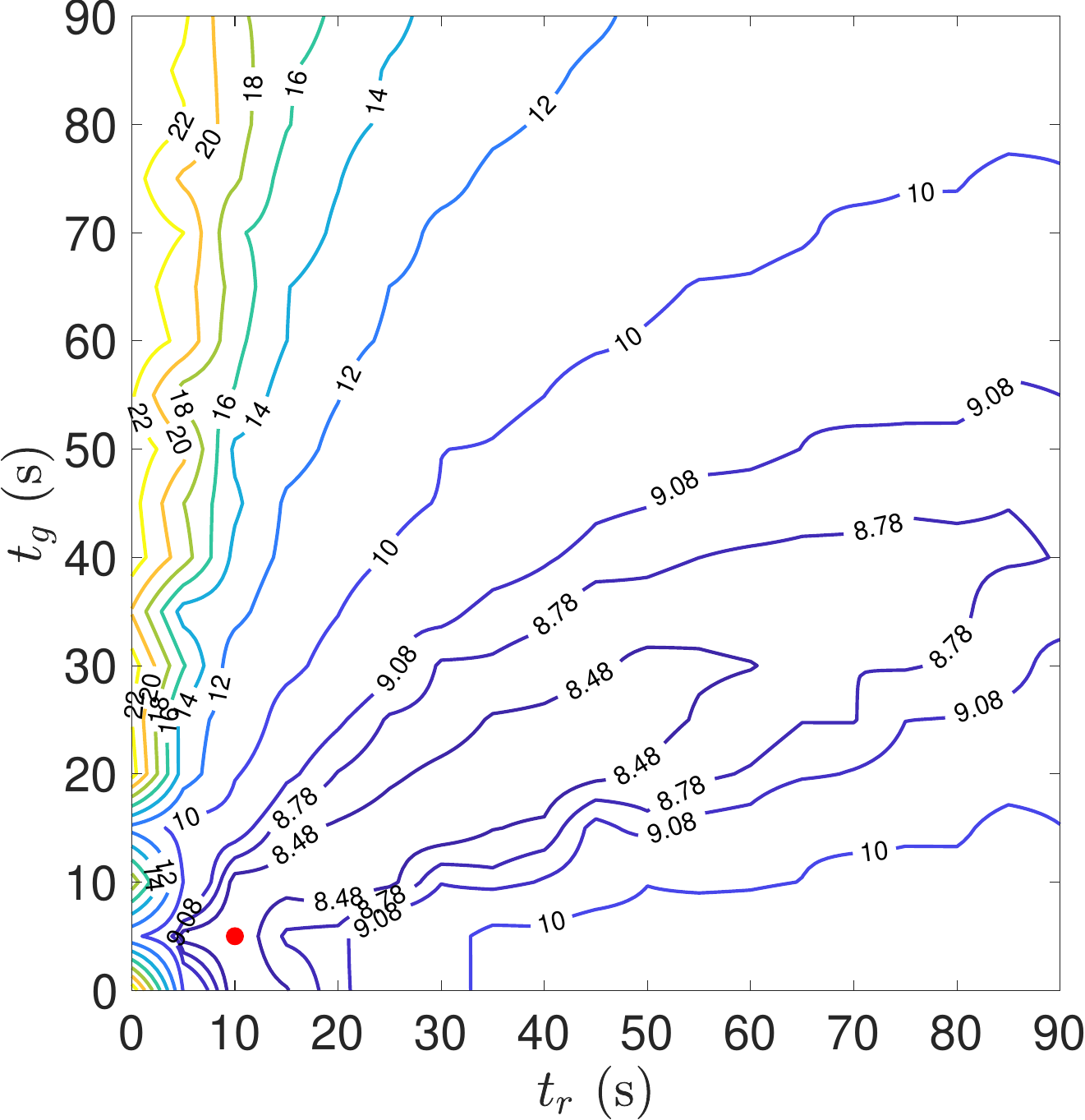}\qquad
\caption{$\funET(t_{g},t_{r})$ as $t_{g}$ and $t_{r}$ vary in $[0,90\,\mysecond]$ with initial data in Table \ref{tab:merge}. In red the optimal point $(t^{opt}_g,t^{opt}_r)=(5\,\mysecond,10\,\mysecond)$.}
\label{fig:testSem}
\end{figure}

\paragraph{Optimal traffic light}
We model a traffic light placed at the end of roads 1 and 2 (see Figure \subref*{fig:sem}) by alternating $\beta=0$ and $\beta=1$ in time.
Specifically, for $\beta=0$ the traffic light is green for road 1 and red for road 2, on the contrary for $\beta=1$ it is red for road 1 and green for road 2. The controls are given by
the green phase duration $t_{g}$ (when $\beta=0$) and red phase duration $t_{r}$ (when $\beta=1$).
The problem \eqref{eq:minProb} is studied for $\Gamma=G \times R\subset\R^{2}$, where $G$ and $R$ are the intervals where $t_g$ and $t_r$ vary, and the cost functional $\funET(\gamma) = \funET(t_{g},t_{r})$.
Fixing $G=R=[0, 90\,\mysecond]$, in Figure \ref{fig:testSem} we plot $\funET(t_{g},t_{r})$ with initial traffic data given in Table \ref{tab:merge}.
The optimal times are $t^{opt}_{g}=5\,\mysecond$ and $t^{opt}_{r}=10\,\mysecond$ and $\funET(t^{opt}_{g},t^{opt}_{r}) = 8.18$. We observe that the 
region bounded by dark-blue lines identifies the points with functional values close to the minimum one. 
Therefore, many couples $(t_g,t_r)$ allow to have low emissions and travel time. 

\bigskip
\noindent In summary, in Table \ref{tab:confrontoBetaSem} we compare the minimum values of $\funE(\gamma)$, $\funT(\gamma)$ and $\funET(\gamma)$ obtained with $\gamma=\beta^{opt}$ and $\gamma=(t^{opt}_{g},t^{opt}_{r})$. The optimal values are very close.
%Moreover, although we are unable to study the profile of $\funET$ with respect to $\gamma$, the numerous tests carried out have shown that 
The numerical tests show that
$\funET$ has a convex shape, both with respect to $\beta^{opt}$ and $(t^{opt}_{g},t^{opt}_{r})$.

\begin{table}[h!]
\centering
\small
\renewcommand{\arraystretch}{1.1}
\begin{tabular}{ccccc}\toprule
\multicolumn{1}{c}{Optimal Control} & Value & $\funE$ & $\funT$ & $\funET$\\\midrule
$\beta^{opt}$ & $0.64$ & 4.45 & 3.66 & 8.05\\\hline
$t^{opt}_{g},t^{opt}_{r}$ &$5\,\mysecond, 10\,\mysecond$ & 4.39  & 3.78 & 8.18\\\bottomrule 
\end{tabular}
\caption{Comparison of $\funE(\gamma)$, $\funT(\gamma)$ and $\funET(\gamma)$ for $\gamma=\beta^{opt}$ and $\gamma=(t^{opt}_{g},t^{opt}_{r})$.}
\label{tab:confrontoBetaSem}
\end{table}

\subsection{Sensitivity to initial data}
Here we investigate numerically  the sensitivity of the minimization problem \eqref{eq:minProb} with respect to the initial traffic states for constant initial data on all three roads.
We consider two different traffic scenarios:
\begin{enumerate}[label=(\roman*),ref=(\roman*)]
\item\label{caso1} $\rho^0_{2,3} < \rhof$, i.e.\ \textit{free flow} traffic conditions on roads 2 and 3. Specifically, we fix $\rho^0_{2} =\rho^0_{3}=15\,\vehkm$ and $w^0_{2}=w^0_{3}=(\wl+\wr)/2$ along the roads;
\item\label{caso2} $\rho^0_{2,3} > \rhof$, i.e.\ \textit{congested} traffic conditions on roads 2 and 3. Specifically, we fix $\rho^0_{2} =\rho^0_{3}=60\,\vehkm$ and $w^0_{2}=w^0_{3}=(\wl+\wr)/2$ be fixed along the roads.
\end{enumerate} 
\noindent The optimal control is computed as function of 
the initial datum on road 1: $(\rho^0_1,w^0_1)\in [0,\rhom]\times [\wl,\wr]$.

\begin{figure}[h!]
\centering
\subfloat[][Free-flow case: $\rho^{2}_{0}=\rho^{3}_{0}=15\,\vehkm$, $w^{2}_{0}=w^{3}_{0}=(\wl+\wr)/2$ (dashed lines).]
{\label{fig:betaFF}
\includegraphics[width=0.4\columnwidth]{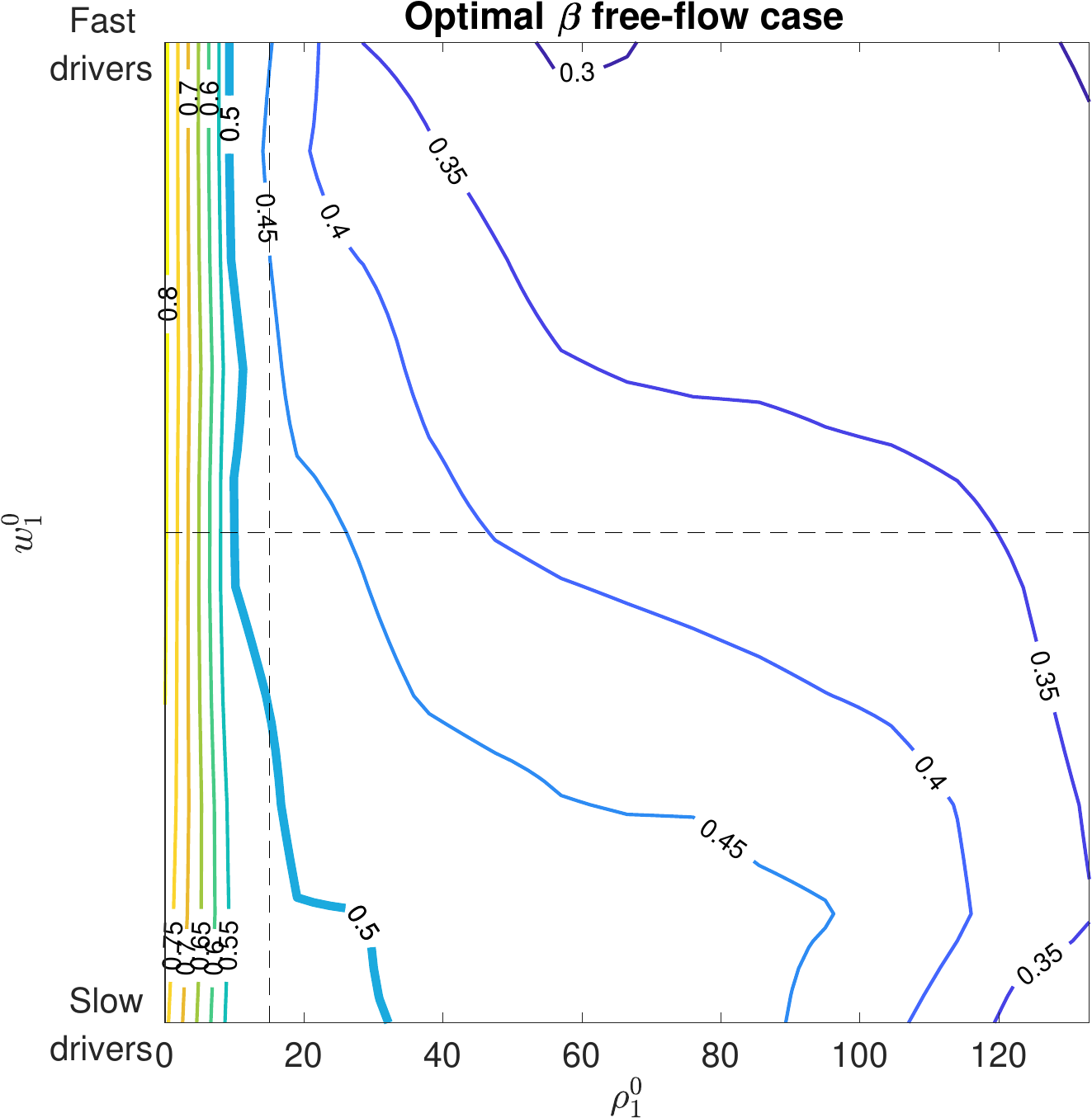}}
\quad
\subfloat[][Congested case: $\rho^{2}_{0}=\rho^{3}_{0}=60\,\vehkm$, $w^{2}_{0}=w^{3}_{0}=(\wl+\wr)/2$ (dashed lines).]
{\label{fig:betaC}
\includegraphics[width=0.4\columnwidth]{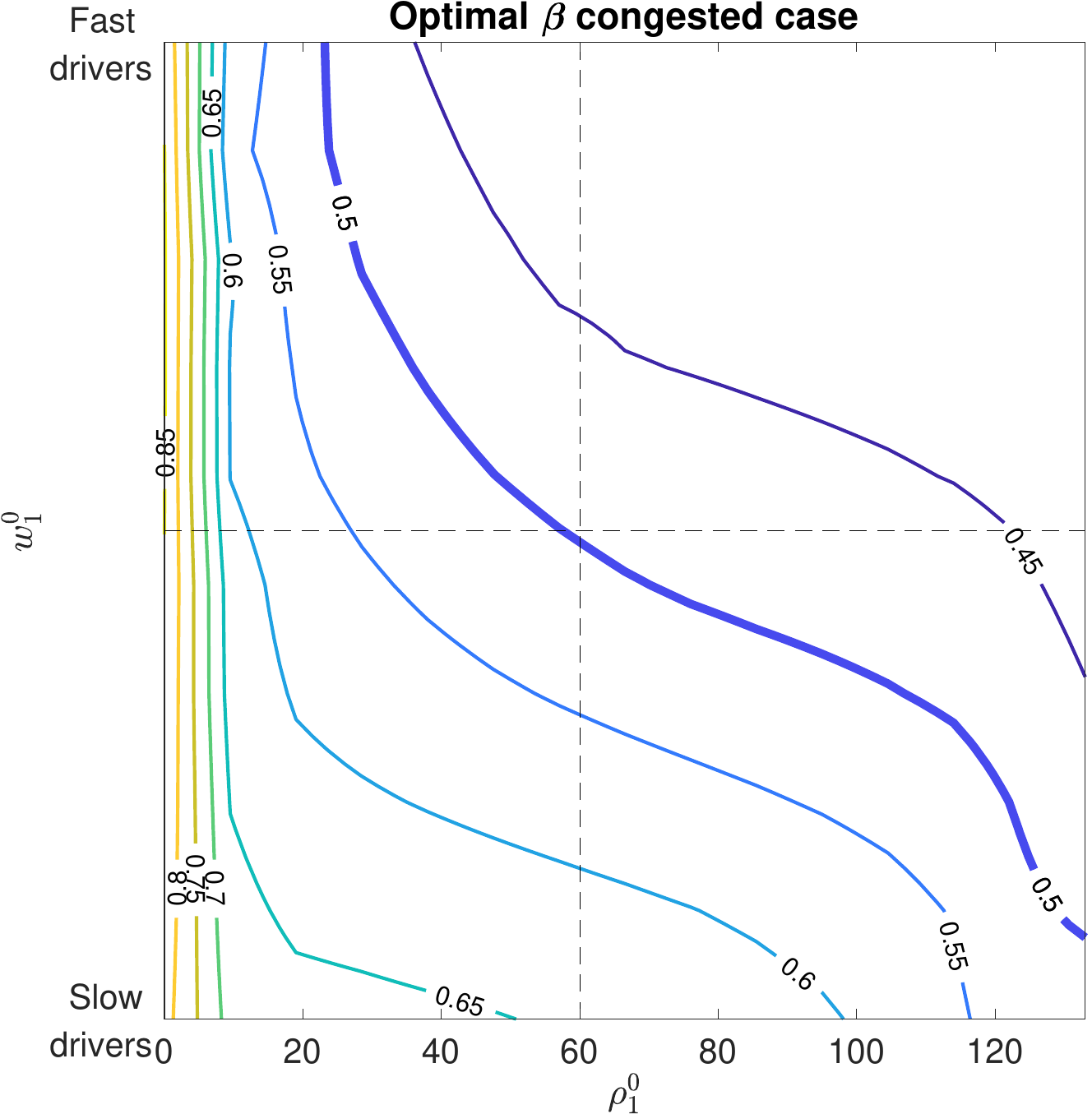}}
\caption{Optimal priority rule as $\rho^{1}_{0}$ and $w^{1}_{0}$ change. In (a) the free-flow case \ref{caso1}; In (b) the congested case \ref{caso2}. When $\beta<0.5$ road 1 has the priority, otherwise road 2 has the priority.}
\label{fig:betaTest}
\end{figure}

\paragraph{Priority rule} We focus on RP algorithm.
%We analyze how the optimal priority rule changes as  $(\rho^{0}_{1},w^{0}_{1})$ varies. 
Recall that values of $\beta<0.5$ give the priority to road 1, % i.e.\ the latter sends more vehicles, 
while values of $\beta>0.5$ give the priority to road 2. 
In Figure \ref{fig:betaTest} we highlight the level curve related to $\beta=0.5$ using a bold line.
In Figure \subref*{fig:betaFF} we show the result for the \textit{free-flow} case \ref{caso1}. The optimal priority $\beta^{opt}$ decreases as $\rho_1^0$ increases.
Specifically, if $\rho^0_1< \rho^0_2=\rho^0_3=15\,\vehkm$ then road 2 should have the priority and  $\beta^{opt}$ is independent of the speed attitude of drivers $w^{0}_{1}$. On the other hand, if $\rho^0_1>15\,\vehkm$ then road 1 should have the priority. In this case, $\beta^{opt}$ depends on $w^{0}_{1}$. In fact, it decreases more rapidly for high values of $w^{0}_{1}$. Hence, vehicles with fast drivers should cross the junction in a higher percentage ($1-\beta$) than vehicles with slow drivers.
In Figure \subref*{fig:betaC} we show the result for the \textit{congested} case \ref{caso2}. As before $\beta^{opt}$ decreases as $\rho_1^0$ increases. 
We observe that road 2 should always have the priority when slow drivers ($w^0_1=w_L$) arrive from road 1. On the other hand, road 1 should have the priority for high values of $\rho^0_1$ and $w^0_1$
(region to the right of the curve $\beta^{opt}=0.5$).

%As we can see in Figure \subref*{fig:betaFF}, in the free-flow case, the optimal priority rule is on average around 0.6, with greater variability for small $\rho^{1}_{0}$ or large $w^{1}_{0}$. When road 2 and 3 are in the congested case, instead, for high density values on road 1 the optimal priority rule stands on 0.5, but we find great variability for $\rho^{1}_{0}<\rhoc$ and for any $w^{1}_{0}$, see Figure \subref*{fig:betaC}.

\begin{figure}[h!]
\centering
\subfloat[][Level curves of $\funET$ computed with $(t^{opt}_g,t^{opt}_r)$.]{\label{fig:FunFF}
\includegraphics[width=0.4\columnwidth]{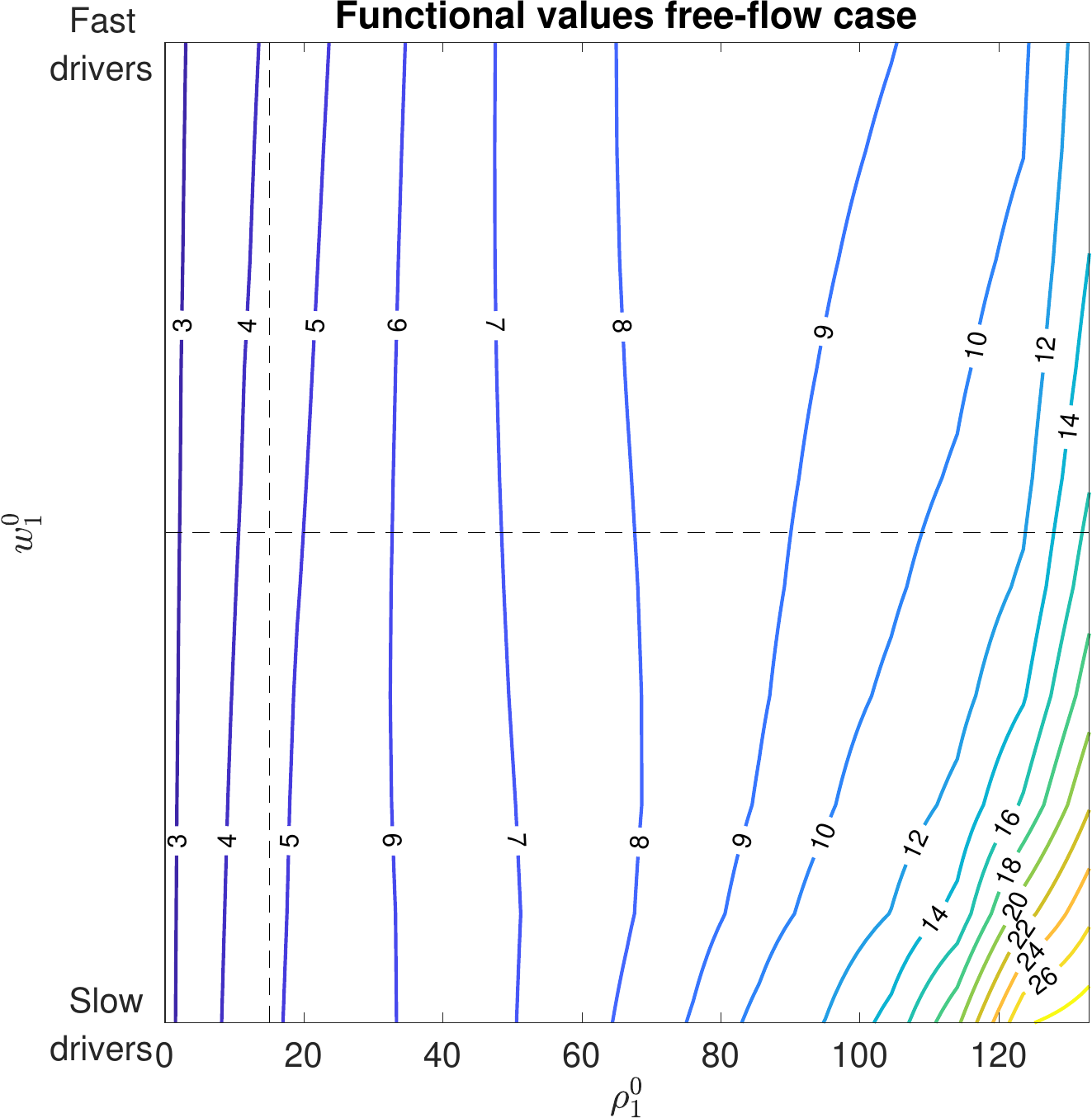}}
\quad
\subfloat[][Level curves of $t^{opt}_g/t^{opt}_r$ ratio.]{\label{fig:RappFF}
\includegraphics[width=0.4\columnwidth]{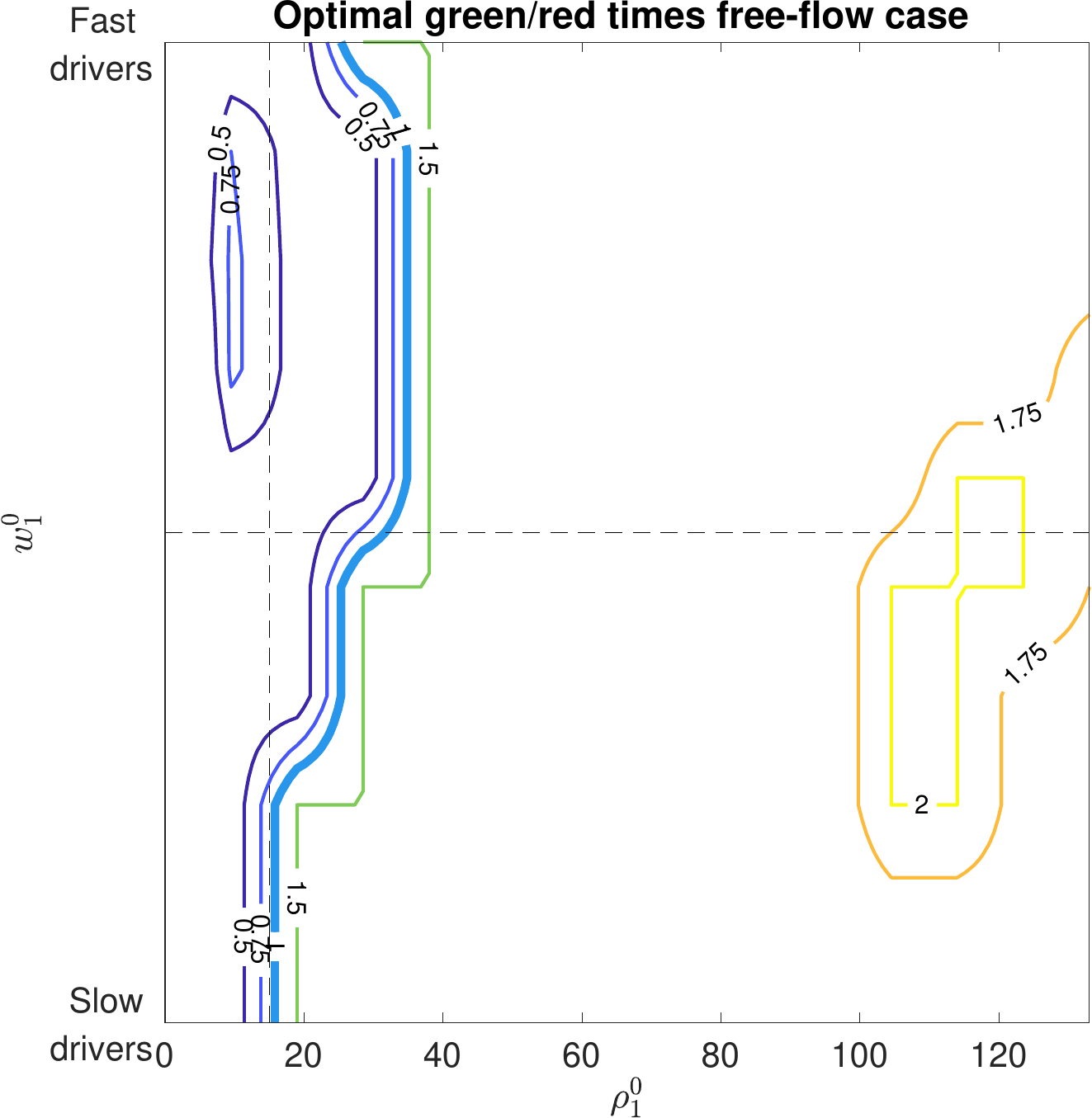}}
\caption{Optimal traffic light timing as $\rho^{0}_{1}$ and $w^{0}_{1}$ change. Roads 2 and 3 start in the free-flow phase:  $\rho^{0}_{2}=\rho^{0}_{3}=15\,\vehkm$ and $w^{0}_{2}=w^{0}_{3}=(\wl+\wr)/2$ (dashed lines).}
\label{fig:FreeFlowSemTest}
\end{figure}

\begin{figure}[h!]
\centering
\subfloat[][Level curves of $\funET$ computed with $(t^{opt}_g,t^{opt}_r)$.]{\label{fig:FunCong}
\includegraphics[width=0.4\columnwidth]{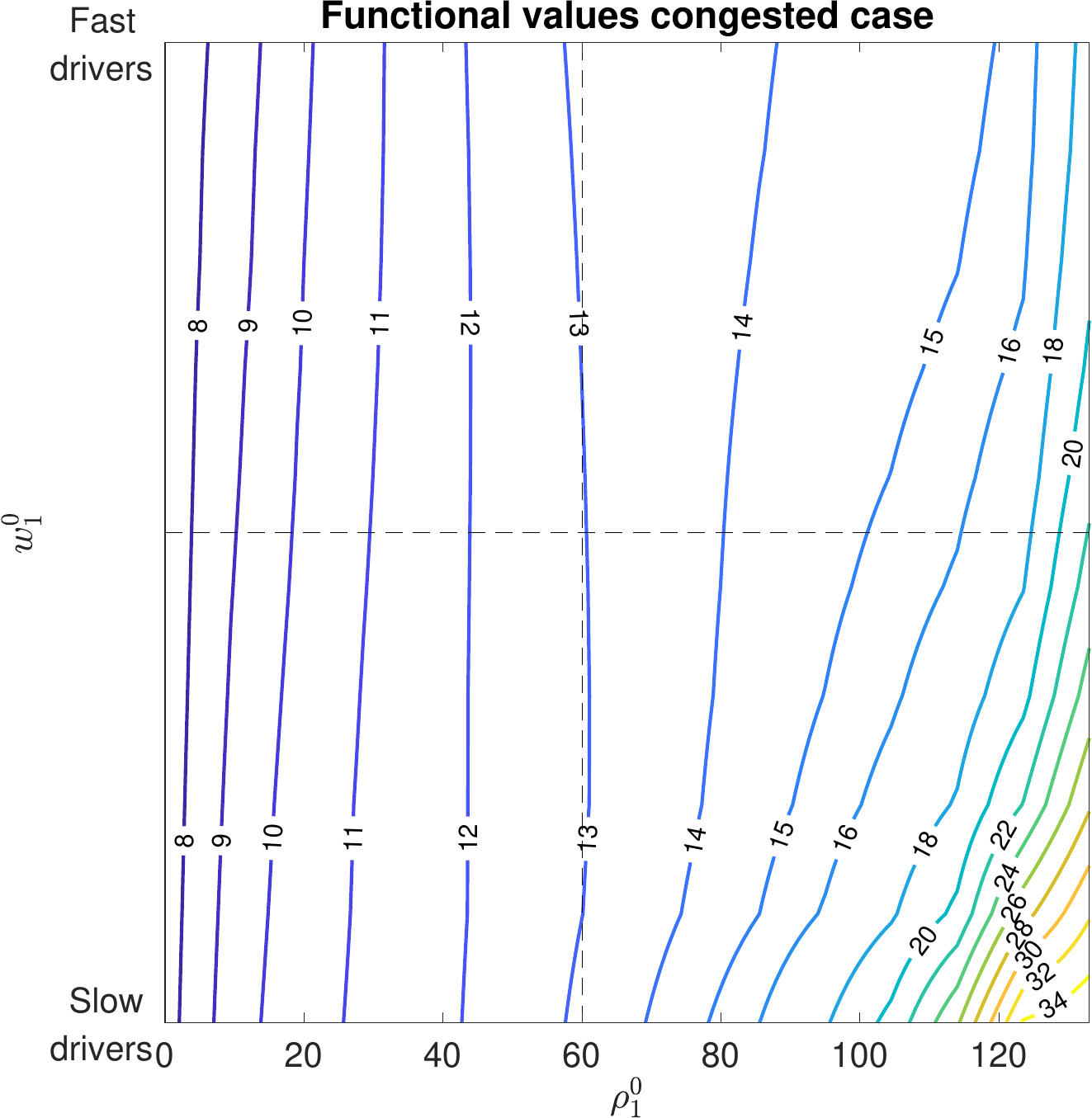}}
\quad
\subfloat[][Level curves of $t^{opt}_g/t^{opt}_r$ rate.]{\label{fig:RappCong}
\includegraphics[width=0.4\columnwidth]{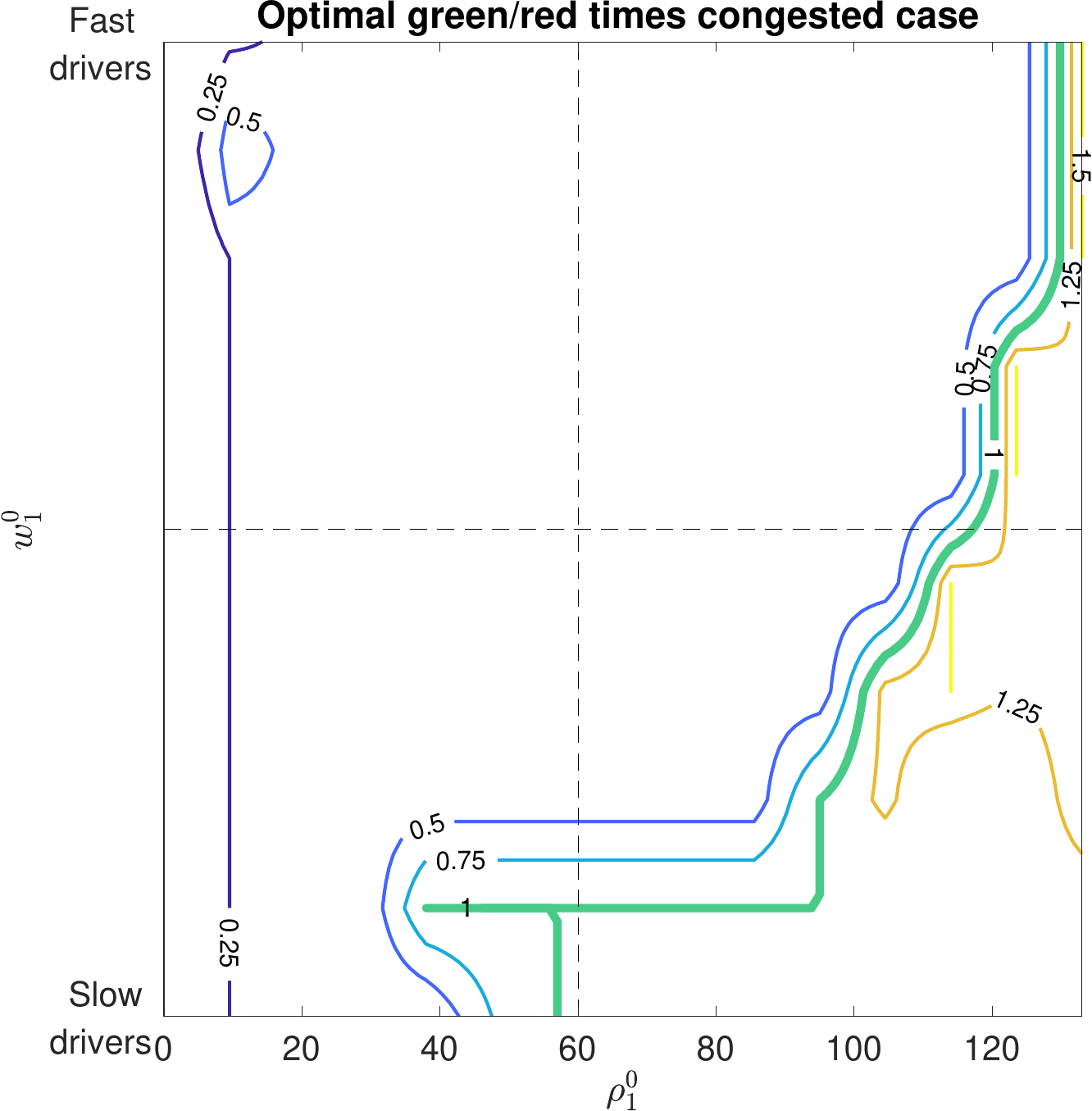}}
\caption{Optimal traffic light timing as $\rho^{0}_{1}$ and $w^{0}_{1}$ change. Roads 2 and 3 start in the congested phase:  $\rho^{0}_{2}=\rho^{0}_{3}=60\,\vehkm$ and $w^{0}_{2}=w^{0}_{3}=(\wl+\wr)/2$ (dashed lines).}
\label{fig:CongestedSemTest}
\end{figure}

\paragraph{Traffic light}
Here we analyze how the ratio between the optimal green and red duration $t^{opt}_g/t^{opt}_r$ varies with respect to $(\rho^0_1,w^0_1)$ for the two traffic scenarios \ref{caso1} and \ref{caso2}.
In Figure \ref{fig:FreeFlowSemTest} we show the result for the \emph{free-flow} case \ref{caso1}. The left plot represents the level curves of $\funET$ computed with the optimal couple $(t^{opt}_g,t^{opt}_r)$: the minimum value is increasing in $\rho^0_1$ independently of $w^0_1$; the dependence on $w^0_1$ only occurs  when many slow vehicles arrive from road 1 (bottom right of the figure). The right plot shows the level curves of the ratio $t^{opt}_g/t^{opt}_r$, where the bold line identifies the curve with $t^{opt}_g/t^{opt}_r=1$. We observe that for small values of $\rho^0_1$, the red phase should be longer than the green one. On the other hand, when $\rho^0_1$ increases, the ratio becomes greater than one, and thus vehicles coming from road 1 should have a longer green phase. 
Again, the solution is not very sensitive to the variations of the speed attitude of drivers $w^0_1$.
In Figure \ref{fig:CongestedSemTest} we show the result for the \emph{congested} case \ref{caso2}. 
The behavior of $\funET(t_g^{opt},t_r^{opt})$ on the left plot is analogous to case \ref{caso1}, while the trend of the ratio $t^{opt}_g/t^{opt}_r$ changes. Indeed, the green phase should be longer than the red one only for high values of $\rho^0_1$ and low values of $w^{0}_{1}$. %Specifically, the ratio grows faster for $\rho^0_1>\rho^0_2=\rho^0_3=60$ and $w^0_1<w^0_2=w^0_3=(\wl+\wr)/2$.
Finally we observe that in both cases, the minimum of the functional is not very sensitive to small perturbations of optimal $(t^{opt}_g,t^{opt}_r)$. 

%%%%%%%%%%%%
%%%%%%%%%%%%
%%
%\begin{figure}[h!]
%\centering
%\subfloat[][$\rho^{2}_{0}=\rho^{3}_{0}=15\,\vehkm$, $w^{2}_{0}=w^{3}_{0}=w_{M}$.]{\label{fig:redFF}
%\includegraphics[width=0.35\columnwidth]{grafici/freeflowRed.pdf}}
%\quad
%\subfloat[][$\rho^{2}_{0}=\rho^{3}_{0}=15\,\vehkm$, $w^{2}_{0}=w^{3}_{0}=w_{M}$.]{\label{fig:greenFF}
%\includegraphics[width=0.35\columnwidth]{grafici/freeflowGreen.pdf}}\\
%\subfloat[][$\rho^{2}_{0}=\rho^{3}_{0}=60\,\vehkm$, $w^{2}_{0}=w^{3}_{0}=w_{M}$.]{\label{fig:redC}
%\includegraphics[width=0.35\columnwidth]{grafici_new/congestedRed.pdf}}
%\quad
%\subfloat[][$\rho^{2}_{0}=\rho^{3}_{0}=60\,\vehkm$, $w^{2}_{0}=w^{3}_{0}=w_{M}$.]{\label{fig:greenC}
%\includegraphics[width=0.35\columnwidth]{grafici_new/congestedGreen.pdf}}
%\caption{Optimal priority rule as $\rho^{1}_{0}$ and $w^{1}_{0}$ change as the density of roads 2 and 3 is in the free-flow phase (left) or in the congested phase (right), with $w_{M}=(\wl+\wr)/2$.}
%\label{fig:semTest}
%\end{figure}

\bigskip

\noindent We can summarize the results as follows.
For the priority-ruled junction, we obtain the minimum of the functional $\funET$ by giving the priority to the incoming road with higher density and favoring fast drivers. For the traffic light too, the road with higher density should have a longer green phase. However, when the three roads are congested, vehicles with slow drivers should have a longer green phase.
As expected, the sensitivity with respect to $w$ is greater when traffic is congested, that is when it is more influenced by $w$.

%%%------------------------------%%%%%%%%
\section{Emissions at roundabouts}%: roundabout vs traffic signal}
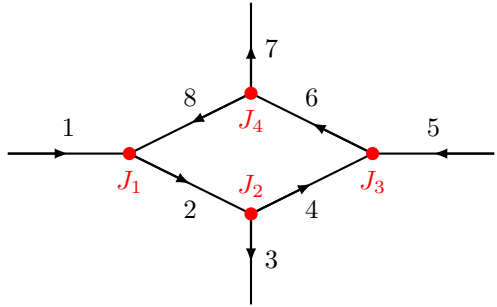
\begin{figure}[h!]
\centering
\begin{tikzpicture}[scale=0.8]
\draw[thick] (0,0) -- node[above = 0.1] {1} (2,0);
\draw[thick,->] (0,0) -- (1,0);
\draw[thick] (2,0) -- node[below=0.1] {2} (4,-1);
\draw[thick,->] (2,0) -- (3,-0.5);
\draw[thick] (4,-1) -- node[right=0.05] {3} (4,-2.5);
\draw[thick,->] (4,-1) -- (4,-1.8);
\draw[thick] (4,-1) -- node[below=0.1] {4} (6,0);
\draw[thick,->] (4,-1) -- (5,-0.5);
\draw[thick] (6,0) -- node[above = 0.1] {5} (8,0);
\draw[thick,->] (8,0) -- (7,0);
\draw[thick] (6,0) -- node[above = 0.1] {6} (4,1);
\draw[thick,->] (6,0) -- (5,0.5);
\draw[thick] (4,1) -- node[right=0.05] {7} (4,2.5);
\draw[thick,->] (4,1) -- (4,1.8);
\draw[thick] (2,0) -- node[above = 0.1] {8} (4,1);
\draw[thick,->] (4,1) -- (3,0.5);
\draw[red,fill=red] (2,0) circle (0.1cm) node[below=0.1] {$J_{1}$};
\draw[red,fill=red] (4,-1) circle (0.1cm) node[above=0.1] {$J_{2}$};
\draw[red,fill=red] (6,0) circle (0.1cm) node[below=0.1] {$J_{3}$};
\draw[red,fill=red] (4,1) circle (0.1cm) node[below=0.1] {$J_{4}$};
\end{tikzpicture}
\caption{Example of roundabout.}
\label{fig:rotatoria}
\end{figure}

\label{sec:rotatoria}
In this section we study emissions and travel times for a roundabout, modeled 
combining merge and diverge junctions as depicted in Figure \ref{fig:rotatoria}.
There are four junctions: $J_{1}$ and $J_{3}$ of type $2\to 1$ (merge); $J_{2}$ and $J_{4}$ of type $1\to 2$ (diverge).
We focus on the AP algorithm to compute the minimum of problem \eqref{eq:minProb}, obtaining the priority parameters $\gamma=(\beta_{J_{1}},\beta_{J_{3}})\in[0,1]^{2}$. 
%We focus on the AP algorithm to compute the optimal priority parameters $\gamma=(\beta_{J_{1}},\beta_{J_{3}})\in[0,1]^{2}$ solving the minimization problem \eqref{eq:minProb}. 
We also compute the optimal timing $\gamma=((t_{g},t_{r})_{J_{1}},(t_{g},t_{r})_{J_{3}})\in [G\times R]^{2}$ 
%which minimizes the functional $\funET$ on 
for the roundabout with traffic lights placed at the two merge junctions $J_{1}$ and $J_{3}$, with $G=R=[25\,\mysecond,90\,\mysecond]$. We exclude traffic light phases smaller than $25\,\mysecond$, and compare the roundabout with priorities with that with traffic lights. 

\begin{figure}[h!]
\centering
\subfloat[][Optimal priority: $\nox$ at $t=5\,\min$.]
{\label{fig:em1}\begin{overpic}[width=0.27\columnwidth]{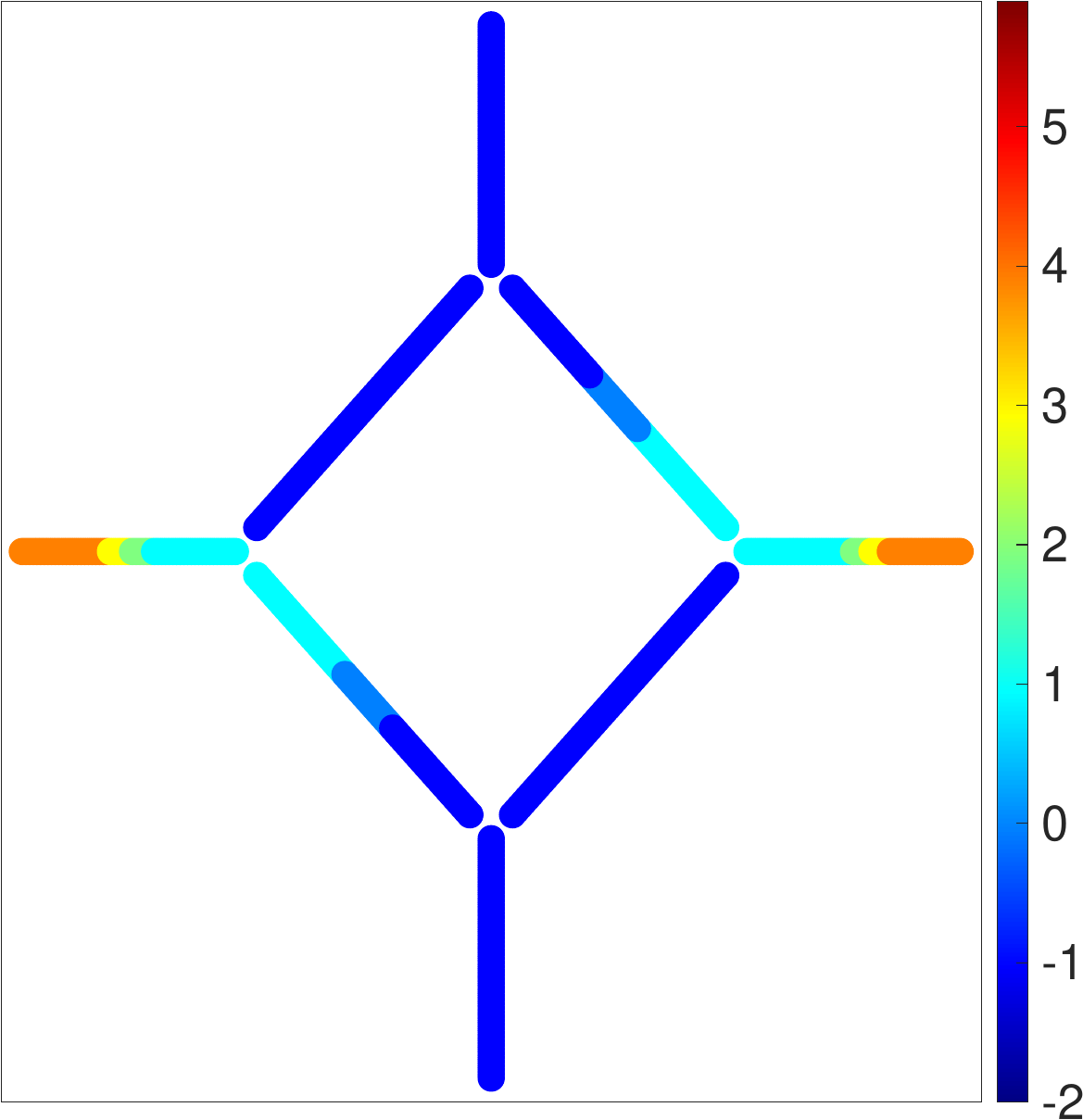}
\put(5,45){\vector(1,0){15}} \put(23,40){\vector(0.95,-1){12.5}}
\put(40,20){\vector(0,-1){12.5}} \put(54,28){\vector(0.95,1){12.5}}
\put(85,54){\vector(-1,0){15}} \put(66,58.5){\vector(-0.95,1){12.5}}
\put(48,78){\vector(0,1){12.5}} \put(34,71){\vector(-0.95,-1){12.5}}
\end{overpic}
} \quad
\subfloat[][Optimal priority: $\nox$ at $t=30\,\minute$.]
{\label{fig:em2}\begin{overpic}[width=0.27\columnwidth]{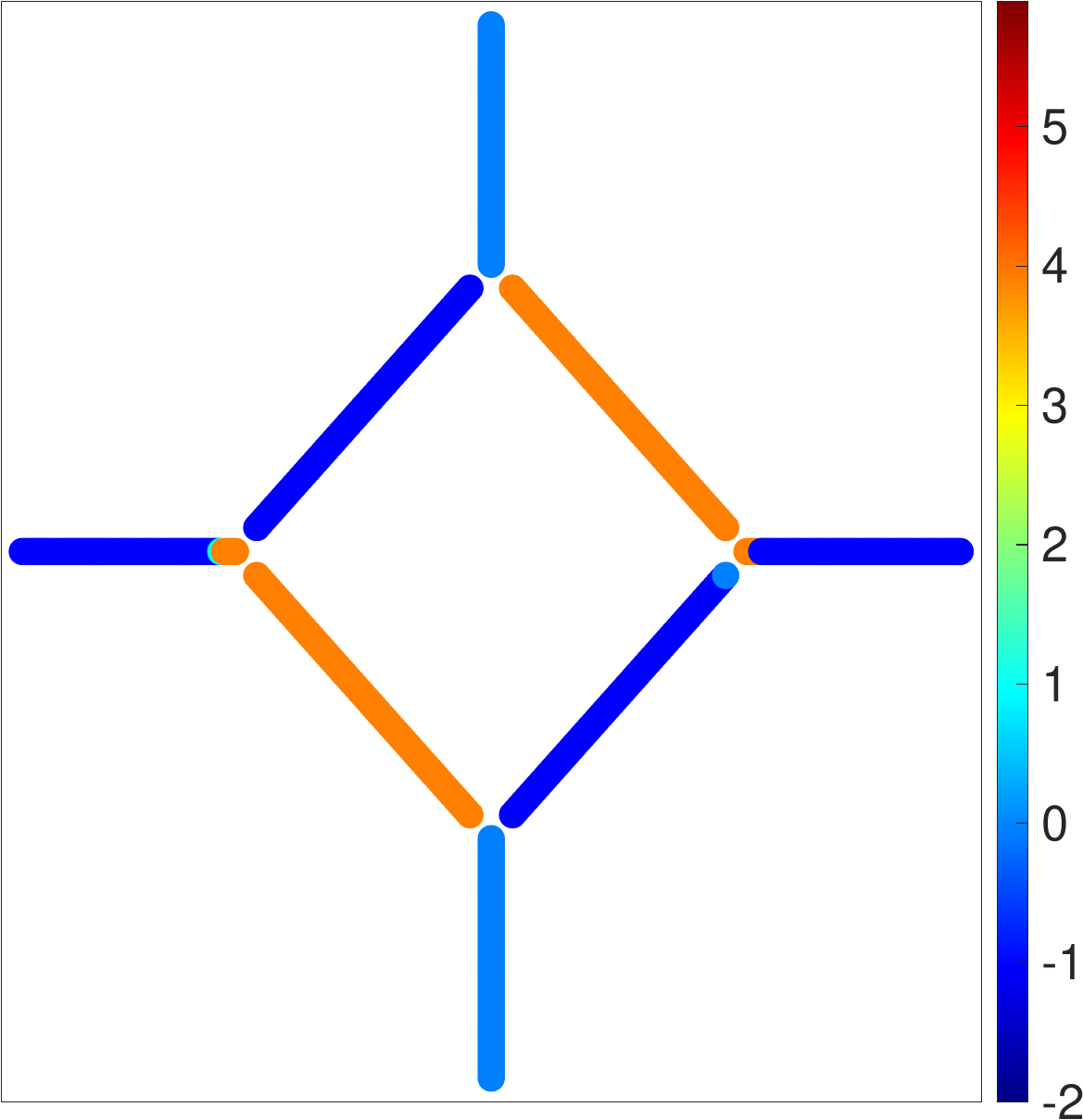}
\put(5,45){\vector(1,0){15}} \put(23,40){\vector(0.95,-1){12.5}}
\put(40,20){\vector(0,-1){12.5}} \put(54,28){\vector(0.95,1){12.5}}
\put(85,54){\vector(-1,0){15}} \put(66,58.5){\vector(-0.95,1){12.5}}
\put(48,78){\vector(0,1){12.5}} \put(34,71){\vector(-0.95,-1){12.5}}
\end{overpic}
} \quad
\subfloat[][Optimal priority: $\nox$ at $t=1\,\myhour$.]
{\label{fig:em3}\begin{overpic}[width=0.27\columnwidth]{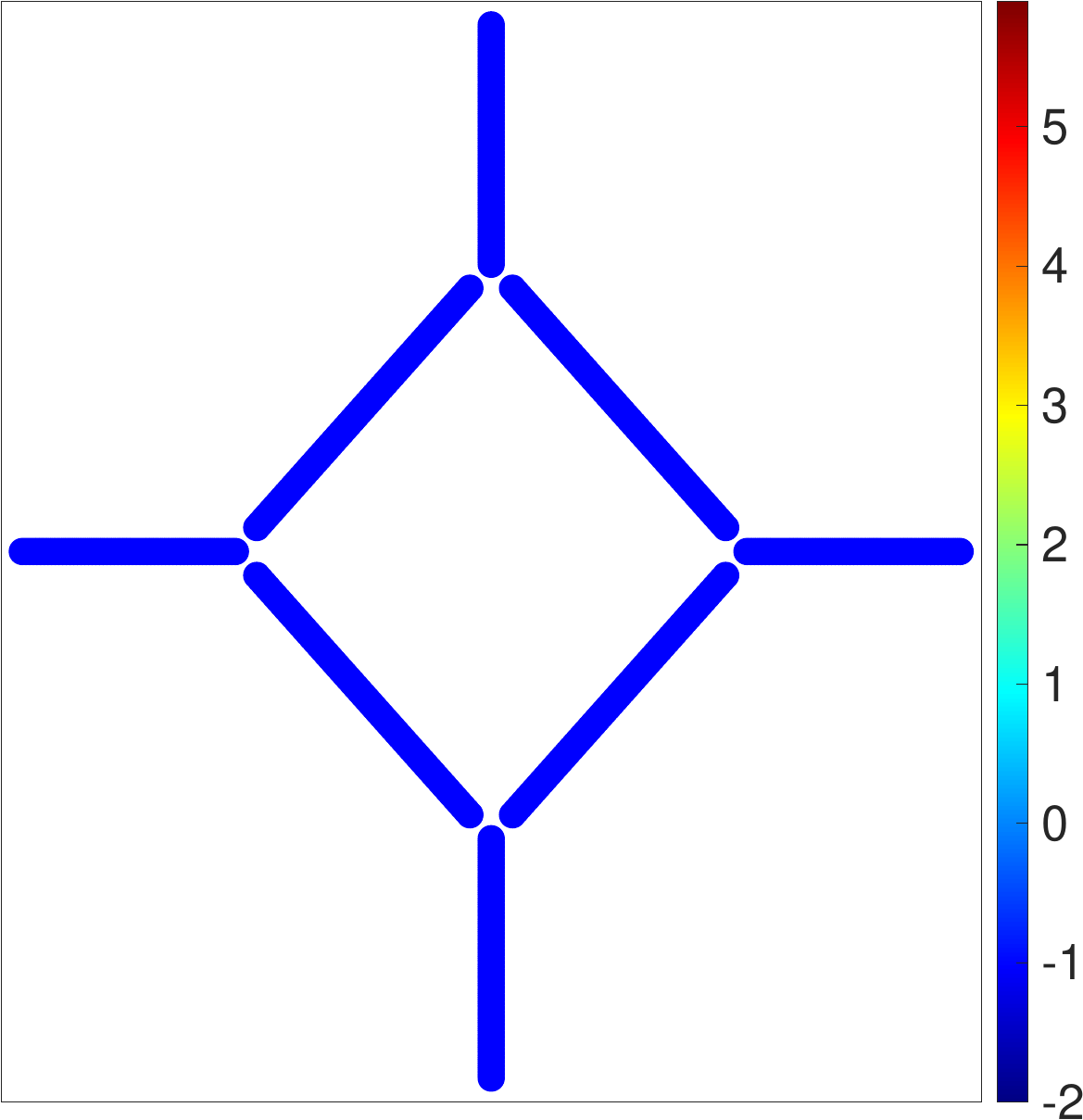}
\put(5,45){\vector(1,0){15}} \put(23,40){\vector(0.95,-1){12.5}}
\put(40,20){\vector(0,-1){12.5}} \put(54,28){\vector(0.95,1){12.5}}
\put(85,54){\vector(-1,0){15}} \put(66,58.5){\vector(-0.95,1){12.5}}
\put(48,78){\vector(0,1){12.5}} \put(34,71){\vector(-0.95,-1){12.5}}
\end{overpic}
} \\
\subfloat[][Optimal traffic light: $\nox$ at $t=5\,\min$.]
{\label{fig:em4}\begin{overpic}[width=0.27\columnwidth]{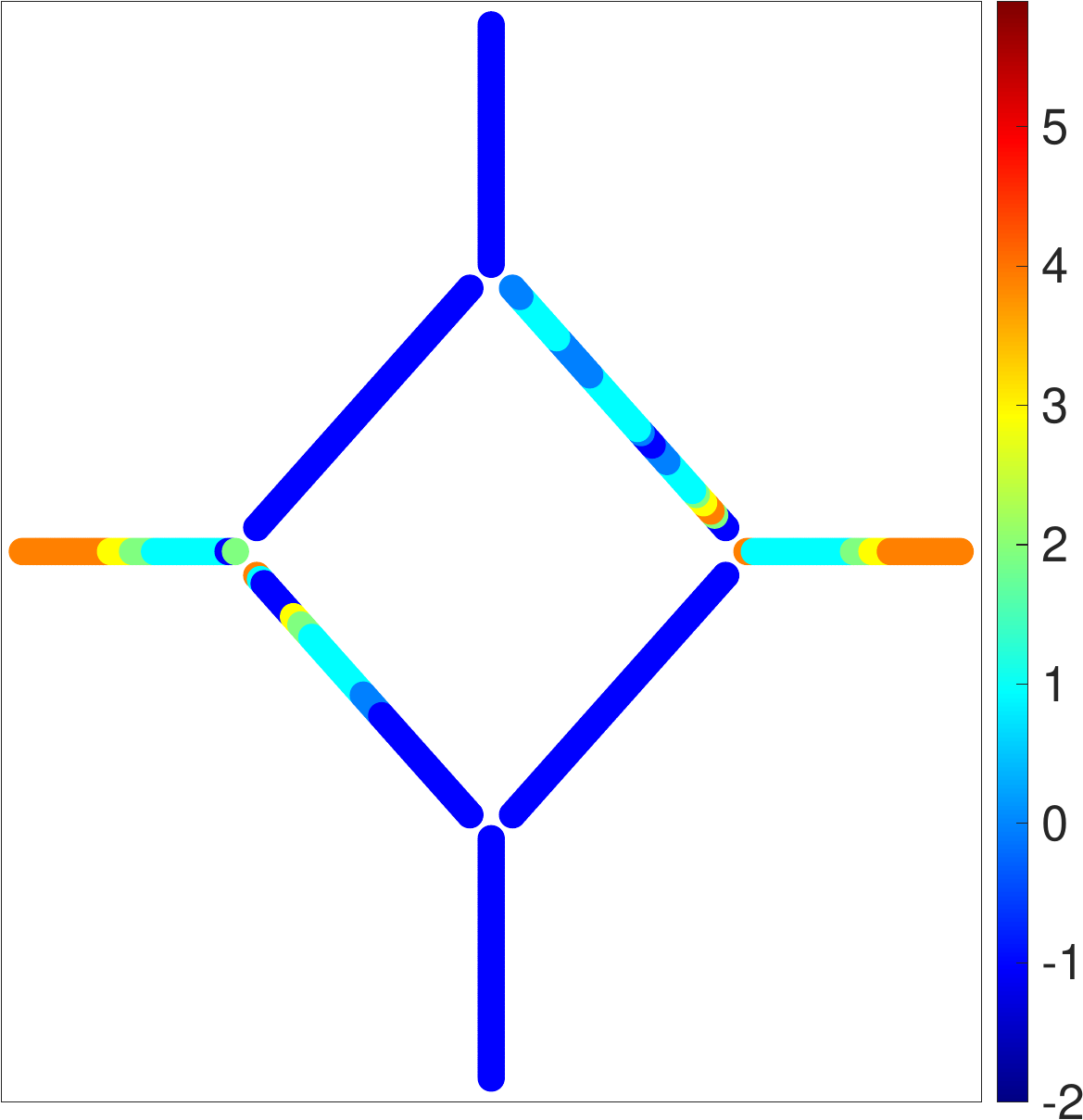}
\put(5,45){\vector(1,0){15}} \put(23,40){\vector(0.95,-1){12.5}}
\put(40,20){\vector(0,-1){12.5}} \put(54,28){\vector(0.95,1){12.5}}
\put(85,54){\vector(-1,0){15}} \put(66,58.5){\vector(-0.95,1){12.5}}
\put(48,78){\vector(0,1){12.5}} \put(34,71){\vector(-0.95,-1){12.5}}
\put(12,53.5){\polygon*(0,-0.3)(0,10.6)(8,10.6)(8,-0.3)}
\put(16,61.3){\color{red}\circle{5}} \put(16,56.1){\color{green}\circle*{5}}
\put(70,36.5){\polygon*(0,-0.3)(0,10.6)(8,10.6)(8,-0.3)}
\put(74,44.3){\color{red}\circle*{5}} \put(74,39.1){\color{green}\circle{5}}
\end{overpic}
}\quad
\subfloat[][Optimal traffic light: $\nox$ at $t=30\,\minute$.]
{\label{fig:em5}\begin{overpic}[width=0.27\columnwidth]{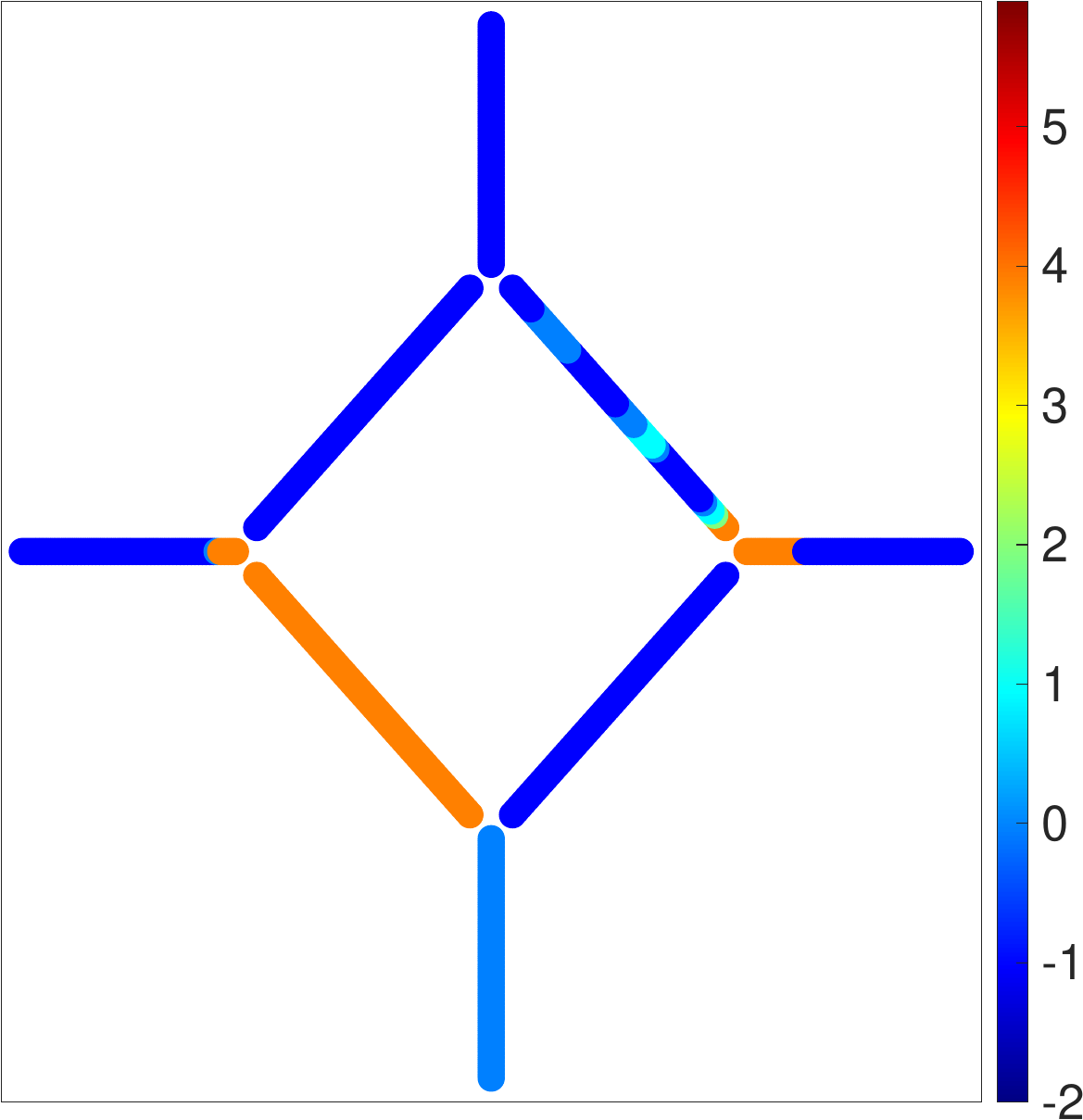}
\put(5,45){\vector(1,0){15}} \put(23,40){\vector(0.95,-1){12.5}}
\put(40,20){\vector(0,-1){12.5}} \put(54,28){\vector(0.95,1){12.5}}
\put(85,54){\vector(-1,0){15}} \put(66,58.5){\vector(-0.95,1){12.5}}
\put(48,78){\vector(0,1){12.5}} \put(34,71){\vector(-0.95,-1){12.5}}
\put(12,53.5){\polygon*(0,-0.3)(0,10.6)(8,10.6)(8,-0.3)}
\put(16,61.3){\color{red}\circle*{5}} \put(16,56.1){\color{green}\circle{5}}
\put(70,36.5){\polygon*(0,-0.3)(0,10.6)(8,10.6)(8,-0.3)}
\put(74,44.3){\color{red}\circle{5}} \put(74,39.1){\color{green}\circle*{5}}
\end{overpic}
}\quad
\subfloat[][Optimal traffic light: $\nox$ at $t=1\,\myhour$.]
{\label{fig:em6}\begin{overpic}[width=0.27\columnwidth]{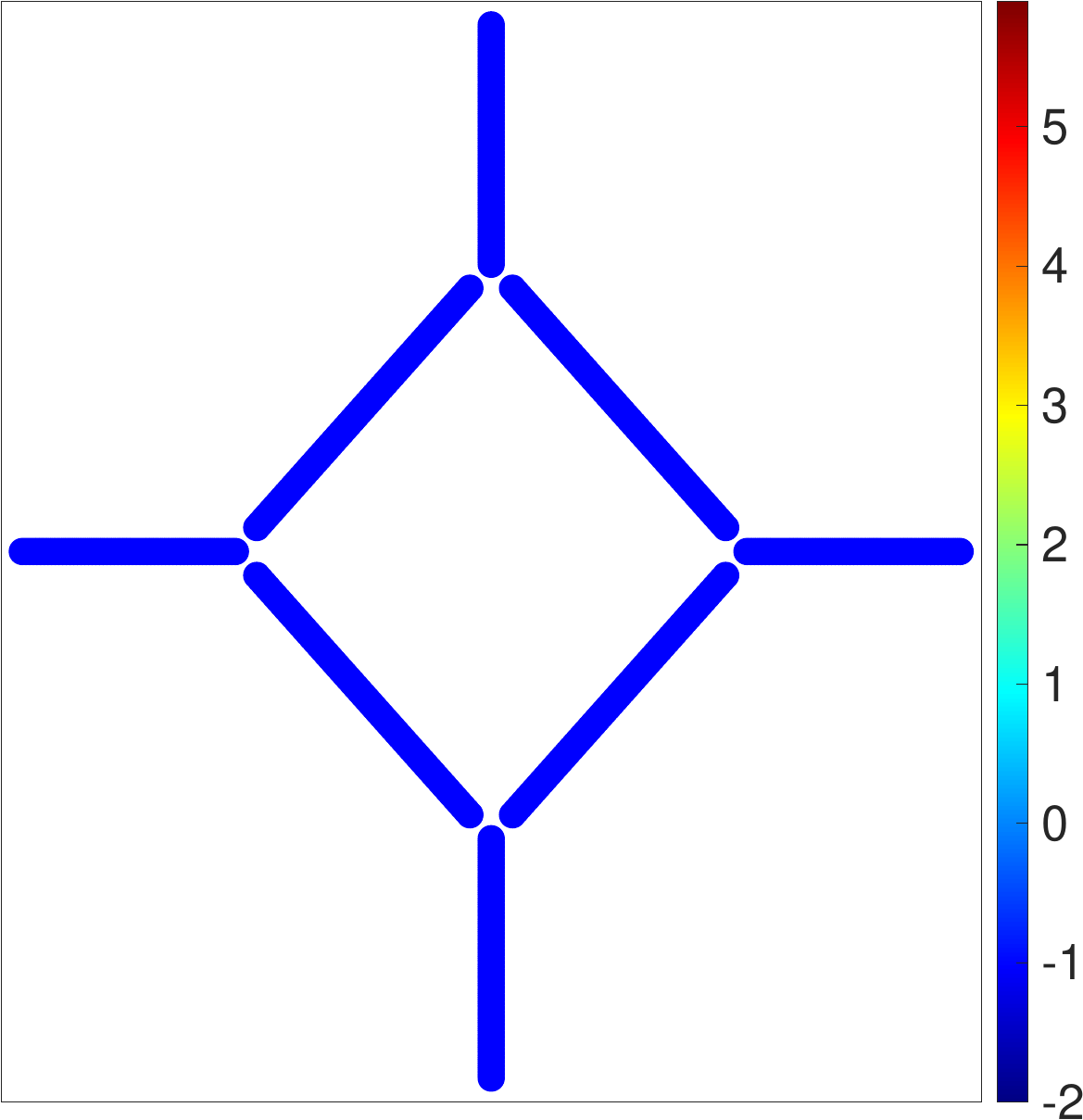}
\put(5,45){\vector(1,0){15}} \put(23,40){\vector(0.95,-1){12.5}}
\put(40,20){\vector(0,-1){12.5}} \put(54,28){\vector(0.95,1){12.5}}
\put(85,54){\vector(-1,0){15}} \put(66,58.5){\vector(-0.95,1){12.5}}
\put(48,78){\vector(0,1){12.5}} \put(34,71){\vector(-0.95,-1){12.5}}
\put(12,53.5){\polygon*(0,-0.3)(0,10.6)(8,10.6)(8,-0.3)}
\put(16,61.3){\color{red}\circle{5}} \put(16,56.1){\color{green}\circle*{5}}
\put(70,36.5){\polygon*(0,-0.3)(0,10.6)(8,10.6)(8,-0.3)}
\put(74,44.3){\color{red}\circle*{5}} \put(74,39.1){\color{green}\circle{5}}
\end{overpic}}
\caption{$\nox$ emission rates $(\mygram/\myhour)$ on a network with priority rules (top) and traffic lights (bottom) in $J_{1}$  and $J_{3}$. }
\label{fig:rotatoriaEm2}
\end{figure}

\begin{figure}[h!]
\centering
\subfloat[][$\bar\rho=15\,\vehkm$]{\label{fig:gammaFisso}
\includegraphics[width=0.3\columnwidth]{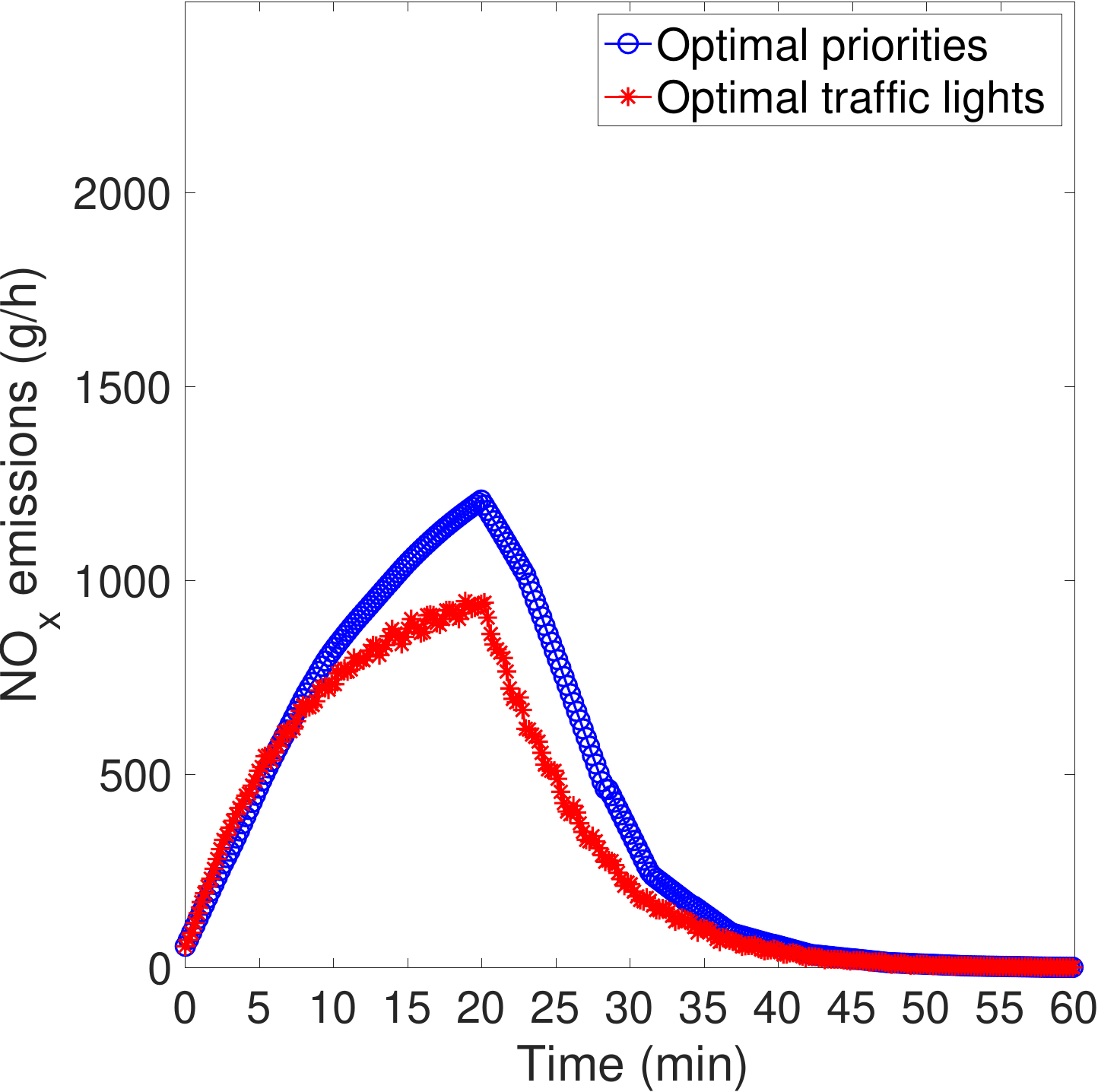}
}\,
\subfloat[][$\bar\rho=40\,\vehkm$]{\label{fig:gammaVariabile}
\includegraphics[width=0.3\columnwidth]{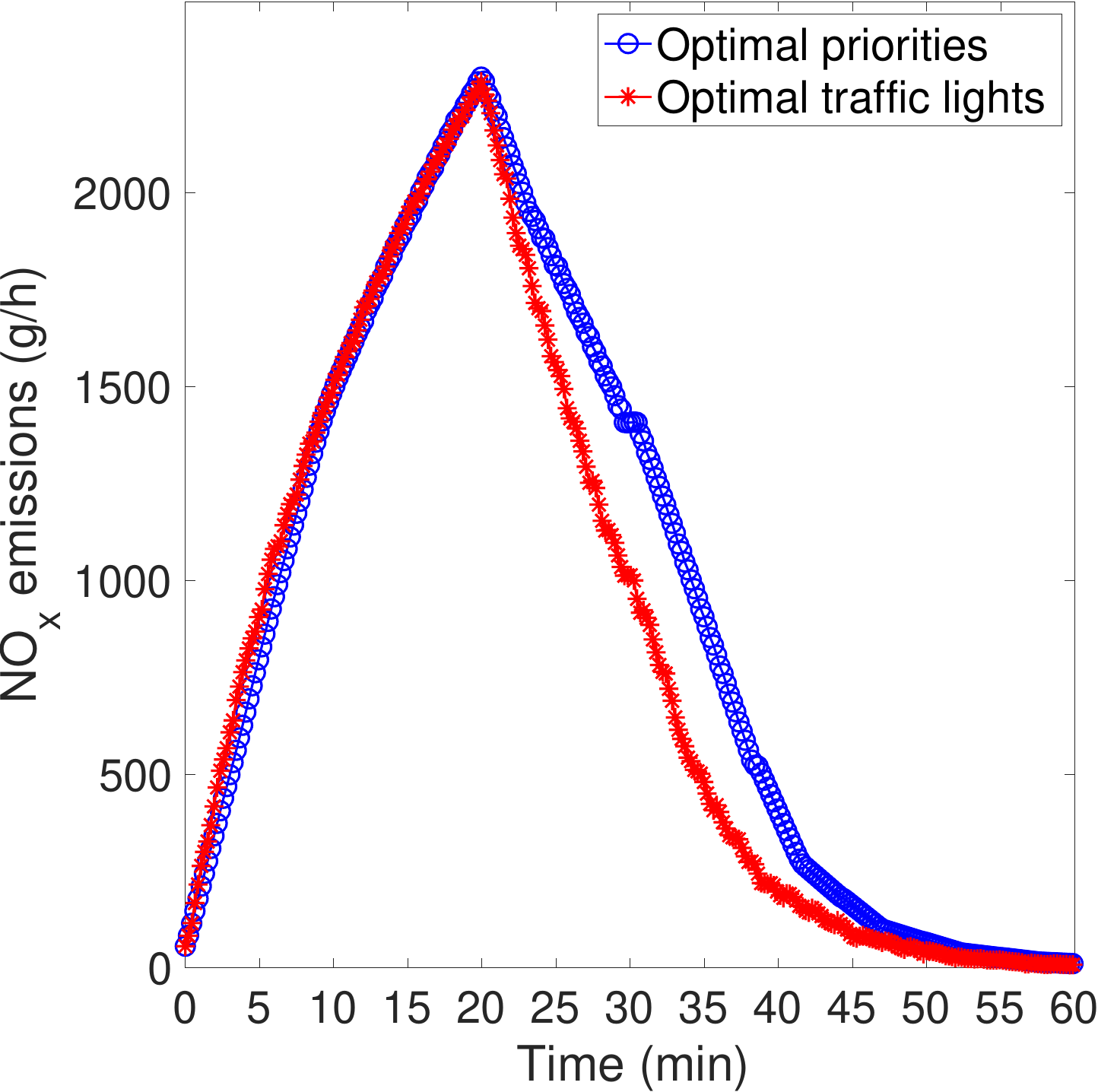}
}\,
\subfloat[][$\bar\rho=80\,\vehkm$]{
\includegraphics[width=0.3\columnwidth]{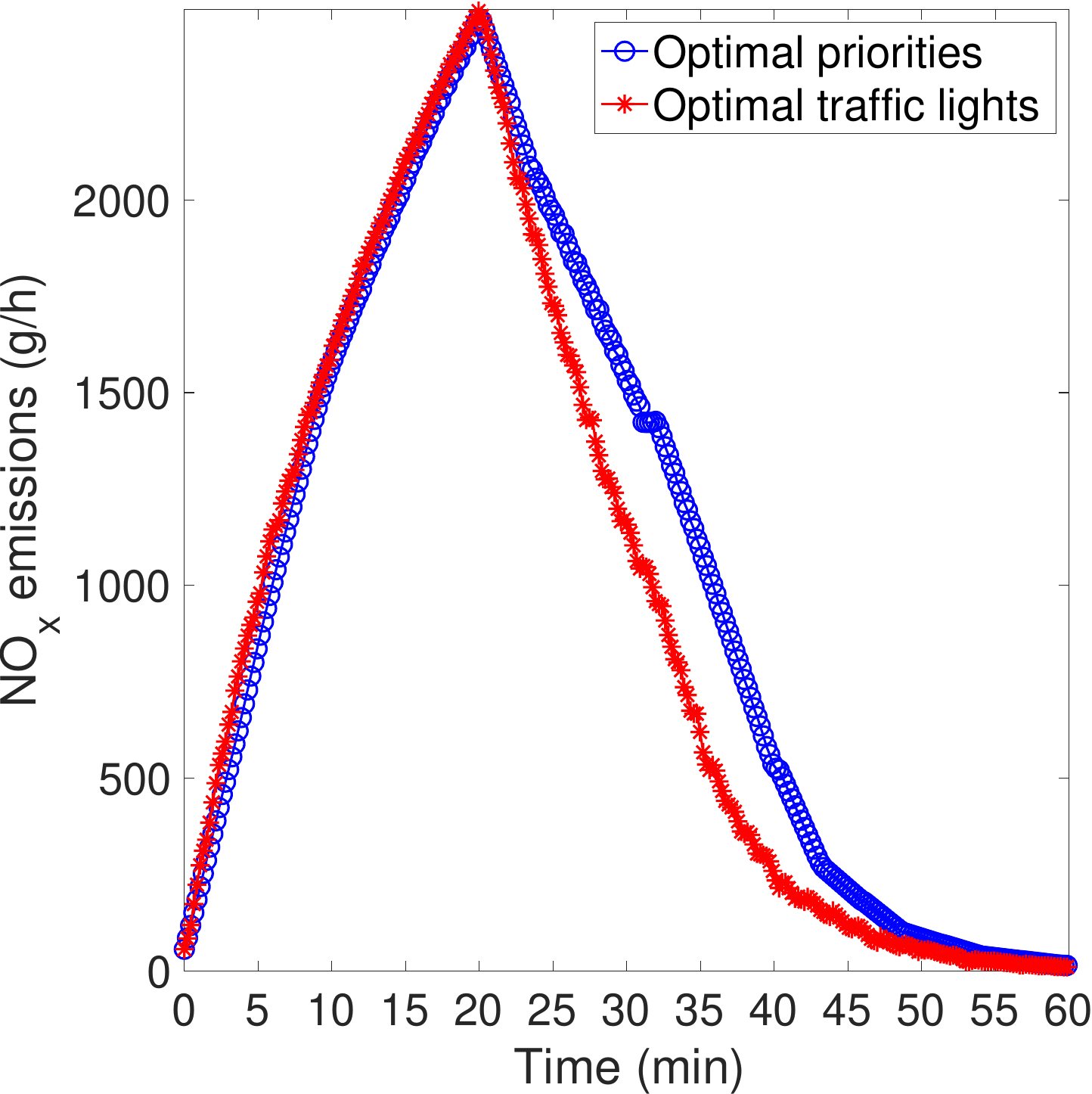}
}
\caption{Total $\nox$ emission rates ($\mygram/\myhour$) along the whole roundabout.}
\label{fig:emTot2}
\end{figure}

The two diverging junctions $J_2$ and $J_4$ have a fixed distribution parameter $\alpha=0.6$.  The model parameters $\rhof$, $\rhom$, $\rhoc$ and $\vmax$, the length of the roads $L$ and the space step $\dx$ are fixed as in Table \ref{tab:parametriTest}. The length of the simulations is $T=1\,\myhour$ and the time step $\dt=2.57\,\mysecond$. 
The initial density is assumed to be null for each road. 
%%%%%%%%%%%%%%
%I think this is irrelevant given the empty network!!
%The variable $w$ at the initial time is assumed to be constant on each road and we fix $w=\wl$ for the roads inside the roundabout (roads 2, 4, 6 and 8), $w=(\wl+\wr)/2$ for the access roads to the network (roads 1 and 5),  and $w=\wr$ for the roads leaving the network (roads 3 and 7). Therefore, we assume a quiet behavior of drivers inside the roundabout, an average behavior in the accesses and an aggressive one when leaving the network.
%%%%%%%%%%%%%%%%%%%%%%%%%
We analyze three traffic scenarios determined by the density of vehicles which enter into the network from roads 1 and 5. On the latter, we used Dirichlet boundary conditions:
\begin{equation}\label{eq:bordoRho}
	\rho_{r,0}^{n} = \begin{cases}
		\bar\rho &\quad\text{if $t^{n}\leq 20\,\min$}\\
		0 &\quad\text{otherwise}
	\end{cases} \quad\quad \bar\rho = 15, 40 \mbox{ or } 80\,\, \vehkm
\end{equation}
%with $\bar\rho$ equal to 15, 40 or 80 $\vehkm$. 
%At the boundary, we set 
and $w^{n}_{0,r}=(\wl+\wr)/2$ for $r=1,5$. We use Neumann boundary conditions for roads 3 and 7,
thus allowing all vehicles to exit the roundabout.
%Moreover, on roads 3 and 7 vehicles are free to leave the road. Therefore, we consider an 
The initially empty network is filled up for the first 20 minutes of simulation, then no more vehicles access the network until the final time $T = 1\,\myhour$. In this way, the emissions are measures both for loading
and unloading of the roundabout.\\
In Table \ref{tab:confrontoRot2} we show the optimal controls and the corresponding functionals values. We observe that $\funE$, $\funT$ and $\funET$ grow as the number of vehicles entering the network increases, both for priorities and traffic lights dynamics. In particular, in the case of $\bar\rho=15\,\vehkm$ in \eqref{eq:bordoRho}, the traffic lights dynamics produce 20\% lower emissions and 2\% lower travel times with respect to priorities. In congested situations, instead, the emissions are reduced by about 11\% in presence of traffic lights, while the travel times are 6\% longer compared to priority-ruled dynamics.
The higher levels of emissions associated to priorities can be observed also in Figure \ref{fig:rotatoriaEm2}, where we plot the emissions on each road of the network at different times. The emissions associated to traffic lights dynamics show an oscillating behavior which is not observed in the priorities case, see plots \subref{fig:em1}, \subref{fig:em2}, \subref{fig:em4} and \subref{fig:em5}. At the final time of the simulation, plots \subref{fig:em3} and \subref{fig:em6}, the emissions are close to 0 as nearly all vehicles have left the network.
Finally, in Figure \ref{fig:emTot2}, we show the change in time of the total emission rates in the whole network. The trend in emission rates is the same for the three cases: emissions rise as vehicles enter the network and then decrease to 0. The peak value grows as $\bar\rho$ increases. In Table \ref{tab:emTot} we report the total number of vehicles that enter the network for the three tests and the corresponding total amount of emissions produced with the two traffic dynamics. We observe that emissions are more than double when $\bar\rho=40\,\vehkm$ compared to $\bar\rho=15\,\vehkm$ and almost triple when $\bar\rho=80\,\vehkm$ with respect to $\bar\rho=15\,\vehkm$, while the difference between the case of $\bar\rho=40\,\vehkm$ and the one of $\bar\rho=80\,\vehkm$ is smaller.\\
%A similar analysis holds for the total number of vehicles accessing the network. 
To check the robustness of our results, we computed the minima of the functional $\funET$ for different values of the weights $c_{1}$ and $c_{2}$ in Appendix \ref{appendice}. The specific values of the functional obviously varies as we change the weights, but not the qualitative and quantitative comparison of priorities with traffic lights. Moreover, the optimal traffic light timing appears to be more robust than the optimal priorities.

\begin{table}[h!]
\centering
\small
\renewcommand{\arraystretch}{1.05}
\begin{tabular}{cccccc}\toprule
$\begin{array}{c}\bar\rho\\(\vehkm)\end{array}$ & \begin{tabular}{c}Optimal\\control\end{tabular} & Value & $\funE$ & $\funT$ & $\funET$\\
\midrule
\multirow{3}{*}{$15$} & $\beta_{J_{1}}, \beta_{J_{3}}$ & 0.50, 0.50 & 0.46 & 1.52 & 1.98\\\cline{2-6} 
& $\begin{array}{c}(t_{g},t_{r})_{J_{1}}\\ (t_{g},t_{r})_{J_{3}}\end{array}$ & $\begin{array}{c}62\,\mysecond, 26\,\mysecond \\27\,\mysecond, 47\,\mysecond \end{array}$ & 0.36 & 1.49 & 1.86 \\ \midrule
%%%%
\multirow{3}{*}{$40$} & $\beta_{J_{1}}, \beta_{J_{3}}$ & 0.34, 0.69 & 1.04 & 1.81 & 2.85\\\cline{2-6} 
& $\begin{array}{c}(t_{g},t_{r})_{J_{1}}\\ (t_{g},t_{r})_{J_{3}}\end{array}$ & $\begin{array}{c}69\,\mysecond, 29\,\mysecond \\27\,\mysecond, 44\,\mysecond \end{array}$ & 0.92 & 1.91 & 2.84 \\ \midrule
%%%%
\multirow{3}{*}{$80$} & $\beta_{J_{1}}, \beta_{J_{3}}$ & 0.34, 0.68 & 1.15 & 1.88 & 3.02\\\cline{2-6}
& $\begin{array}{c}(t_{g},t_{r})_{J_{1}}\\ (t_{g},t_{r})_{J_{3}}\end{array}$ & $\begin{array}{c}69\,\mysecond, 29\,\mysecond \\27\,\mysecond, 44\,\mysecond \end{array}$ & 1.03 & 1.99 & 3.02 \\ 
\bottomrule 
\end{tabular}
\caption{Comparison of $\funE(\gamma)$, $\funT(\gamma)$ and $\funET(\gamma)$ for $\gamma$ chosen as the optimal controls on the junctions $J_{1}$ and $J_{3}$ of the network for different Dirichlet boundary conditions. }
\label{tab:confrontoRot2}
\end{table}

\begin{table}[h!]
\centering
\small
\begin{tabular}{cccc}\toprule
$\begin{array}{c}\bar\rho\\(\vehkm)\end{array}$ & \begin{tabular}{c}Total\\vehicles\end{tabular} & \begin{tabular}{c}Total emissions\\ priorities (g/h)\end{tabular} &\begin{tabular}{c}Total emissions\\ traffic lights (g/h)\end{tabular} \\\midrule
$15$ &  \phantom{1}620 & \phantom{1}576328  & \phantom{1}458563 \\
$40$ &\phantom{1}961 & 1316544  & 1169322 \\
$80$ &1012 & 1450627 & 1299261 \\
\bottomrule
\end{tabular}
\caption{Total number of vehicles entering the network and total amount of emissions produced for the three cases analyzed.}
\label{tab:emTot}
\end{table}

%%%%--------------------------------%%%
\section{Conclusions}\label{sec:conclusioni}
In this work, we have extended the Generic Second Order Model to a road network with merge and diverge
junctions and proposed a tool to estimate and minimize traffic emissions by regulating traffic
dynamics. Such regulation corresponds to the choice of suitable model parameter $\gamma$ that governs the distribution of traffic in a diverge and priorities in a merge.\\
%Moreover, the use of second order models allowed us to study the sensitivity of the minimization problem with respect to the property of the drivers $w$. 
Different scenarios have been considered, such as: a traffic policeman who strictly enforces the priority rule (RP algorithm), an uncontrolled intersection where drivers tend to maximize the flow (AP algorithm), and the presence of a traffic light. A functional measuring emissions and travel times was tested numerically
on a single merge junction, showing that the minimum is achieved by giving the priority and a longer green traffic light to the incoming road with higher density and fast drivers.  
On the other hand, the test performed on a roundabout has pointed out that traffic lights appear to be convenient with respect to priorities for emissions, especially at low densities. This indicates that the increasingly common roundabouts may benefit from the installation of traffic lights at entrances.
We conclude by stating that our approach is very flexible and can easily be used as a decision support for traffic management.

\appendix 
\section{Senstivity of $\funET$ to weights $c_{1}$ and $c_{2}$}\label{appendice}

In this appendix we investigate the sensitivity of the functional $\funET$ with respect to the weights $c_{1}$ and $c_{2}$ in \eqref{eq:c1c2} for the roundabout. 
Our aim is to compare the optimal controls obtained by giving more importance once to emissions and once to the travel time. Therefore, we define $\funET_{c_{1}}=\kappa c_{1}\funE+c_{2}\funT$ and $\funET_{c_{2}}= c_{1}\funE+\kappa c_{2}\funT$ with $\kappa=10,100$.\\
%Consider again the network described in Section \ref{sec:rotatoria}. 
In Tables \ref{tab:10} and \ref{tab:100} we report the optimal controls computed for $\funET_{c_{1}}$ and $\funET_{c_{2}}$, using the Dirichlet boundary conditions in \eqref{eq:bordoRho} for different $\bar\rho$ as in Section \ref{sec:rotatoria}. 
First, we observe that the values of the functional $\funET_{c_{1}}$ are lower than those of the functional $\funET_{c_{2}}$. Therefore, giving more importance to emissions rather than to travel time allows to reduce the total cost. Analogously to the case of functional $\funET$ studied in Section \ref{sec:rotatoria}, in all cases traffic lights dynamics are convenient in terms of emissions production, while the travel time is shorter when traffic is ruled by priorities. Finally, note that the optimal priorities are influenced by the choice of the functional, while the optimal traffic light timing is always the same for all the tests.

\begin{table}[h!]
\centering
\small
\setlength{\tabcolsep}{2pt}
\renewcommand{\arraystretch}{1.05}
\subfloat[][$\funET_{c_{1}}=10\funE+ \funT$]{
\begin{tabular}{cccccc}\toprule
$\bar\rho$ & \begin{tabular}{c}Optimal\\control\end{tabular} & Value & $\funE$ & $\funT$ & $\funET_{c_{1}}$\\
\midrule
\multirow{3}{*}{$15$} & $\beta_{J_{1}}, \beta_{J_{3}}$ & 0.50, 0.50 & 0.46 & 1.52 & 6.08\\\cline{2-6} 
& $\begin{array}{c}(t_{g},t_{r})_{J_{1}}\\ (t_{g},t_{r})_{J_{3}}\end{array}$ & $\begin{array}{c}62\,\mysecond, 26\,\mysecond \\27\,\mysecond, 47\,\mysecond \end{array}$& 0.36 & 1.49 & 5.12 \\\midrule 
%%%%
\multirow{3}{*}{$40$} & $\beta_{J_{1}}, \beta_{J_{3}}$ & 0.28, 0.77 & 0.10 & 1.83 & 12.13\\\cline{2-6} 
& $\begin{array}{c}(t_{g},t_{r})_{J_{1}}\\ (t_{g},t_{r})_{J_{3}}\end{array}$ & $\begin{array}{c}69\,\mysecond, 29\,\mysecond \\27\,\mysecond, 44\,\mysecond \end{array}$& 0.92 & 1.91 & 11.16 \\\midrule 
%%%%
\multirow{3}{*}{$80$} & $\beta_{J_{1}}, \beta_{J_{3}}$ & 0.45, 0.74 & 1.14 & 1.88 & 13.30\\\cline{2-6} 
& $\begin{array}{c}(t_{g},t_{r})_{J_{1}}\\ (t_{g},t_{r})_{J_{3}}\end{array}$ & $\begin{array}{c}69\,\mysecond, 29\,\mysecond \\27\,\mysecond, 44\,\mysecond \end{array}$& 1.03 & 1.99 & 12.26 \\

\bottomrule 
\end{tabular}
}
%%%%
%\,
%%%%%%%
\subfloat[][$\funET_{c_{2}}=\funE+10 \funT$]{
\begin{tabular}{cccccc}\toprule
$\bar\rho$ & \begin{tabular}{c}Optimal\\control\end{tabular} & Value & $\funE$ & $\funT$ & $\funET_{c_{2}}$\\
\midrule
\multirow{3}{*}{$15$} & $\beta_{J_{1}}, \beta_{J_{3}}$ & 0.50, 0.50 & 0.46 & 1.52 & 15.68\\\cline{2-6} 
& $\begin{array}{c}(t_{g},t_{r})_{J_{1}}\\ (t_{g},t_{r})_{J_{3}}\end{array}$ & $\begin{array}{c}62\,\mysecond, 26\,\mysecond \\27\,\mysecond, 47\,\mysecond \end{array}$& 0.36 & 1.49 & 15.29 \\\midrule 
%%%%
\multirow{3}{*}{$40$} & $\beta_{J_{1}}, \beta_{J_{3}}$ & 0.33, 0.67 & 1.04 & 1.81 & 19.17\\\cline{2-6} 
& $\begin{array}{c}(t_{g},t_{r})_{J_{1}}\\ (t_{g},t_{r})_{J_{3}}\end{array}$ & $\begin{array}{c}69\,\mysecond, 29\,\mysecond \\27\,\mysecond, 44\,\mysecond \end{array}$& 0.92 & 1.91 & 20.06 \\\midrule 
%%%%
\multirow{3}{*}{$80$} & $\beta_{J_{1}}, \beta_{J_{3}}$ & 0.34, 0.14 & 1.15 & 1.88 & 19.94\\\cline{2-6} 
& $\begin{array}{c}(t_{g},t_{r})_{J_{1}}\\ (t_{g},t_{r})_{J_{3}}\end{array}$ & $\begin{array}{c}69\,\mysecond, 29\,\mysecond \\27\,\mysecond, 44\,\mysecond \end{array}$& 1.03 & 1.99 & 20.94 \\
\bottomrule 
\end{tabular}
}
\caption{Comparison of $\funE(\gamma)$, $\funT(\gamma)$, $\funET_{c_{1}}(\gamma)$ and $\funET_{c_{2}}(\gamma)$ for $\gamma$ chosen as the optimal controls on junctions $J_{1}$ and $J_{3}$ for different boundary $\bar\rho\, (\vehkm)$ and $\kappa=10$.}
\label{tab:10}
\end{table}

\begin{table}[h!]
\centering
\small
\setlength{\tabcolsep}{1.8pt}
\renewcommand{\arraystretch}{1.05}
\subfloat[][$\funET_{c_{1}}=100\funE+ \funT$]{
\begin{tabular}{cccccc}\toprule
$\bar\rho$ & \begin{tabular}{c}Optimal\\control\end{tabular} & Value & $\funE$ & $\funT$ & $\funET_{c_{1}}$\\
\midrule
\multirow{3}{*}{$15$} & $\beta_{J_{1}}, \beta_{J_{3}}$ & 0.50, 0.50 & 0.45 & 1.52 & 47.08\\\cline{2-6} 
& $\begin{array}{c}(t_{g},t_{r})_{J_{1}}\\ (t_{g},t_{r})_{J_{3}}\end{array}$ & $\begin{array}{c}62\,\mysecond, 26\,\mysecond \\27\,\mysecond, 47\,\mysecond \end{array}$& 0.36 & 1.49 & 37.74 \\\midrule 
%%%%
\multirow{3}{*}{$40$} & $\beta_{J_{1}}, \beta_{J_{3}}$ & 0.26, 0.98 & 1.01 & 2.06 & 103.53\\\cline{2-6} 
& $\begin{array}{c}(t_{g},t_{r})_{J_{1}}\\ (t_{g},t_{r})_{J_{3}}\end{array}$ & $\begin{array}{c}69\,\mysecond, 29\,\mysecond \\27\,\mysecond, 44\,\mysecond \end{array}$& 0.92 & 1.91 & 94.34 \\\midrule 
%%%%
\multirow{3}{*}{$80$} & $\beta_{J_{1}}, \beta_{J_{3}}$ & 0.27, 0.98 & 1.12 & 2.17 & 114.10\\\cline{2-6} 
& $\begin{array}{c}(t_{g},t_{r})_{J_{1}}\\ (t_{g},t_{r})_{J_{3}}\end{array}$ & $\begin{array}{c}69\,\mysecond, 29\,\mysecond \\27\,\mysecond, 44\,\mysecond \end{array}$& 1.03 & 1.99 & 104.69 \\
\bottomrule 
\end{tabular}
}
%%%%
%\,
%%%%%%%
\subfloat[][$\funET_{c_{2}}=\funE+100 \funT$]{
\begin{tabular}{cccccc}\toprule
$\bar\rho$ & \begin{tabular}{c}Optimal\\control\end{tabular} & Value & $\funE$ & $\funT$ & $\funET_{c_{2}}$\\
\midrule
\multirow{3}{*}{$15$} & $\beta_{J_{1}}, \beta_{J_{3}}$ & 0.50, 0.50 & 0.46 & 1.52 & 152.67\\\cline{2-6} 
& $\begin{array}{c}(t_{g},t_{r})_{J_{1}}\\ (t_{g},t_{r})_{J_{3}}\end{array}$ & $\begin{array}{c}62\,\mysecond, 26\,\mysecond \\27\,\mysecond, 47\,\mysecond \end{array}$& 0.36 & 1.49 & 149.68 \\\midrule 
%%%%
\multirow{3}{*}{$40$} & $\beta_{J_{1}}, \beta_{J_{3}}$ & 0.33, 0.67 & 1.04 & 1.81 & 182.35\\\cline{2-6} 
& $\begin{array}{c}(t_{g},t_{r})_{J_{1}}\\ (t_{g},t_{r})_{J_{3}}\end{array}$ & $\begin{array}{c}69\,\mysecond, 29\,\mysecond \\27\,\mysecond, 44\,\mysecond \end{array}$& 0.92 & 1.91 & 192.32 \\\midrule 
%%%%
\multirow{3}{*}{$80$} & $\beta_{J_{1}}, \beta_{J_{3}}$ & 0.33, 0.56 & 1.15 & 1.88 & 188.93\\\cline{2-6} 
& $\begin{array}{c}(t_{g},t_{r})_{J_{1}}\\ (t_{g},t_{r})_{J_{3}}\end{array}$ & $\begin{array}{c}69\,\mysecond, 29\,\mysecond \\27\,\mysecond, 44\,\mysecond \end{array}$& 1.03 & 1.99 & 200.13 \\
\bottomrule 
\end{tabular}
}
\caption{Comparison of $\funE(\gamma)$, $\funT(\gamma)$, $\funET_{c_{1}}(\gamma)$ and $\funET_{c_{2}}(\gamma)$ for $\gamma$ chosen as the optimal controls on junctions $J_{1}$ and $J_{3}$ for different boundary $\bar\rho\, (\vehkm)$ and $\kappa=100$.}
\label{tab:100}
\end{table}

%\section*{Acknowledgments}
%We would like to acknowledge the assistance of volunteers in putting
%together this example manuscript and supplement.

\bibliographystyle{siam}
\bibliography{references_complete}
\end{document}